\documentclass[article,longauth]{aa} 
\usepackage{graphicx}
\usepackage{natbib}
\usepackage{multirow}
\usepackage{amsmath}
\usepackage{txfonts}  
\usepackage{orcidlink}
\usepackage{placeins}

\def\deg{$^{\circ}$ }
\def\degb{^{\circ}}

\newcommand{\anthony}[1]{\textcolor{black}{#1}}

\def\filterCnt{CntH$\mathrm{\alpha}$ }
\def\filterHa{NH$\mathrm{\alpha}$ }
\def\Ha{H$\mathrm{\alpha}$ }
\begin{document}

\author{
    Anthony Boccaletti\orcidlink{0000-0001-9353-2724}\inst{\ref{lira}}, 
    Emmanuel Di Folco\orcidlink{0009-0009-9618-4927}\inst{\ref{lab}},
    Anne Dutrey\orcidlink{0000-0003-2341-5922}\inst{\ref{lab}},
    Tang Ya-Wen, \orcidlink{0000-0003-4118-8751}\inst{\ref{taipei}},
    Stephane Guilloteau\orcidlink{0000-0003-3773-1870}\inst{\ref{lab}},
    Thomas Collin-Dufresne\orcidlink{0009-0008-9128-0330}\inst{\ref{lab}},
    Anne-Marie Lagrange\orcidlink{0000-0002-2189-2365}\inst{\ref{lira}}, 
    Eric Pantin\orcidlink{0000-0001-6472-2844}\inst{\ref{cea}}, 
    Jeffrey S. Bary\orcidlink{0000-0001-8642-5867}\inst{\ref{colgate}},
    Nuria Huélamo\orcidlink{0000-0002-2711-8143}\inst{\ref{cabinta}}, 
    József Varga\orcidlink{0000-0003-4989-575X}\inst{\ref{konkoly}},
    Julien Milli\orcidlink{0000-0001-9325-2511}\inst{\ref{ipag}}, 
    Vincent Piétu\orcidlink{0009-0006-3497-397X}\inst{\ref{iram}}, 
    William Danchi\orcidlink{0000-0002-9209-5830}\inst{\ref{nasag}}, 
    Bin Ren\orcidlink{0000-0002-9821-5864}\inst{\ref{oca},\ref{xiamen}}, 
    Clément Baruteau\orcidlink{0000-0002-2672-3456}\inst{\ref{irap}}, 
    Mickael Bonnefoy\orcidlink{0000-0001-5579-5339}\inst{\ref{ipag}},
    Tracy Beck \orcidlink{0000-0002-6881-0574}\inst{\ref{stsci}},
    Maud Langlois \orcidlink{0000-0003-1079-8346}\inst{\ref{cral}},
    Sylvestre Lacour\orcidlink{}\inst{\ref{lira}}, 
    Bruno Lopez\orcidlink{}\inst{\ref{oca}},
    Alexis Matter\orcidlink{}\inst{\ref{oca}},
    Julien Woillez\orcidlink{}\inst{\ref{eso}},
    Florentin Millour\orcidlink{}\inst{\ref{oca}},
    Matthis Houllé\orcidlink{}\inst{\ref{ipag}},
    Philippe Berio\orcidlink{}\inst{\ref{oca}}
}
\institute{
LIRA, Observatoire de Paris, Universit{\'e} PSL, CNRS, Sorbonne Universit{\'e}, Univ. Paris Diderot, Sorbonne Paris Cit{\'e}, CY Cergy Paris Université, 5 place Jules Janssen, 92195 Meudon, France\label{lira}
\and Laboratoire d'Astrophysique de Bordeaux, Université de Bordeaux, CNRS, B18N, Allée Geoffroy Saint-Hilaire, 33615, Pessac, France\label{lab}
\and Academia Sinica, Institute of Astronomy and Astrophysics, 11F of AS/NTU Astronomy-Mathematics Building, No.1, Sec. 4, Roosevelt Rd, Taipei, Taiwan \label{taipei}
\and  Laboratoire CEA, IRFU/DAp, AIM, Université Paris-Saclay, Université Paris Diderot, Sorbonne Paris Cité, CNRS, F-91191 Gif-sur-Yvette, France \label{cea}
\and Colgate University, 13 Oak Dr., Hamilton, NY 13346, USA)\label{colgate}
\and Centro de Astrobiología (CAB), CSIC-INTA, ESAC Campus, Camino bajo del Castillo s/n, 28692, Madrid, Spain\label{cabinta}
\and  Konkoly Observatory, HUN-REN Research Centre for Astronomy and Earth Sciences, MTA Centre of Excellence, Konkoly-Thege Miklós út 15-17, 1121 Budapest, Hungary\label{konkoly}
\and  CNRS, IPAG, Univ. Grenoble Alpes, F-38000 Grenoble, France\label{ipag}
\and IRAM, 300 rue de la piscine, Domaine Universitaire, 38406 Saint-Martin d'H\`eres, France\label{iram}
\and NASA Goddard Space Flight Center, Astrophysics Division, Greenbelt, MD, 20771, USA\label{nasag}
\and Observatoire de la Côte d'Azur, 96 Bd de l'Observatoire, 06304 Nice, France\label{oca}
\and  Department of Astronomy, Xiamen University, Xiamen, China.\label{xiamen}
\and IRAP, Université de Toulouse, CNRS, Université Paul Sabatier, CNES, Toulouse, France\label{irap}
\and  Space Telescope Science Institute, 3700 San Martin Drive, Baltimore, MD, 21218, USA\label{stsci}
\and Centre de recherche astrophysique de Lyon (CRAL), Université Claude Bernard Lyon 1, CNRS, ENS, 9 avenue Charles Andre, 69561 Saint Genis Laval, France
 \label{cral}
\and European Southern Observatory, Karl-Schwarzschild-Straße 2, 85748, Garching, Germany \label{eso} 
}

 \abstract
   {The young disk around AB\,Aur features a complex assembly of spiral arms, several compact structures, and a protoplanet candidate, AB\,Aur\,b, suggesting ongoing planet formation in this young system. Because of its brightness and spatial extent, AB\,Aur represents a perfect laboratory for investigating the conditions under which planets start to form around intermediate-mass stars.}
   {In this paper, we present near-IR polarized images of the AB\,Aur disk at three epochs spanning 3.85 years with SPHERE/IRDIS, as well as \Ha images obtained with SPHERE/ZIMPOL at a single epoch. The purpose of this study is to analyze the dynamics of the entire disk and of the various structures in near-IR polarimetry, and to identify sources of \Ha emission to derive constraints on their mass accretion rate. }
   {We developed a method to measure the rotation of the disk as a function of the radius, covering physical separations from as close as $\sim$25\,au up to 400\,au. We applied this method to the global structure of the disk as well as to specific features of interest, including both extended or compact sources. For the compact sources, we performed orbital analyses. We also studied the variability of shadows seen as thin radial streaks. For the \Ha data, we extracted photometric measurements of several features and derived estimations of the accretion luminosities and mass accretion rates, assuming three different accretion models. }
   {The dynamical study in the near-IR shows that the disk globally follows Keplerian rotation, but we observe a departure from this behavior at radii smaller than $\sim60$\,au. At the smallest radius of $\sim$25\,au, we measure a deviation from Keplerian rotation as large as $\sim$12\deg over 3.85 years, demonstrating sub-Keplerian rotation. 
   The two bright spirals within the millimeter cavity have different dynamic trends, and we discuss their possible link with the identified planet candidates.
   We also discuss the implications of the non-Keplerian behavior, and we posit that it could be related to interactions with multiple protoplanets orbiting out of the disk plane on elliptical orbits.    
   Furthermore, the orbital analysis of the compact sources (labeled \texttt{f1}, \texttt{f2}, and \texttt{f3}) suggests that their orbital planes are significantly inclined with respect to the disk plane by several tens of degrees. 
   The variability of the shadows suggests that they are produced by optically thick regions located within $\sim$60\,au. 
   For the photometric analysis in H$\mathrm{\alpha}$, we derive a flux of about $8.22\times10^{-15}$ erg/s/cm$^2$ for the \anthony{entire} feature \texttt{f1}, but only $6.46\times10^{-16}$ erg/s/cm$^2$ at the location of AB\,Aur\,b, consistent with non-detection. If \texttt{f1} were a point source and the accretion remained constant for 1\,Myr, it would correspond to $\sim5-20$ Jupiter masses according to the magnetospheric accretion model or $\sim6-10$ Jupiter masses according to the boundary layer accretion model. We further discuss the non-detection of \Ha emission on AB\,Aur\,b. Finally, we discuss the binarity of the host star, in particular using Gaia measurements.}
   {AB\,Aur is a rare system in which the morphology and dynamics can be studied at a very high level of detail, contrasting with the generic picture of a young planet-forming disk. The excellent image quality of SPHERE, both in the near-IR and in the visible, allows us to track the disk rotation with unprecedented precision thanks to the stability of the instrument across several years and to study localized \Ha emissions in the disk.  Overall, these observations strongly argue for an active and complex phase of planet formation in this system.}
   

\authorrunning{A. Boccaletti et al.}
\title{Destructuring the disk of AB Aurigae: Dynamics and accretion}
\titlerunning{AB Aur: Dynamics and accretion}

\date{Received  6 March 2026 / Accepted  27 April 2026}

\maketitle

\section{Introduction}

The mechanisms that lead to the formation of protoplanets in the first million years of a planetary system lack observational data that would allow robust constraints on models proposing various formation pathways. However, in these early evolutionary stages ($<$5\,Myr) planets are still embedded in their parent protoplanetary disk and presumably are surrounded by their own circumplanetary material \citep{szulagyi_meridional_2022}. For this reason, previous attempts to detect protoplanets and their circumplanetary disks (CPDs) in the near-IR were significantly impacted by contamination of the circumstellar disk (CSD), which implies high levels of dust opacities.

Indeed, while planets reside in disk midplanes, direct imaging in the visible and near-IR detects mostly scattered light from the central star bouncing off dust grains in the upper layers of the disk. To overcome this issue, most searches with ground-based adaptive optics (AO) systems and high-contrast imagers have focused on transition disks with dust cavities. Cavities not only suggest the presence of planets or chains of planets carving the disk, but also guarantee a level of dust opacity which allows the detection of protoplanets orbiting in the disk midplane.
This strategy has proven effective for the planets PDS\,70\,b,c \citep{muller_orbital_2018, haffert_two_2019}, and more recently, WISPIT\,2\,b \citep{van_capelleveen_wide_2025}. Several other protoplanet candidates are still awaiting confirmation or remain debated because they are found inside complex disk structures or do not appear as unambiguous and reproducible point sources \citep{sallum_accreting_2015,wagner_thermal_2019,gratton_blobs_2019,currie_images_2022}. However, considering that disk morphology and dynamics are considerably influenced by the presence of planets, these systems are valuable for studying the conditions and environments in which planets form and for testing different planet formation scenarios. 

In this context, due to its brightness and spatial extension, which make it an ideal target for high contrast imaging observations,
AB\,Aurigae (AB Aur) stands as a unique system with strong potential  for studying planet formation. It has been studied  with the most advanced instruments at high spatial resolution and across a large spectral range, delivering observations that are sensitive to the properties of gas and dust. 
On the one hand, there is particular interest in analyzing the complex spiral structures in this disk, both in thermal emission at submillimeter wavelengths \citep{tang_circumstellar_2012,tang_planet_2017, fuente_probing_2017} and in scattered light in the near-IR \citep{boccaletti_possible_2020, jorquera_large_2022, dykes_scexaocharis_2024}. On the other hand, identifying planetary-mass objects responsible for the disk structures is an active field of research, which has led to the detection of the protoplanet candidate AB\,Aur b \citep{currie_images_2022}. 

One of the striking characteristics of the AB\,Aur disk is its morphology, which harbors many individual spiral arms that wind from north to west. Several types of spirals have been identified, including those originally imaged by \citet{fukagawa_spiral_2004}, which lie outside the cavity defined by the submillimeter ring located at {about 1\,arcsec (or $\sim 160$\,au)}, as measured in the dust continuum at 1.3\,mm \citep{tang_planet_2017}. Some of these spirals are clearly located outside the disk plane and trace material falling onto the disk \citep{tang_circumstellar_2012, speedie_mapping_2025}. 
In contrast, the spirals located inside the cavity have attracted more interest, as they may have been caused by gravitational disturbances resulting from the presence of orbiting companions. In that respect, \citet{tang_planet_2017} propose two protoplanet candidates based on the morphology of the spirals observed for the very first time in the gas with 
ALMA, \anthony{the Atacama Large Millimeter/submillimeter Array}; 
one of these could be related to AB Aur\,b, later detected by \citet{currie_images_2022}, while the closer one coincides with the feature \texttt{f1} reported in \cite{boccaletti_possible_2020}. \citet{dong_how_2016} also proposed a planet candidate in a gap at about 0.5'' eastward of the star using hydrodynamical simulations based on earlier scattered-light images \citep{hashimoto_direct_2011}, but so far such an object has not been identified in observations with deeper contrast capabilities.

In the protoplanet phase the gas is assumed to accrete at free-fall velocity onto the central object, creating a shock at the photosphere that in turn produces emission in hydrogen lines, while the accretion geometry defines the flux of these lines \citep{marleau_accreting_2022}. The most popular is the \Ha line, which has motivated the construction of dedicated instruments or observing modes, including dual-band imagers such as ZIMPOL \anthony{\citep[Zurich IMaging POLarimeter,]{schmid_spherezimpol_2018}}, VAMPIRES \anthony{\citep[Visible Aperture Masking Polarimetric Imager for Resolved Exoplanetary Structures,][]{norris_vampires_2015}}, MagAOX \anthony{\citep[Magellan Adaptive Optics eXtreme,][]{males_magao-x_2024}}, as well as spectro-imagers such as MUSE \anthony{\citep[Multi Unit Spectroscopic Explorer,][]{bacon_muse_2010}}. 
Although several surveys have yielded null results \citep{cugno_search_2019, zurlo_widest_2020, xie_searching_2020}, a few objects show clear signatures of  \Ha emission, including PDS\,70\,b,c \citep{haffert_two_2019}, WISPIT\,2\,b \citep{close_wide_2025}, and Delorme\,1\,b \citep{eriksson_strong_2020}, while again many other \Ha candidates remain debated. 
AB\,Aur\,b \citep{currie_images_2022}, detected both in the near-IR with CHARIS/SCExAO 
\anthony{\citep[Coronagraphic High Angular Resolution Imaging Spectrograph at Subaru Coronagraphic Extreme Adaptive Optics,][]{groff_first_2017}} and in H$\alpha$, falls into this category. 
In contrast to other objects, it is spatially extended, which is interpreted as scattered light scattered from the CSD of an embedded protoplanet. Its variability in \Ha suggests that accretion is ongoing \citep{bowler_h_2025}.
However, UV and optical observations with \anthony{the Wield Field Planetary Camera 3 on the Hubble Space Telescope (HST/WFPC3)} 
favor the case of a circumstellar structure \citep{zhou_hstwfc3_2022,zhou_uv-optical_2023}. 
Finally, \citet{currie_vltmuse_2025} obtained a spectrum of AB\,Aur\,b revealing an inverse P Cygni profile for the \Ha line (blueshifted emission component, redshifted absorption component) which they interpret as material infall, so compatible with accretion onto a protoplanet.

The present study provides additional observations along two lines: the morphology and dynamics of the spiral structures and the accretion rate extracted from \Ha observations. In section \ref{sec:obsIR} we describe the observations with \anthony{the IR Dual-band Imager and Spectrograph (IRDIS) in SPHERE (Spectro-Polarimetric High-contrast Exoplanet REsearch)} 
in near-IR polarimetry, including data reduction, disk dynamics and features, shadows, and a discussion. Section \ref{sec:zimpol} presents the results of SPHERE/ZIMPOL observations in the \Ha filters, including photometric measurements, accretion rates, and a discussion. 
Section \ref{sec:conclusions} concludes this work.
\begin{table*}   
\caption{SPHERE observation log} 
\begin{tabular}{l l l l c c c c c }     
\hline\hline                             
Date UT & prog. ID & instrument & filter(s) & Pol. cycle & Field rot. ($\degb$) & DIT (s)  & T$_{\mathrm{exp}}$ (s) & Seeing ($''$)   \\ 
\hline                    

   2019-12-17 & 0104.C-0157  & IRDIS-DPI & BB\_H  & 10 &25.1  & 32 & 5120 & $0.46-0.79$   \\  
   2021-11-04 & 108.227A.001 & IRDIS-DPI & BB\_J  & 7 & 17.3 & 32  & 3584 & $0.32-0.63$  \\ 
   2021-11-24 & 108.227A.002 & IRDIS-DPI & BB\_H  & 7 & 17.6 & 32 & 3584 & $0.42-0.68$\\  
   2022-01-02 & 108.227A.003 & IRDIS-DPI & BB\_Ks & 2 & 5.1 & 32 & 1024 & $0.55-0.768$\\ 
   2023-10-23 & 112.25CD.001 & IRDIS-DPI & BB\_H  & 10 & 25.0 & 32 & 5120 & $0.31-0.73$\\ \hline
   2021-10-29 & 108.227A.005 & ZIMPOL imaging  & CntH$\alpha$/NH$\alpha$ & -  &  28.6         &   15   & 5760 &   $0.35 - 0.68$  \\

\hline   
\end{tabular}
\tablefoot{From left to right: observation date, program ID, instrument mode, filter combination,  number of polarimetric cycles, field rotation in degrees, individual integration time per frame (DIT) in seconds, total exposure time in seconds ($T_{exp}$), and DIMM \anthony{(Differential Image Motion Monitor)}
seeing variations in arcseconds.}
\label{table:obs_log}
\end{table*}

\section{SPHERE/IRDIS polarimetric imaging}
\label{sec:obsIR}
\subsection{Observations and data reduction}

We observed AB\,Aur (V=7.05, H=5.06, K=4.23) with SPHERE/IRDIS in polarimetry \anthony{with the Differential Polarimetric Imaging mode} (DPI,12.25\,mas/pixel) using the BB\_H filter at three epochs in 2019 (period 104), 2021-2022 (period 108), and 2023 (period 112), approximately every two years. 
In the second epoch we also obtained data in two other filters : BB\_J and BB\_Ks. We used the pupil-tracking mode to optimize coronagraphic performance (the pupil is fixed with respect to the Lyot stop) and to minimize variations in pupil aberrations. The coronagraph for the shortest-wavelength filters, BB\_J and BB\_H, has a diameter of 185\,mas, while it is slightly larger for BB\_Ks (230\,mas). It uses a combination with an apodizer in the pupil following the principle of the apodized pupil Lyot coronagraph \citet{soummer_apodized_2005}, and \citet{guerri_apodized_2011}. In the DPI mode of IRDIS, data acquisition proceeds through several polarimetric cycles. A ``QU'' cycle corresponds to four rotations  of the half-wave plate to switch the orientation of the polarization states, which allows reconstruction of the Stokes Q and U vectors from the two beams on the detector (0$\degb$ and 45$\degb$ to derive Q$^+$ and Q$^-$, and 22.5$\degb$ and 67.5$\degb$ to derive U$^+$ and U$^-$). We obtained several data sets with the $K_s$ filter in P108, but here we present the one with the best quality. 

We reduced the data with the publicly available software IRDAP
\footnote{https://irdap.readthedocs.io} \citep[\anthony{IRDIS Data reduction for Accurate Polarimetry,}][]{holstein_polarimetric_2020}, which follows the procedure of \citet{boer_polarimetric_2020} and is adapted from \citet{schmid_limb_2006} to reject the unpolarized stellar signal through double-differencing. Instead of using the Q and U Stokes parameters, we calculated the azimuthal Stokes parameters $Q_{\phi}$ and $U_{\phi}$. In this convention, $Q_{\phi}$  represents the polarized intensity and $U_{\phi}$ the noise in the approximation of single scattering events. However, the $U_{\phi}$ image (Fig. \ref{fig:dpi_uphi}) indicates that multiple scattering is certainly taking place in AB\,Aur's disk. We assumed that this had no impact on disk dynamics, provided that we performed relative measurements. 

\begin{figure*}
    \centering
    \includegraphics[width=18cm]{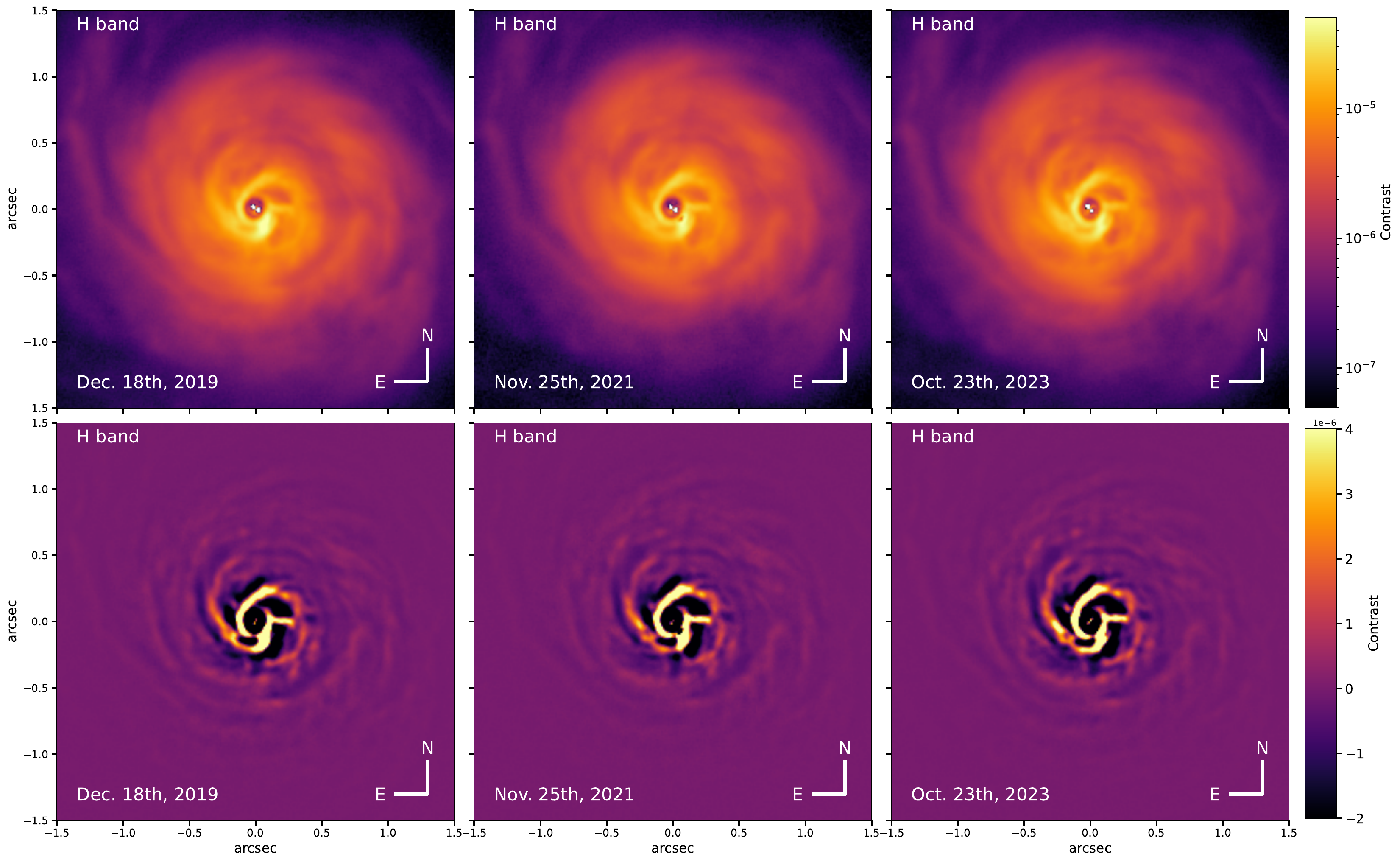}
    \caption{H-band images of AB Aur Q$_{\phi}$ in a $3''\times3''$ field of view at three epochs (December 2019, November 2021, and October 2023). The top row is displayed in logarithmic scale, while the bottom row is a high-pass filtered version in linear scale. The color bar shows the contrast obtained by normalizing with the out-of-mask point spread function image. North is up and east is left.}
    \label{fig:irdap_3epochs_Hband}
\end{figure*}

\subsection{General description}
\begin{figure}
    \centering
    \includegraphics[width=9cm]{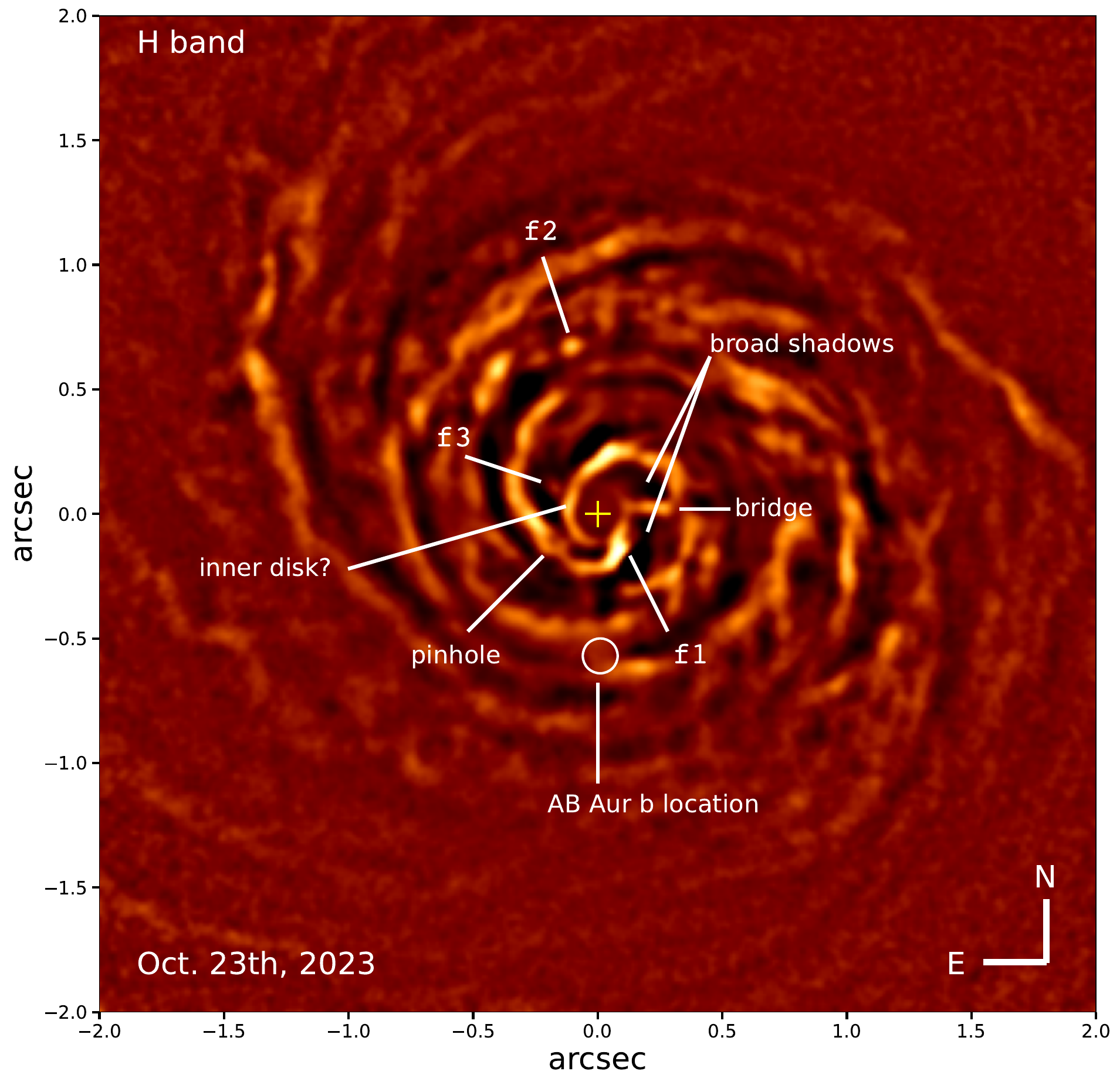}
    \includegraphics[width=9cm]{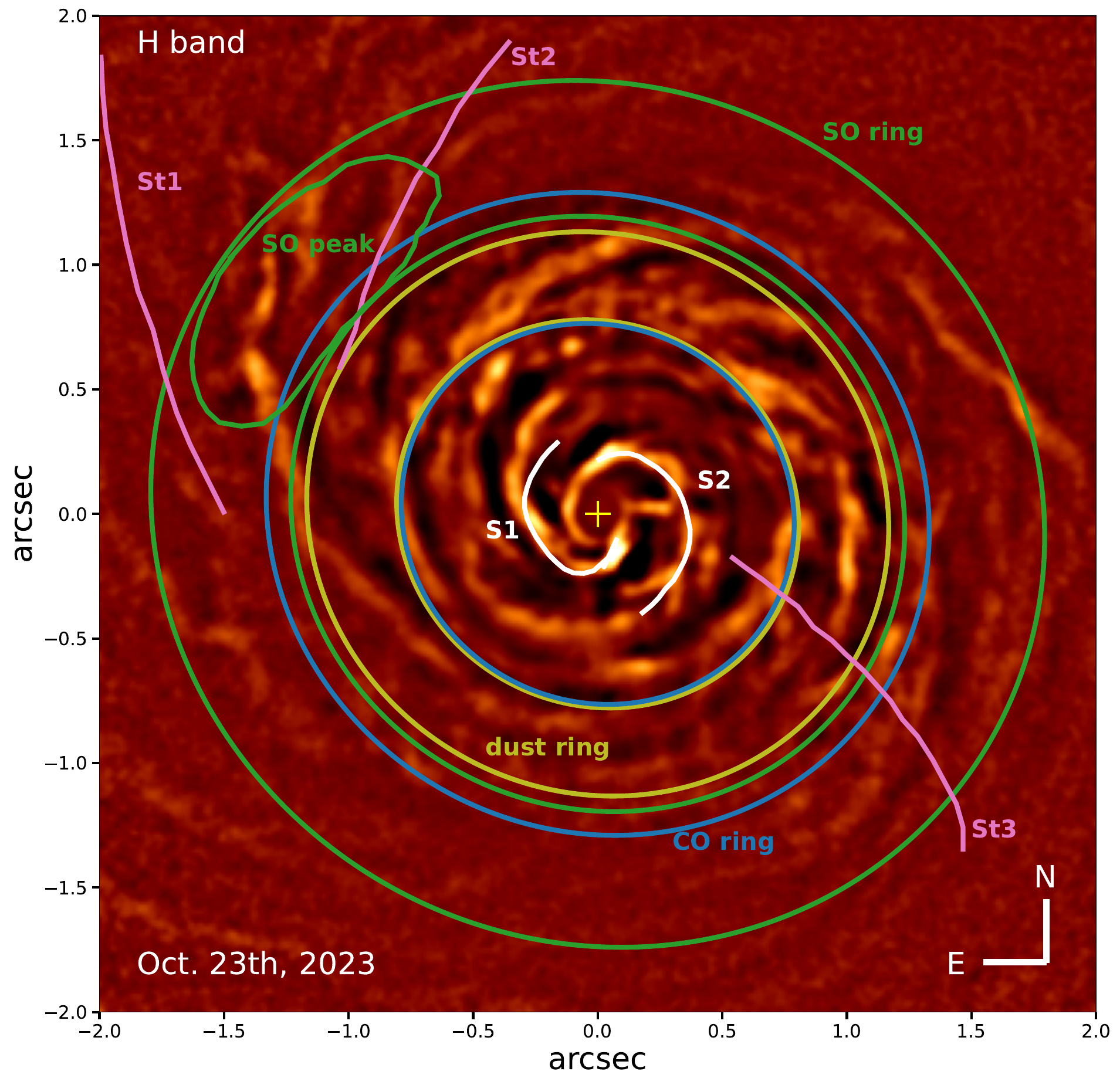}
    \caption{H-band images of AB Aur (October 2023), showing the main structures. Top panel: Scattered light features. Bottom panel: Millimetric features (yellow: dust ring; white: $^{12}$CO spirals S1 and S2 from \citet{tang_planet_2017}; green: SO ring and SO peak from \cite{dutrey_sulfur_2024}; blue: C$^{18}$O ring from \cite{dutrey_sulfur_2024} and pink: out of plane spirals (or streamers) St1, St2, and St3 from \citet{speedie_mapping_2025}).
    The central yellow cross indicates the star's position.  Images are normalized by $r^2$ and spatially filtered. North is up and east is left.}
    \label{fig:irdap_labels}
\end{figure}

Figure \ref{fig:irdap_3epochs_Hband} presents the $Q_{\phi}$ images for the three epochs in the H band, while Fig. \ref{fig:irdap_1epoch_JHKband} shows the second epoch only (P108), in the three bands. The top rows are displayed on a logarithmic scale, and the bottom rows are spatially filtered versions on a linear scale (obtained by subtracting a blurred image using a boxcar method with a width of 12 pixels to perform high-pass filtering and enhance the finest disk structures). Figure \ref{fig:irdap_labels} displays the spatially filtered image of epoch 3 (October 2023) multiplied by the square of the stellocentric distance ($r^2$) for visualization purposes (at this stage the image is not deprojected, so the $r^2$ multiplication does not rigorously represent the density variation). 

While the disk appears continuous in the submillimeter, it is more structured in the near-IR. To provide a synthetic view of the system, Fig. \ref{fig:irdap_labels} (bottom panel) overlays the SPHERE image with the main components observed with ALMA: the ``millimeter'' ring \citep{tang_circumstellar_2012}, the CO inner spirals \citep{tang_planet_2017}, the CO and SO rings \citep{dutrey_sulfur_2024}, and the out-of-plane infalling spirals or streamers \citep{speedie_mapping_2025}.

As already presented in \citet{boccaletti_possible_2020}, the most striking features in the SPHERE image are the inner spiral arms, labeled S1 (eastern spiral) and S2 (western spiral), and also the ``bridge,'' which extends to the west and links the inner region to S2 (see Fig. \ref{fig:irdap_labels} for labels). At the smallest angular separations, the two spirals wrap around the coronagraphic mask, possibly indicating the presence of the outer edge of the innermost disk detected in the mid-IR \citep{di_folco_flared_2009}. 
The regions between the spiral arms are relatively dark, especially in the west, reminiscent of the submillimetric dust continuum cavity and the contrast of the spiral arms in CO \citep{tang_planet_2017}. 

These spirals are rather clumpy, as seen in the spatially filtered images in Fig. \ref{fig:irdap_labels}, in particular with the brightest feature labeled \texttt{f1}, ``the twist,'' at the root of S1 (PA $\sim212\degb$), as previously reported in \citet{boccaletti_possible_2020}. 
Zoom-in images in Fig. \ref{fig:irdap_zoomin} highlight the evolution of the various features discussed below. 
From a twisted feature in 2019, \texttt{f1} evolves into a two-component structure, which merges the inward extremity of S1 with the edge of the inner disk. 
The H-band image is saturated near the position of \texttt{f1} in the second epoch.  
The S1 spiral arm is thicker at a position angle $PA=\sim123\degb$ due to a  pinhole-like feature. As demonstrated in \cite{boccaletti_possible_2020}, S1 strongly deviates from an Archimedean spiral in the proximity of \texttt{f1}. The S2 spiral arm is composed of two parts (S2a and S2b), a bright component at the shortest separations, and an abrupt change in intensity further out along the spiral. 
A dark region is visible beyond the bridge,
which may be consistent with a shadow cast on S2. 
We also recover the nearly point-like feature \texttt{f2} at all epochs. 
Compared with \citet{boccaletti_possible_2020}, we identify another source of interest, labeled \texttt{f3}, located at approximately the same radius as \texttt{f1}, but diametrically opposite with respect to the star. 

Finally, we do not detect the candidate protoplanet AB Aur\,b in polarimetry, consistent with the findings of \citet{currie_images_2022}. Figure \ref{fig:irdap_labels} shows its position using the average separation and position angle reported by \citet{zhou_uv-optical_2023}, which is the closest in time. More recently, \citet{kozdon_12_2026} also claimed the detection of a source based on $^{12}$CO spectroscopic observations, located at about 65\,au and PA\,$\approx 147\degb$, which is not coincident with the location of AB Aur b and has no obvious counterpart in near IR.

\subsection{Accuracy of the image orientation}
\begin{figure*}
    \centering
    \includegraphics[width=18cm]{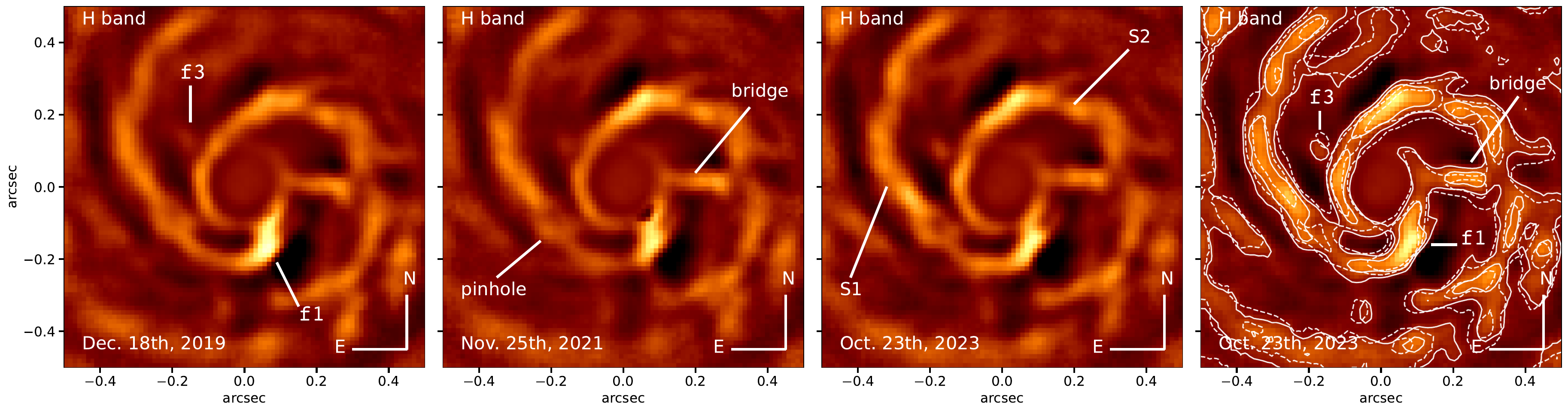}
    \caption{H-band images of AB Aur at three epochs in a $1''\times1''$ field of view. The rightmost panel displays the contours of epoch 1 (dashed line) and epoch 3 (solid line) superimposed on the image from epoch 3.
    Images are spatially filtered. North is up and east is left. }
    \label{fig:irdap_zoomin}
\end{figure*}

The quality of the data and the stability of the instrument allow a detailed dynamical analysis, particularly of the rotation of disk structures. We therefore first assessed the accuracy of the image orientation. 
\cite{maire_sphere_2016} demonstrated that the true north is oriented at $-1.75\pm0.08\degb$ with IRDIS in pupil-tracking mode. The pipeline corrects this absolute misalignment. To measure the relative uncertainty between the three epochs, we calculated the position of a background star (detected in total intensity but not in DPI) located at $\Delta \mathrm{\alpha}=-6.326''$ and $\Delta \mathrm{\delta} = 0.646''$, while the proper motion of AB Aur is $\mu\alpha=4.018$\,mas/yr and $\mu\delta=-24.027$\,mas/yr. We calculated a difference between the expected and the measured position angle of 0.03$\degb$ between epochs 1 and 2, and less than 0.01$\degb$ between epochs 2 and 3, so we adopted 0.03$\degb$ as the global absolute error on the image orientation.
Owing to this accurate calibration of the image orientation, disk rotation becomes an obvious characteristic when visually comparing the three epochs. To illustrate the counterclockwise motion, we overlay the contours of epoch 1 and epoch 3 in Fig. \ref{fig:irdap_zoomin} and we provide an animation online.

\subsection{Dynamics of the disk}
\label{sec:dynamical}
We first analyze the global dynamics of the disk. To this end, we deprojected the images to measure the dynamical evolution in the plane of the disk. To estimate the dispersion of the measurements, we considered  
variations of the disk $PA$ ($55\degb$ and $60\degb$) and disk inclination ($20\degb$, $25\degb$, and $30\degb$), based on the literature \citep{di_folco_flared_2009, tang_planet_2017, betti_detection_2022}. 
\citet{ren_dynamical_2020} devised a method to estimate the velocity of spiral arms that involves precisely locating the spiral spine  and fitting a spiral model. With the current data, this approach would be impractical because of the complex morphology of AB Aur, including the irregularity of the spirals, possible confusion between all structures, and uncertainty regarding the connection between features. Therefore, we propose an alternative approach in the following, which also has the advantage of measuring the velocity as a function of separation from the star. 

Because we expect differential rotation due to the Keplerian motion of the disk material, we divided the deprojected images into several rings. Their corresponding angular radii range from 0.1$''$ to 0.6$''$ in steps of 0.1$''$, and from 0.6$''$ to 2.4$''$ in steps of 0.2$''$, spanning physical distances of 15 to 370\,au in the disk plane. 
For each annulus, we solved for the optimal rotation and flux ratio between each pair of epochs that minimizes their quadratic difference. 
Before the minimization, the $Q_\phi$ images were high-pass filtered by subtracting a blurred boxcar image (width 12 pixels; 1 pixel = 12.25\,mas), and multiplied by $r^2$, the square of the stellocentric distance (this normalization has little effect because of the ring-wise breakdown). We plotted the angle of rotation for each separation bin and compared this quantity to the expected Keplerian rotation in the disk plane as a function of distance, assuming $M_*=2.4M_\odot$ \citep{dewarf_intrinsic_2003}. We verified that the scale height of the disk has little impact on this Keplerian profile. We calculated the error bars from the standard deviation of the six realizations of the parameter pairs (PA and inclination), quadratically added to the minimization error derived from the covariance matrix, and we forced the minimal error to be $0.03\degb$, corresponding to the image-orientation accuracy. 

\begin{figure}
    \centering
    \includegraphics[width=8cm]{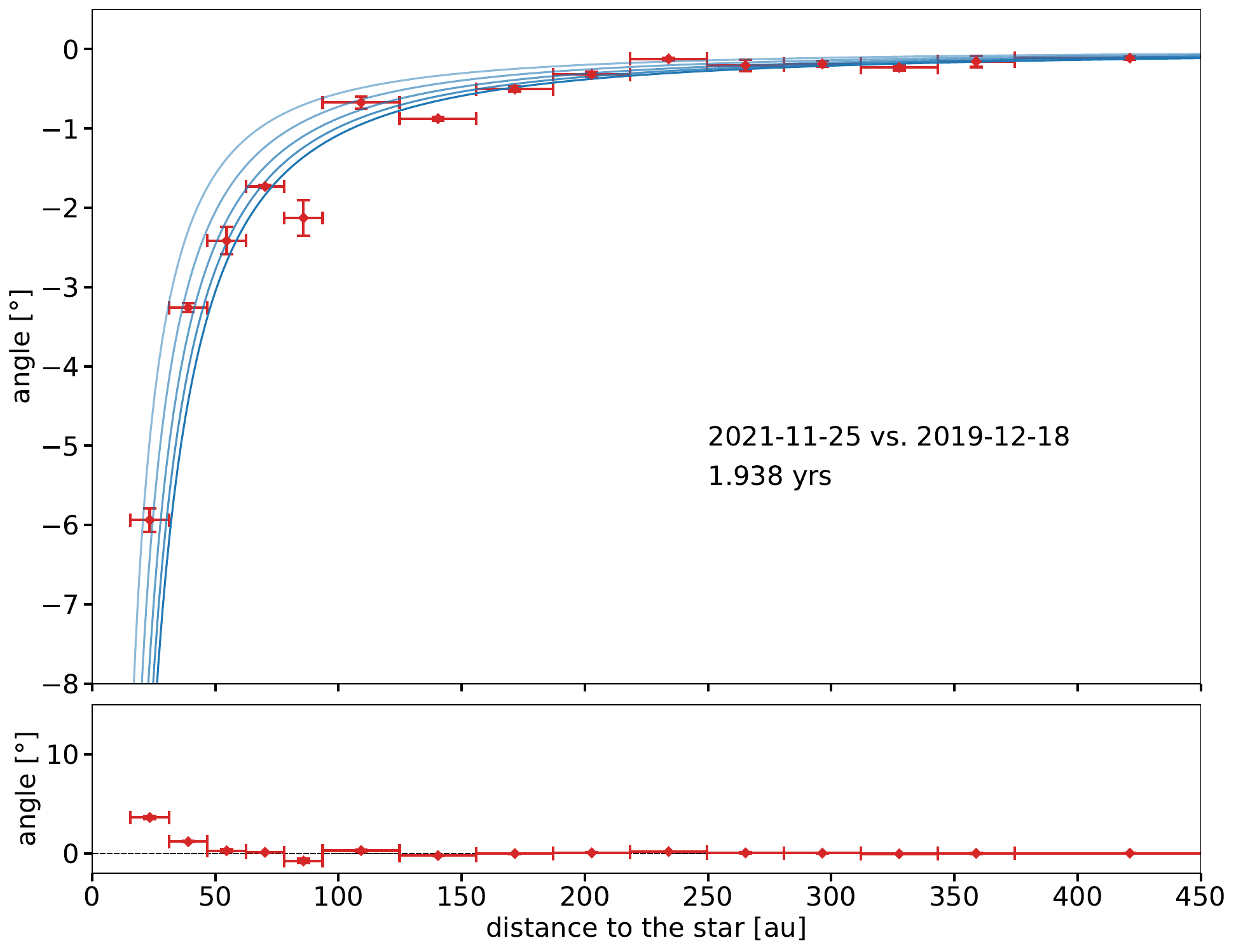}
    \includegraphics[width=8cm]{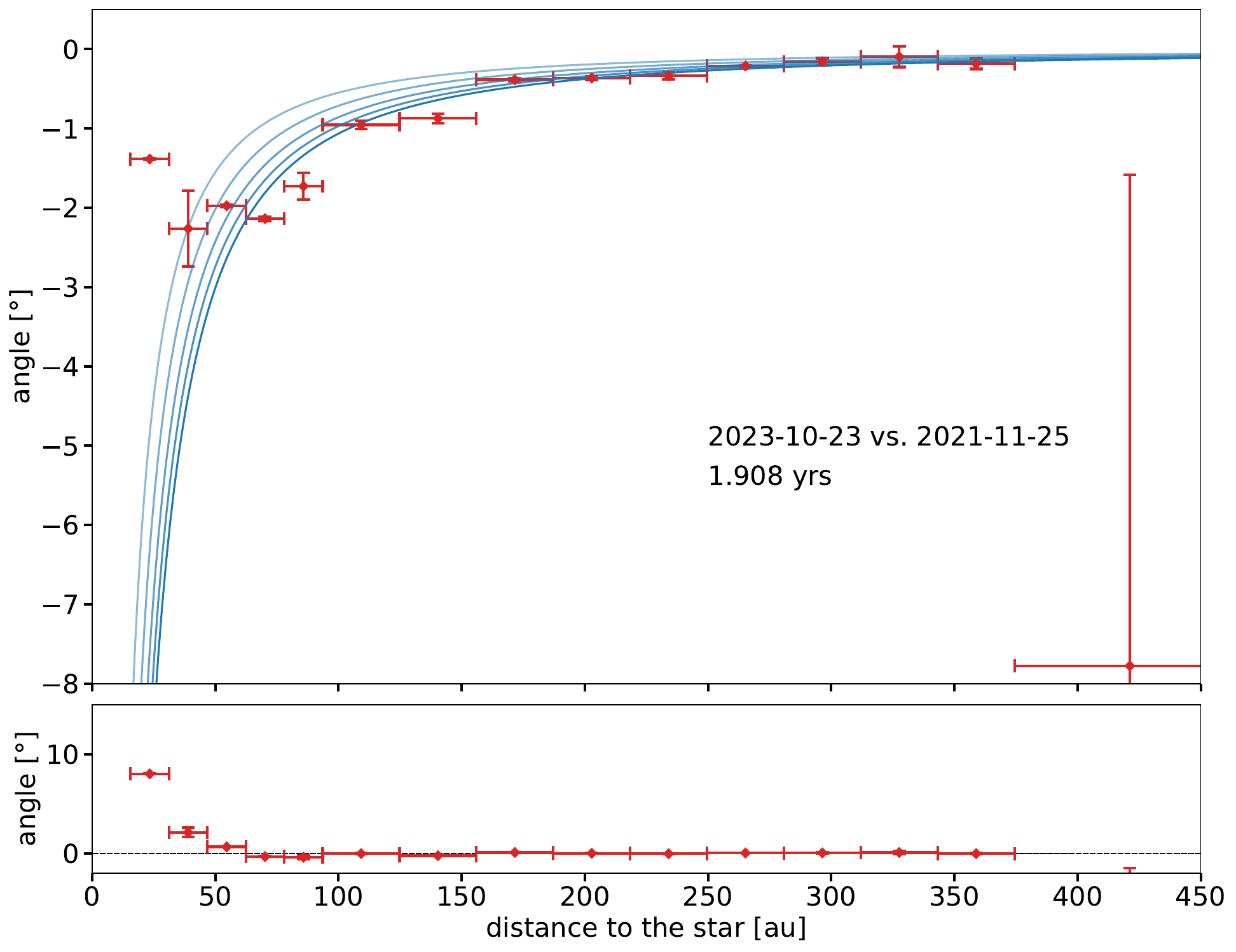}
    \includegraphics[width=8cm]{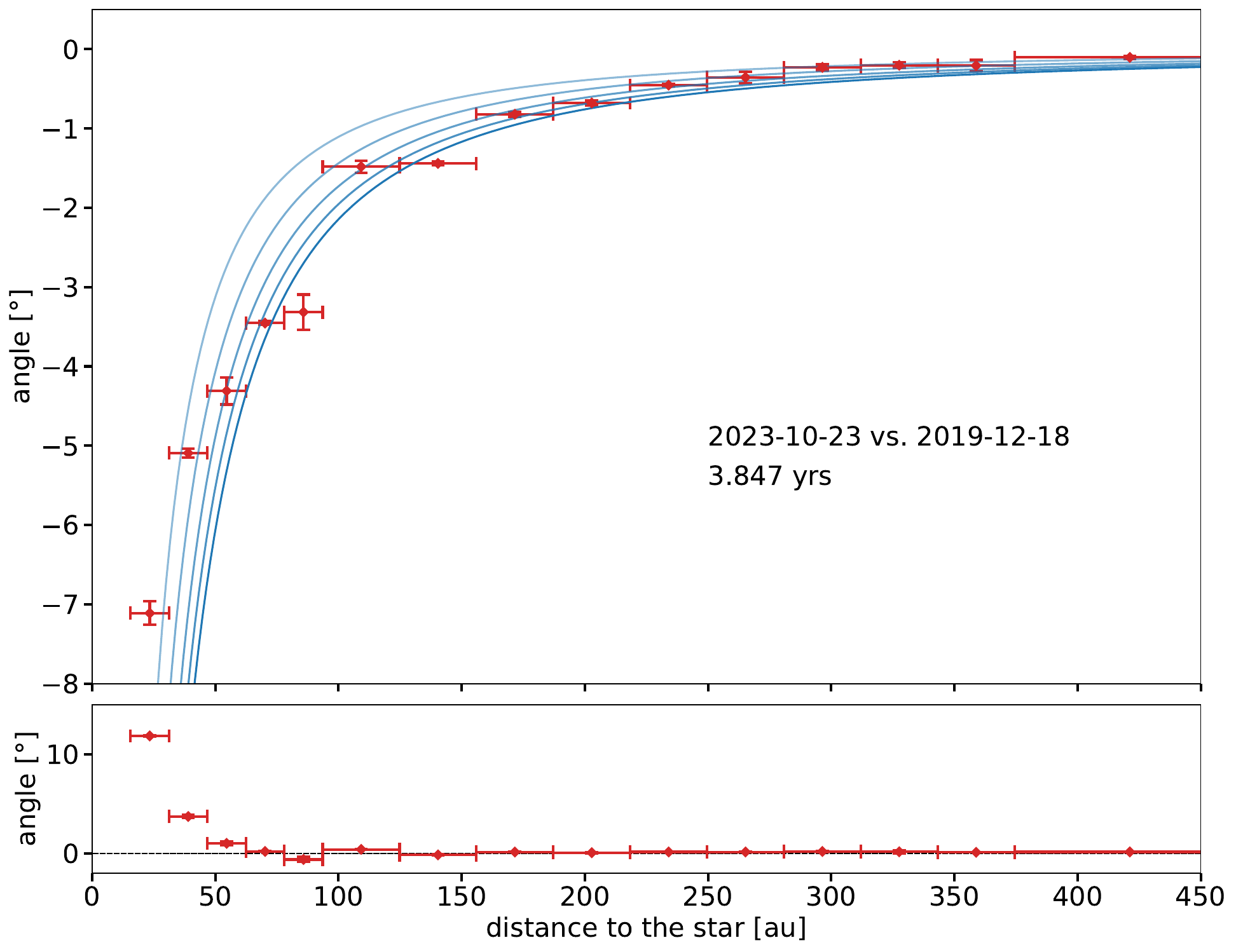}
    \caption{Disk rotation as a function of radius for each pair of epochs. The data points are shown in red, while the blue line represents the expected Keplerian rotation in the disk plane (solid) and for inclinations of 20$\degb$, 30$\degb$, 40$\degb$, and 50$\degb$ \anthony{(darkest to lightest lines)}. 
    The residuals between the data and the Keplerian are shown below each plot. From top to bottom: Epoch 2 vs. epoch 1, epoch 3 vs. epoch 2, and epoch 3 vs. epoch 1. The time interval is given in years.}
    \label{fig:rotation_nomask}
\end{figure}

Figure \ref{fig:rotation_nomask} presents the angle of rotation between two epochs. This analysis unambiguously demonstrates that the AB\,Aur disk globally follows counterclockwise rotation in the disk plane (hence the negative values). The amplitude of the rotation reaches about $-7\degb$ at the smallest achievable separations (about $15 - 30$\,au, equivalent to $0.1-0.2''$ in the sky plane) and converges to 0$\degb$ at the largest separations.  
The measurements agree well with the expected rotation amplitude for Keplerian motion (shown by the blue lines in Fig. \ref{fig:rotation_nomask}), particularly beyond a typical distance of $\sim$50\,au. However, the agreement is poorer inward, as seen in the residuals, indicating either that the disk rotates more slowly than the local Keplerian velocity or that temporal morphological changes do not allow for a sufficiently accurate measurement of the velocity. We performed the same analysis on each spiral arm separately using a mask to isolate the relevant pixels (results are shown in Fig. \ref{fig:rotation_withmasks}). Although we only probed the inner part of the disk ($<100$\,au) where the spirals S1 and S2 are located, we measure the same kind of behavior in the full image, with a deviation at physical distances smaller than 50\,au. This similarity between the whole disk behavior and that of the spirals can be explained by the high-pass filtering of the data, which already selects the spirals as the main structures in the image.

In addition to this ring-wise approach, we also measured the rotation of features as if they were each undergoing a solid-like rotation, using the same minimization methodology between two epochs. Tab. \ref{tab:rotation_spirals} reports the values, and the error bars correspond to the dispersion of the measurements for the six deprojection parameters. The feature names correspond to the masks shown in Fig. \ref{fig:rotation_withmasks}. 
We find that the spirals S1 and S2 behave differently, with the rotation of S2 progressing more linearly with time (about $-2.1\degb$ between the closest epochs in time and about $-4.5\degb$ over the longest temporal baseline), while it is more difficult to identify a clear trend for S1. Again, such measurements can be biased by the morphological evolution of the features, but they can also indicate different dynamical behaviors and possibly different origins for S1 and S2. 

\begin{table}   
\caption{Rotation of the main features.} 
\begin{tabular}{l l l l}     
\hline\hline                             
features & 2021 vs. 2019 & 2023 vs. 2021 & 2023 vs. 2019   \\ 
\hline                    
   S1a & $-3.48 \pm 0.02\degb$& $-1.07 \pm 0.74\degb$ & $-3.48 \pm 1.55\degb$  \\
   S1b & $-0.69 \pm 0.67\degb$ & $-2.16 \pm 0.04\degb$ & $-3.07 \pm 0.09\degb$  \\  \hline
   S2  & $-2.10 \pm 0.93\degb$ & $-2.36 \pm 0.06\degb$ & $-4.74 \pm 0.05\degb$  \\ 
   S2a & $-2.77 \pm 0.14\degb$ & $-1.88 \pm 0.83\degb$ & $-5.08 \pm 0.25\degb$  \\ 
   S2b & $-1.65 \pm 0.03\degb$ & $-2.00 \pm 0.02\degb$ & $-3.57 \pm 0.02\degb$  \\ \hline 
   bridge  & $-3.19 \pm 0.07\degb$ & $-2.30 \pm 0.04\degb$ & $-5.47 \pm 0.07\degb$  \\ \hline
\end{tabular}
\tablefoot{Rotation for each pair of epochs. See Fig. \ref{fig:rotation_withmasks} for the regions corresponding to the features.}
\label{tab:rotation_spirals}
\end{table}

\subsection{Dynamics of individual features}
\label{sec:dynamic_f1f2f3}
\begin{figure}
    \centering
    \includegraphics[width=8cm]{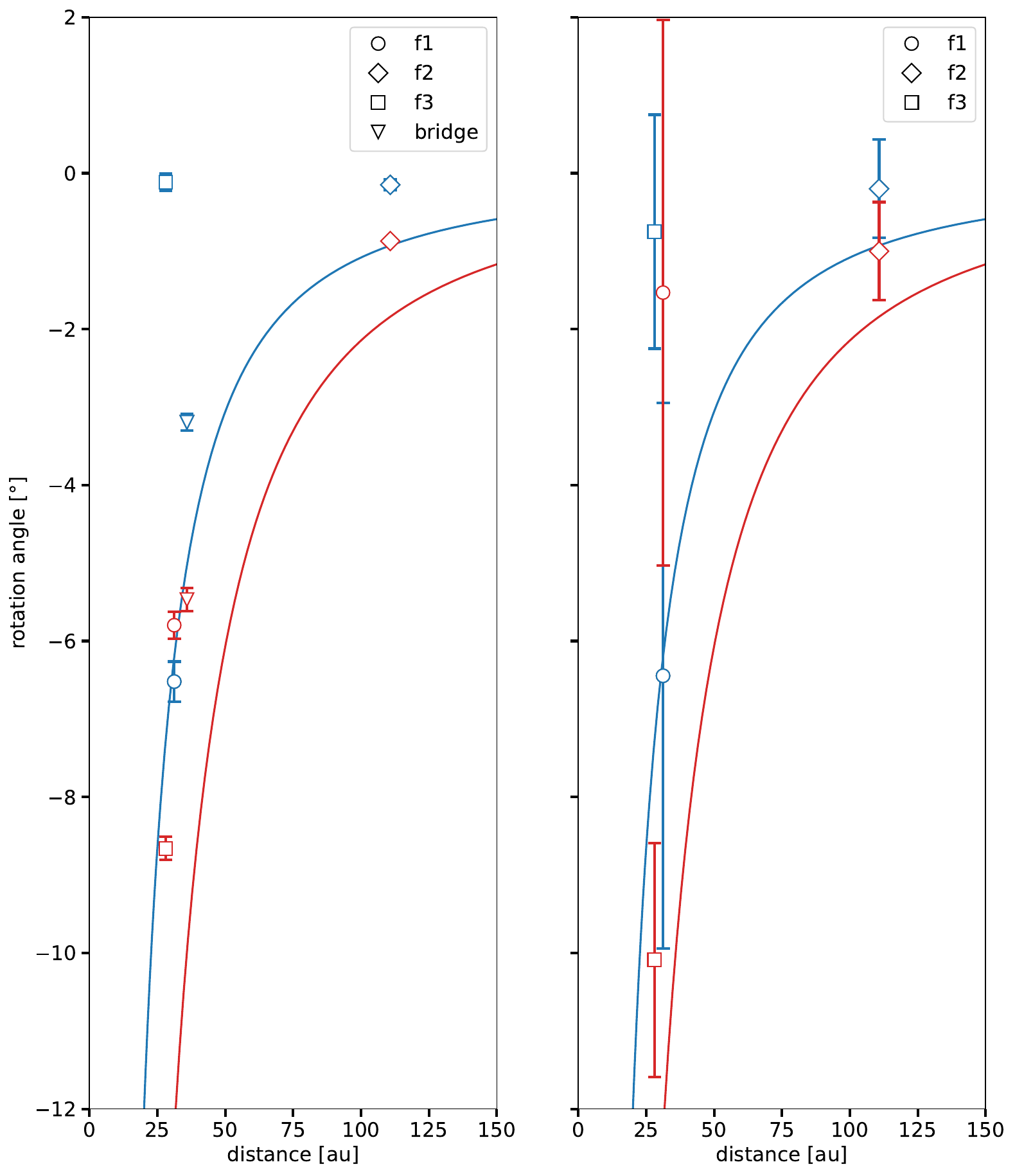}
    \caption{Rotation of the main disk features for two epoch pairs (2021 vs. 2019 in blue and 2023 vs. 2019 in red). The symbols correspond to \texttt{f1} (circle), \texttt{f2} (diamond), \texttt{f3} (square), and the bridge (triangle). Each curve shows the expected Keplerian rotation using the same color coding as the symbols. The left panel shows the results from the optimization method, while the right panel shows results from Gaussian fitting (only for \texttt{f1}, \texttt{f2,} and \texttt{f3}).}
    \label{fig:rotation_features}
\end{figure}

The disk of AB\,Aur is composed of several characteristic features that may have a different dynamical behavior than the disk itself. Measuring  their evolution is important to ultimately derive constraints on perturbating bodies. In this section, we specifically analyze the motion of \texttt{f1}, \texttt{f2}, and \texttt{f3} (located at 0.17'', 0,67'', and 0.19'' from the star, respectively), and the bridge. We proceeded similarly as for the disk: we isolated each feature with a numerical mask and computed the angle of rotation that minimizes the residuals between two epochs. For \texttt{f1}, \texttt{f2}, and \texttt{f3}, we also measured their position in each individual epoch by Gaussian fitting. This required tighter masking to avoid contamination with other extended spiral features of the disk, which inevitably yielded larger dispersions. However, we emphasize that both methods can be biased by the disk environment, even when the dispersion is small. Nevertheless, both methods are consistent, although the error bars are larger for the latter. Fig. \ref{fig:rotation_features} reports the measurements.

Visually, all features show a clear unambiguous counterclockwise rotation, except \texttt{f2}, which appears almost static. However, according to the analysis, none of the features perfectly follows the Keplerian rotation of the disk plane at all epochs. Although \texttt{f1} is compatible with the Keplerian motion for the \anthony{2021-2019} 
epoch pair (blue circle), it significantly deviates for the \anthony{2023-2019} pair (red circle). 
This behavior can be explained by changes in the shape of the feature itself, which appears more structured in 2023, with a more localized intensity peak. 
As for \texttt{f3}, which appears radially symmetrical to \texttt{f1}, its rotation is marginally compatible with the Keplerian motion for the \anthony{2023-2019} 
pair (red square), but appears static for the \anthony{2021-2019} 
pair (blue square). The latter result is very likely an issue with the isolation of the feature in the 2021 image. The bridge (triangles) exhibits a clear rotation at all epochs, but the speed is slower than the Kelperian motion. Finally, \texttt{f2} is the most puzzling because it is almost static, while the surrounding disk features actually vary in rotation. The measurements (diamonds) are also consistent with zero or small rotation amplitudes, although they are marginally compatible with Keplerian when using the Gaussian-fitting method because of larger error bars.      

We also performed orbital analyses for \texttt{f1}, \texttt{f2}, and \texttt{f3}. As described in Appendix \ref{appendix:dynam_features} and moderated by the small orbital coverage, the main outcome is that the features would likely orbit out of the disk plane with inclinations of $75\degb$, $38\degb$, and $34\degb$, or $32\degb$, $77\degb$, and $78\degb$, for \texttt{f1}, \texttt{f2}, and \texttt{f3}, respectively. The two families of solutions arise from the 180$\degb$ uncertainty in the position angle of the nodes. For each value, the dispersion of the posterior distribution can reach about 15$\degb$ to 20$\degb$, as shown in the corner plots (Fig. \ref{fig:cornerf1f2f3}).

\subsection{Radial shadows}
\begin{figure*}
    \centering
    \includegraphics[width=18cm]{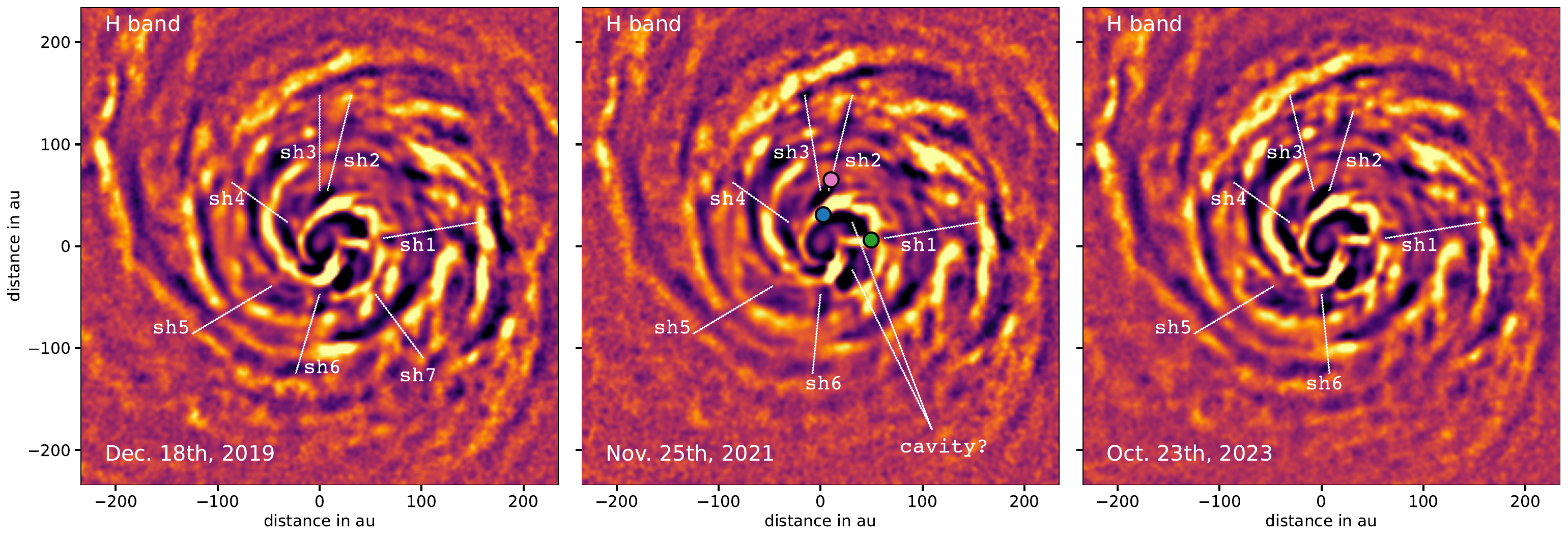}
    \caption{Deprojected H-band images of AB Aur for the three epochs. Dotted lines approximately trace the main shadow patterns identified at each epoch. Images are multiplied by $r^2$ and spatially filtered. The blue, green, and pink circles trace the potential location of clumps that could be casting the shadows.}
    \label{fig:irdap_shadows}
\end{figure*}

The disk of AB\,Aur exhibits several shadow patterns with different morphologies, which presumably correspond to different origins. In Fig. \ref{fig:irdap_labels}, we indicate two distinct  broad shadows in the west part of the image, located on either side of the bridge and bounded by spiral arm S1 and the root of S2 (also visible in Fig. \ref{fig:irdap_3epochs_Hband} prior to filtering). These may be connected to the inner cavity observed in the submillimeter. In this section, we focus instead on the radial, thin shadows that are barely visible in Figs. \ref{fig:irdap_3epochs_Hband} and \ref{fig:irdap_1epoch_JHKband}. They undergo azimuthal rotations, mostly revealed visually by blinking between epochs \anthony{(see the online animation)}, from which we can constrain the distance at which they originate. We label these nearly (but not exactly) radial shadows in Fig. \ref{fig:irdap_shadows}. Some are consistent across epochs but show measurable rotation, not necessarily centered on the star, while others show faster variability. 

\begin{figure*}
    \centering
    \includegraphics[width=18cm]{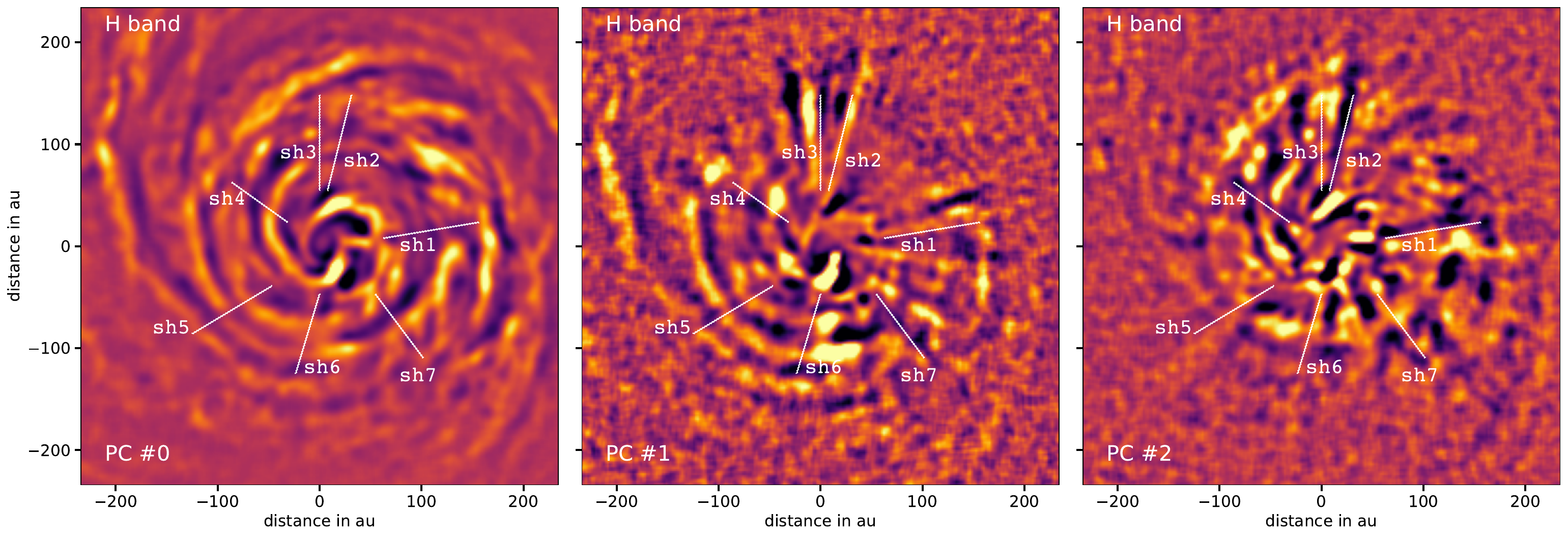}
    \caption{Principal component analysis (PCA) of the three images shown in Fig. \ref{fig:irdap_shadows}. Principal components 0, 1, and 2 are shown from left to right. The shadow positions at the first epoch (December 18, 2019) are superimposed.}
    \label{fig:irdap_shadows_pca}
\end{figure*}

To visually enhance the shadows, we applied a principal component analysis (PCA) to the set of three H-band images to identify static and moving features across the epochs. Qualitatively, we obtained the best visibility using deprojected, $r^2$ normalized, high-pass filtered images, as shown in Fig. \ref{fig:irdap_shadows_pca}, which displays the three principal components. While the first mode (PC\#0) represents the quasi-static main structures across the three epochs, the two other modes (PC\#1 and PC\#2) emphasize radial structures. We note that since the principal components are built on the temporal dimension, PC\#1 and PC\#2 do not directly trace the locations of the radial shadows, but provide an indication of the areas covered by the shadows over time. We describe the observed properties of the shadows in the following. 
\begin{itemize}
    \item The four features \texttt{sh1}, \texttt{sh4}, \texttt{sh5}, and \texttt{sh6} are broadly aligned with the bridge, the feature \texttt{f3}, the pinhole, and \texttt{f1}, respectively.
    \item The feature \texttt{sh3} shows a clear azimuthal rotation of about 12$\degb$ between the first and last epochs (3.847 years). Assuming Keplerian motion in the disk plane and a circular orbit, this is consistent with a shadow cast by an object or a local optically thick clump at an orbital distance of about 31\,au. Fig. \ref{fig:irdap_shadows} shows the approximate position of such a clump, marked by a blue circle.
    \item Feature \texttt{sh3} also overlaps with \texttt{f2}, producing a dimming of the feature that is strongest in 2021. Relative to its mean flux measured in spatially filtered images (unfiltered images are disk-dominated), \texttt{f2} varies by -14\%, -20\%, and +33\% in 2019, 2021, and 2023. This places a strong constraint on the nature of \texttt{f2}, which therefore cannot be a self-luminous object in the near-IR, and instead is a scattered-light feature.
    \item The shadow \texttt{sh1}, which extends the bridge, varies by 6$\degb$, equivalent to 50\,au, corresponding exactly to the distance where we detect the outer edge of the bridge (green circle in  Fig. \ref{fig:irdap_shadows}). 
    \item The shadow \texttt{sh2} has a lower rotation amplitude of about 4$\degb$, corresponding to a distance of 66\,au (pink circle in  Fig. \ref{fig:irdap_shadows}).    
    \item Feature \texttt{sh6} is located in the radial direction away from the hot spot mentioned by \citet{tang_planet_2017} and the feature \texttt{f1} identified by \citet{boccaletti_possible_2020}. It is also sweeping in the direction of AB Aur b. Whether it is related to its H$\alpha$ variability remains to be investigated \citep{bowler_h_2025}.
\end{itemize}

While Fig. \ref{fig:irdap_shadows} shows the possible locations of opaque clumps that are consistent with the azimuthal variation of the shadows, there are no associated clumpy features in the image. However, shadows may also be cast by a structure with some vertical extent, such as a spiral arm crossing the midplane. This is the case for sh3 (blue circle connected to the northern arm of the spiral S2) and for sh1 (green circle connected to the bridge).
Similar radial shadows have been observed in MWC\,758. Based on 3D hydrodynamical simulations, \citet{calcino_are_2020} conclude that the shadows were cast by optically thick regions of the spiral arms rather than by an inclined inner disk, as in HD\,142527 \citep{marino_shadows_2015}, consistent with what is observed in AB\,Aur. 
Shadows observed in scattered light can also be related to the local vertical extent of the disk, where a feature such as \texttt{f1}, or even the entire spiral arm, traces bumpiness on top of the disk average surface, thereby casting a shadow behind it.

Additionally, using the framework developed by \citet{akansoy_modelling_2025}, we tentatively estimated the mass of objects that could cast radial shadows. As the results are not fully conclusive, we report this analysis in Appendix \ref{appendix:shadows}.

\subsection{Discussion on dynamics}
\subsubsection{Keplerian rotation}

Our multi-epoch data set allows us to trace rotation patterns for a variety of disk structures. To our knowledge, we detect the global Keplerian rotation of the disk here for the first time through the investigation of scattered-light variations. Our analysis assumes that small-scale structures in the spatially filtered images persist across epochs. Our estimate of the rotation is broadly consistent with a Keplerian pattern around a 2.4\,M$_{\odot}$ central star down to a stellocentric distance of about 60\,au (Fig. \ref{fig:rotation_nomask}), whereas all three epoch combinations consistently show slower rotation in the inner regions of the disk cavity, down to the edge of the coronagraphic mask (also in Fig. \ref{fig:rotation_withmasks}).  \citet{tang_planet_2017} report Keplerian patterns for the inner eastern and western CO spirals S1 and S2 using Doppler velocity measurements with ALMA (see their Fig. 3), whereas our IR observations trace motions projected onto the sky plane. Although planet-induced spirals are expected to rotate at the orbital speed of the planetary companion, Keplerian motion is instead expected for spirals induced by gravitational instabilities \citep{cossins_characterizing_2009, yoshida_winding_2025}. This may be particularly relevant for the large-scale spirals beyond $\sim 150$\,au. 

 \citet{speedie_gravitational_2024} recently proposed a gravitational instability scenario to account for wiggles occurring in the velocity field of outer CO spirals beyond $\sim160$\,au (outside the cavity) and argue that the disk mass may represent a third of the stellar mass. In that case the estimate of the Keplerian velocity would be biased, as the disk mass would be smaller at smaller stellocentric distances, which could effectively slow the rotation of the disk. However,  \citet{calcino_anatomy_2025} dispute this interpretation and find that late infall produces signatures similar to those of gravitational instability. This scenario may apply to the outermost spiral arms of the disk. 
Moreover, the distance at which the disk rotation starts to depart from a Keplerian motion corresponds to the inner disk cavity, where both gas and millimeter-size grains are depleted. 
Furthermore, most spiral motions investigated through multi-epoch measurements have so far favored the companion-disk interaction scenario over gravitational instability \citep[e.g.,][]{xie_dynamical_2023}. As a sanity check, we used the approach in Fig. \ref{fig:rotation_nomask} to estimate the stellar mass that can cancel out the residuals at the smallest physical separations of about 50\,au. We find values close to 1 M$_{\odot}$ (or even lower) for the longest temporal baseline, which is unrealistic given the spectral type of the star. Therefore, the measured disk rotation does not seem to support a strong impact from the disk mass.

\subsubsection{Deviations from Keplerian rotation}

In the central regions of the millimeter cavity, \anthony{the 
velocity} discrepancy may arise either from a true variation in the angular frequency or from artifacts. If the observed slower rotation has a physical origin, it may be linked to interactions with one or several gravitational perturbers located within the cavity. 
In the case of a companion on a circular orbit, the entire spiral arm rotates at the perturber's frequency \citep[e.g.,][]{xie_disk_2024}. Because the positive shifts of the residuals of the rotation angles in Fig. \ref{fig:rotation_nomask} have different amplitudes for the three innermost annuli (the largest deviation is usually  found for the innermost annulus around 22\,au), a single gravitational perturber can hardly account for the observed trends. Within 60\,au, our measurement is greatly influenced by the presence of three main bright structures: S1, S2, and the bridge. We therefore investigated the motion of the inner spirals S1 and S2 separately using different numerical masks (Fig. \ref{fig:rotation_withmasks}), and we also performed the analysis assuming a single rotation speed for each structure (Tab. \ref{tab:rotation_spirals}). 
If the S1 spiral were generated by interactions with a perturber located at \texttt{f1} on a circular orbit, it should rotate much faster than the local Keplerian value, in contrast to our findings. Conversely, if S2 were generated by interactions with a perturber located at AB\,Aur\,b, it should rotate slower, at the Keplerian speed corresponding to $a\simeq 75$\,au 
\citep[according to the preliminary orbital fit proposed in][]{currie_images_2022}. Only S2 yields a consistent angular velocity across the three epochs, with a mean value of about 1.2\,deg.yr$^{-1}$ (Tab. \ref{tab:rotation_spirals}). Therefore, we cannot completely exclude a dynamical link between S2 and the AB~Aur b candidate, whereas the motion of S1 appears too slow to be connected with the \texttt{f1} candidate.
We also note that we did not identify a protoplanet candidate in the wake of S2 that could correspond to the \citet{dong_how_2016} prediction.

The situation is more complex if the perturber's orbit is eccentric. 
A planet on an eccentric orbit can generate multiple spiral arms with significant deviations from Keplerian speed and with angular frequencies that depend on the orbital phase 
\citep[Fig. 11 in ][]{zhu_simple_2022}. 
In this respect, the presence of detached or broken spiral arms and bifurcations in AB~Aur is also reminiscent of the structures expected from eccentric planets, but the complicated patterns make their identification quite challenging.

Alternatively, deviations from Keplerian speed could also be attributed to the following artifacts: (1) the ability to measure rotation close to the star at very short angular separations; (2) structural or illumination changes with a shorter timescale closer to the star, which would bias our measurement process (although this would hardly explain the systematic positive shift in the measured rotation angles); and (3) deprojection effects. 
We emphasize that any structure that is not coplanar with the outer disk midplane would appear closer to the star due to simple projection effects and may artificially produce a slower apparent rotation. Indeed, an inclined Keplerian curve provides a slightly better agreement with our measurements (\anthony{lines with various shades} in Fig.\ref{fig:rotation_nomask} and Fig. \ref{fig:rotation_withmasks}), but it cannot account for the largest deviations at the smallest angular separations. For simplicity, we used a geometrically thin disk approximation in the deprojection step, which does not account for the altitude or flaring of the scattered-light surface; however,  the modest disk inclination should mitigate such an artifact. Inclined structures up to 25-45$\degb$ with respect to the disk midplane have previously been reported and interpreted as infalling material \citet{tang_circumstellar_2012,speedie_mapping_2025}, but these are located in the outer disk region.

Moreover, the bright radial feature labeled the bridge has a motion roughly consistent with a solid rotation pattern, with a constant speed of $\sim 1.3$\,deg.yr$^{-1}$ across the three annuli (Fig. \ref{fig:rotation_withmasks}). It is tentatively associated with a low signal to noise (S/N) feature reported in the HCO$^+$ emission by \citet{riviere-marichalar_gas_2019} with NOEMA \anthony{(NOrthern Extended Millimeter Array)}. 
It casts a deep and broad shadow onto the S2 spiral (Figs. \ref{fig:irdap_labels}, \ref{fig:irdap_zoomin}, \ref{fig:irdap_shadows} with the label \texttt{sh1}), which suggests either a larger vertical thickness or a tilted, optically thick component for this filament with respect to the surrounding material.

\section{SPHERE/ZIMPOL H{$\alpha$} imaging}
\label{sec:zimpol}
\subsection{Observations and data reduction}
We obtained observations with ZIMPOL on October 29, 2021 (P108) in imaging configuration using the pupil-tracking mode. The dichroic was set to the DIC\_HA position to optimize the photon sharing, sending the R band to ZIMPOL and the remaining bandwidth to the wavefront sensor. The coronagraph configuration corresponds to V\_CLC\_M\_WF, which is a suspended 155\,mas diameter opaque Lyot mask in the focal plane, combined with a 78\% transmission pupil stop. 
We used the \filterHa and \filterCnt filters simultaneously for each ZIMPOL arm, centered in ($\lambda_0$=656.34\,nm) and out ($\lambda_0$=644.90\,nm) of the H$\alpha$ line, but have different bandwidths \citep[0.75\,nm and 3.83\,nm, respectively;][]{schmid_spherezimpol_2017}. Science frames with the source on the coronagraph total an integration time of 5760s (392 frames of 15s each). We also obtained out-of-mask PSFs for flux normalization. The parallactic angle varies by a total of 28.6$\degb$.

\begin{figure*}
    \centering
    \includegraphics[width=18cm]{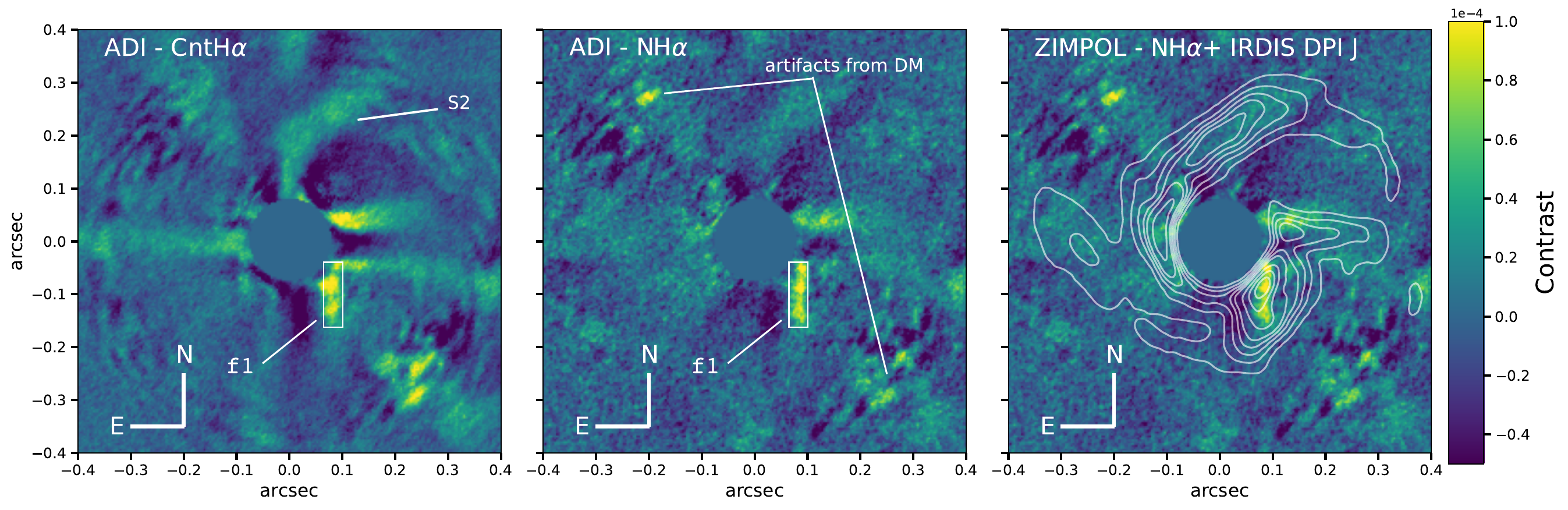}
    \caption{Principal component analysis (PCA) reduction in ADI 
    using  ZIMPOL for the two filters CntH${\alpha}$ (left) and N\_H$\alpha$ (middle and right). 
    The rectangle indicates the aperture used for photometry. Structures oriented along the diagonal of the images at about 0.35$''$ from the star are produced by the SPHERE deformable mirror (DM).
    The right subpanel shows the IRDIS DPI J-band contours overlaid on the ZIMPOL NH$\alpha$ PCA image. 
    The intensity scale is in contrast relative to the star. North is up and east is left.}
    \label{fig:zim_pca}
\end{figure*}

The High Contrast Data Center \citep{delorme_sphere_2017} processed the data using standard cosmetic reduction steps, including dark subtraction, flat-field correction, and bad-pixel removal. We further rejected remaining bad pixels using sigma clipping and subtracted the median of each column to remove detector artifacts, masking the central 200-pixel radius region to avoid biasing the median with the star flux. Fig. \ref{fig:zim_raw} shows images of the target on and off the coronagraph. Next, we processed the data cubes using a PCA-based  angular differential imaging (ADI) method following \citet{soummer_detection_2012}. Fig. \ref{fig:zim_pca} presents the results for two-mode truncation. 

The ZIMPOL PCA images reveal astrophysical signals that are coincident with the position of \texttt{f1}, the northern part of spiral S2, and the bridge, as well as diffraction patterns, that are either radial (attributed to the mask support and the deformable mirror correction) or located at the AO correction radius (at $0.28''-0.42''$, Fig. \ref{fig:zim_abaurb}), which is strongest at $PA=30-60\degb$. Investigating whether real sources exist (in particular the faint source circled in yellow southeast of the star at $[\delta RA, \delta Dec]=[0.30'',-0.17'']$) in the region contaminated by the correction radius would require additional observations. These possible artifacts 
are clearly less dominant in the N\_H$\alpha$ filter image, while the brightest feature coincides with the position of \texttt{f1}. 
In contrast to the smooth profile  observed for \texttt{f1} in the near-IR, its shape appears elongated north-south, and patchy with three distinct knots, in both \filterCnt and \filterHa. Due to the nearly two times smaller bandwidth, the S/N in the \filterHa filter is worse than in the continuum. We note that \texttt{f2} and \texttt{f3} are undetected. We discuss the detection of AB Aur b in section \ref{sec:testabaurb}, and we also report in Fig. \ref{fig:zim_abaurb} the recent detection of a companion candidate in $^{12}$CO ro-vibrational transitions \citep{kozdon_12_2026}, which again results in a non-detection in \Ha with ZIMPOL in a region that is not attenuated by the coronagraph spiders.

To verify the reliability of the features detected in \Ha, we overplotted contour lines from the 2021 IRDIS DPI J-band image onto the ZIMPOL \filterHa image in Fig. \ref{fig:zim_pca} (right). While this  unambiguously confirms the connection between the visible features and near-IR features at the location of \texttt{f1}, the recovery of the bridge is more questionable. Given that the coronagraph mask in the focal plane is suspended by four spiders in a fixed orientation while the field rotates in pupil-tracking mode, it is important to verify the potential obscuration of real disk features. Figure \ref{fig:zim_corospider} shows the area swept by the spiders in the field of view. As a result, the feature \texttt{f1} is not impacted by the spiders, in contrast to the bridge and, to a lesser extent, AB\,Aur\,b. Therefore, in the following, we consider that the bridge signal is altered too much by the coronagraphic mask's spiders to extract reliable photometry. 
We note that despite the attenuation due to these spiders, some regions of the disk have no \Ha counterpart, like for instance the east-northern part of S1. Although located at a similar physical distance to \texttt{f1}, the regions should have a brightness ratio similar to that measured in the near-IR if it were dominated by scattered light. This observational fact possibly argues in favor of a distinct \Ha emission at \texttt{f1}.
 
Attempts to combine angular and spectral differential imaging by subtracting the images in the two filters were not satisfactory and are therefore not shown. We suspect that the difference in the filter bandwidths prevents such a process. We describe the procedure used to estimate the H$\alpha$ flux emission in the next section. 

\subsection{H$\mathrm{\alpha}$ flux and accretion rate}
We estimated the flux in the \Ha line following \citet{schmid_spherezimpol_2017} and \citet{cugno_search_2019}. The method converts the count rates measured in the data into \anthony{flux} 
in physical units using the tabulated zero points for ZIMPOL in the \filterCnt and \filterHa filters. We then estimated the continuum contribution in the \filterHa filter to isolate the \Ha flux alone. Appendix \ref{appendix:fluxdensityHa} details the calculations of $F_{\mathrm{H{\alpha}}}$.
Next, we derived the luminosity directly from the H$\alpha$ line flux, 

\begin{equation}
    L_{\mathrm{H{\alpha}}} = \frac{F_{\mathrm{H{\alpha}}}.4\pi.d^2}{ L_{\odot}}.
\end{equation}

To estimate the accretion luminosity (L$_{acc}$), we considered two approaches. The first approach
is based on empirical calibrations of the relationship between $L_{\mathrm{H{\alpha}}}$ and $L_{acc}$. 
Most studies have so far used the calibration for classical T Tauri stars \citep[CTTS; ][]{rigliaco_x-shooter_2012}, which involves strong magnetic fields and was updated by \citet[Al2017]{alcala_x-shooter_2017}. More recently \citet[][Ao2021]{aoyama_comparison_2021} provided an updated model for shock emission at the planetary surface that is more relevant for protoplanets.
The second approach is based on 3D thermohydrodynamical simulations of planet formation \citep[][SE2020]{szulagyi_hydrogen_2020}. The former case implies magnetospheric accretion which operates when the magnetic field exceeds approximately 60\,G, whereas the latter involves boundary-layer accretion occurring for weaker magnetic fields. 
We provide the corresponding relationships, L$_{acc}$ versus $L_{\mathrm{H{\alpha}}}$, for the different models in Appendix \ref{appendix:fluxdensityHa}.

In the CTTS and shocks framework, the accretion rate follows
\begin{equation}
\dot{M}_{acc}^{Al2017/Ao2021} = \left( 1-\frac{R}{R_{in}}\right)^{-1} \frac{L_{acc}^{Al2017/Ao2021}.R} {G.M},
 \label{eq:Macc} 
\end{equation}
where $R$ and $M$ are the radius and the mass of the accreting object, $R_{in}$ is the truncation radius of the circumplanetary disk, and $G$ is the gravitational constant. 
Most terms in this equation are undefined, so we adopted standard assumptions: $R= 1\,R_{Jup}$, $M = 1\,M_{Jup}$, and $R_{in}=5\,R$. We then derived a tentative estimate of the object mass at the system age under the assumption of a constant accretion rate. This assumption neglects accretion variability. 

In contrast, the model of \citet{szulagyi_hydrogen_2020} provides a direct relationship between the line luminosity and the planet mass, 
\begin{equation}
M^{Sz2020} = \frac{\mathrm{log}(L_{\mathrm{H{\alpha}}})+22.76}{2.23\pm0.53}.
\end{equation}

\subsection{Photometric measurements}
\subsubsection{Stellar accretion}
We first estimated the stellar accretion rate. We integrated the signal within an aperture radius of 1.5$''$ to encompass the full PSF. We measured $F_{\mathrm{CntH}\alpha}= 7.10\times10^{-11}$\,erg/s/cm$^2$ and $F_{\mathrm{NH}\alpha}= 5.78\times10^{-11}$\,erg/s/cm$^2$, which yield a line-to-continuum ratio of 241\% and therefore a flux for $F_{\mathrm{H}\alpha}= 4.09\times10^{-11}$\,erg/s/cm$^2$. We used the relation from \citet{mendigutia_accretion_2011} (similar to Eq. \ref{eq:lacc_r2012} but with different coefficients) calibrated for the $\mathrm{H}\alpha$ line in HAeBe stars instead of \citet{alcala_x-shooter_2017} which holds for CTTS.
The corresponding mean accretion luminosity is $L_{acc}=4.32\,L_{\odot}$, with a mass accretion rate of $\dot{M}_{acc}=1.80\times10^{-7}\,M_{\odot}$/yr, which is consistent, despite the large dispersion, with other estimates based on Br$_{\gamma}$ \citep[$L_{acc}=4.26\,L_{\odot}$;][]{garcia_lopez_accretion_2006} and Pf$_{\beta}$ \citep[$L_{acc}=3.16\,L_{\odot}$;][]{salyk_measuring_2013}.

\subsubsection{Photometry and accretion at \texttt{f1}}
We now focus on \texttt{f1}, the main feature in the \filterHa filter. 
However, applying accretion models to extended features is questionable, as protoplanets or a circumplanetary disk should exhibit a more compact morphology. However, \texttt{f1} appears to be substructured with knots, although their reliability remains to be confirmed with further observations. We therefore considered the brightest knot identified in Fig. \ref{fig:zim_pca}, located in the middle of \texttt{f1}. 
We first estimated the attenuation of the PCA processing on \texttt{f1}, given that the true morphology and intensity distribution of the structure are unknown. For this purpose, we used forward modeling and for simplicity we assumed a fake feature with a Gaussian \anthony{shape}, the size of which ($2\times2$ pixels) is representative of the knot. We projected this model onto the same eigenbasis and with the same mode truncation as that used for the data. This modeling allowed us to estimate an attenuation of a factor of 2.37 (assumed to be constant for both filters) in the photometric aperture due to PCA, which we used to compensate for the count rate loss.
In addition, we accounted for coronographic transmission. For this, we built a data cube with a binary mask representing the central spot and the spiders of the coronagraphic mask, rotated according to the parallactic angle variation. Summing these frames and normalizing yields an estimate of the transmission projected on the sky. In the photometric aperture defined for the knot \texttt{f1}, we measured a transmission of unity.
We then estimated the flux in each filter: $F_{\mathrm{CntH}\alpha}= 4.38\times10^{-15}$\,erg/s/cm$^2$ and $F_{\mathrm{NH}\alpha}= 2.63\times10^{-15}$\,erg/s/cm$^2$, which yields a line-to-continuum ratio of 151\% and therefore a flux of $F_{\mathrm{H}\alpha}= 1.58\times10^{-15}$\,erg/s/cm$^2$.

Using the relation from \citet{alcala_x-shooter_2017} and assuming a truncation radius of five times the radius of the accreting object, we find  $L_{acc}=3.65\times10^{-6} -  3.42\times10^{-5}\,L_{\odot}$ and $\dot{M}_{acc}=8.02\times10^{-8} - 7.52\times10^{-7}$\,M$_{Jup}$/yr. 
The ranges of these values correspond to the dispersions of the empirically calibrated relationships of the luminosities described in sec. \ref{appendix:accretionmodels}.
Therefore, the time to accrete 1\,M$_{Jup}$ is 1.4-13\,Myrs, assuming a constant accretion rate.
Assuming the emission originates from the shocks at the planetary surface \citep{aoyama_comparison_2021}, the accretion luminosity is much higher, with $L_{acc}=4.84\times10^{-5} -  1.93\times10^{-4}\,L_{\odot}$, and the corresponding mass accretion rate is $\dot{M}_{acc}=1.06\times10^{-6} - 4.23\times10^{-6}$\,M$_{Jup}$/yr. Over 1 Myr, the accretion rate would correspond to 1-4\,M$_{Jup}$, significantly higher than in the CTTS framework.

Although extended, \texttt{f1} appears as a series of knots in H$\mathrm{\alpha}$, which could correspond to multiple objects. We therefore estimated the total luminosity and accretion rate of the whole structure assuming it was composed of several point sources.
We defined an aperture of $10\times34$ pixels optimized to encompass \texttt{f1} in both the \filterHa and \filterCnt filters. 
Assuming a forward model of an elliptical Gaussian profile with the same extension and position as \texttt{f1}, we derive an attenuation of the PCA processing of 3.75. As previously, we also estimated the attenuation from the coronagraph, which is 0.978.
The measured fluxes are $F_{\mathrm{CntH}\alpha}= 2.00\times10^{-14}$\,erg/s/cm$^2$ and $F_{\mathrm{NH}\alpha}= 1.30\times10^{-14}$\,erg/s/cm$^2$, which yields a line-to-continuum ratio of 172\% and therefore a flux of $F_{\mathrm{H}\alpha}= 8.22\times10^{-15}$\,erg/s/cm$^2$. 
Using the relation from \citet{alcala_x-shooter_2017}, we obtain $L_{acc}=2.55\times10^{-5} -  2.03\times10^{-4}\,L_{\odot}$ and $\dot{M}_{acc}=5.09\times10^{-6} - 2.02\times10^{-5}$\,M$_{Jup}$/yr.
Therefore, the time to accrete 1 \, M$_{Jup}$ is less than 1\ Myr, which implies that, for the age of the system, the H$\alpha$ emission from the whole f1 feature still corresponds to \anthony{typical masses of giant planets}. 
Using the model of \citep{aoyama_comparison_2021}, the accretion luminosity reaches $L_{acc}=2.31\times10^{-4} -  9.22\times10^{-4}\,L_{\odot}$  and thus the mass accretion rate is $\dot{M}_{acc}=5.08\times10^{-6} - 2.02\times10^{-5}$\,M$_{Jup}$/yr. Over 1 Myr, this accretion rate would correspond to 5-20\,M$_{Jup}$, potentially consistent with several planetary-mass objects. Although we cannot exclude the possibility that the detected emission originates from stellar H$\alpha$ emission scattered off clumps located along the twisted spiral S1.

\subsection{Testing the detection of AB\,Aur\,b}
\label{sec:testabaurb}
\begin{figure*}
    \centering
    \includegraphics[width=18cm]{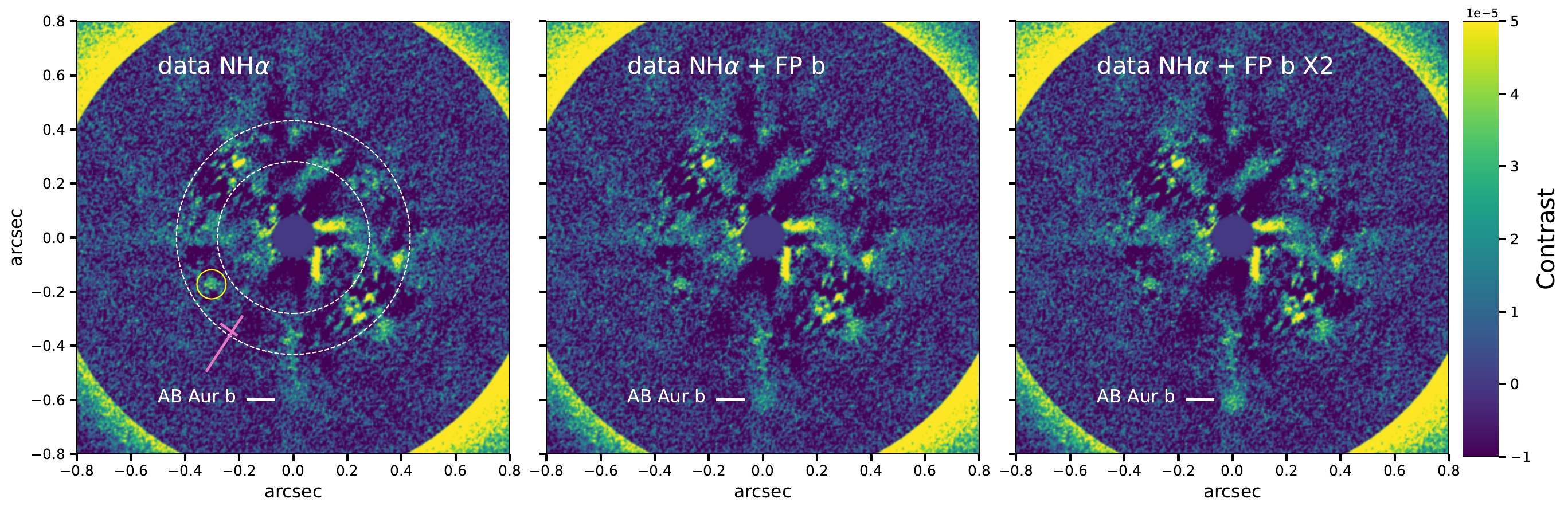}
    \caption{Images in \filterHa (left), with a fake planet (FP) injected at the flux of AB\,Aur\,b (middle) and with twice the flux (right). Labels indicate the position of the candidate planet. The dashed circles trace the region contaminated by the AO correction radius, while the pink cross represents the companion candidate reported in \citet{kozdon_12_2026} in $^{12}$CO ro-vibrational transitions. The yellow circle at $\rho=0.35''$, $PA=119.7\degb$ marks a new signal that remains to be investigated.}
    \label{fig:zim_abaurb}
\end{figure*}

The protoplanet candidate AB\,Aur\,b has been detected in H$\alpha$ filters \citep{currie_images_2022, zhou_hstwfc3_2022} and at UV wavelengths  \citep{zhou_uv-optical_2023}, contrasting with the ZIMPOL data, which show a faint signal at the expected location (Fig. \ref{fig:zim_abaurb}). To assess whether this signal could be related to AB\,Aur\,b, we used fake planet injection, starting from the flux measured with HST/WFPC3, $1.03\times10^{-15}$\,erg/s/cm$^2/\AA$, or
$7.725\times10^{-15}$\,erg/s/cm$^2$, which we converted to count rate by inverting Eq. \ref{eq:fluxNHa}. 
However, we note that the WFPC3 H$\alpha$ filter is more than twice as broad as the  \filterHa filter, so using this flux for injecting a fake planet would be valid only if the H$\alpha$ line dominates over other sources.
To account for the spatial extent of AB\,Aur\,b we generated the fake planet as a $90\times60$\,mas 2D-Gaussian \anthony{shape}, 
scaling the flux to the exposure of one frame, shifting at 0.6$''$ south, and rotating with the parallactic angles to account for the field rotation.  
We reprocessed this new data cube with PCA ADI.  Fig. \ref{fig:zim_abaurb} shows the results for the expected planet's flux and a flux twice as large, both compared to the original image. Therefore, the faint signal observed at the position of AB\,Aur\,b is marginally consistent with a detection. 

Assuming that the detected signal corresponds to AB\,Aur\,b in the \filterHa filter, we measured the counts and applied the same calculation as for \texttt{f1}. We defined an elliptical aperture of $90\times60$\,mas to match the extension of AB\,Aur\,b. 
The PCA-induced attenuation factor is 1.44, as measured from the fake planet injection step, while the coronagraphic transmission is 0.84. 
We measure $F_{\mathrm{CntH}\alpha}= 8.82\times10^{-16}$\,erg/s/cm$^2$ and $F_{\mathrm{NH}\alpha}= 8.57\times10^{-16}$\,erg/s/cm$^2$, hence $F_{\mathrm{H}\alpha}= 6.47\times10^{-16}$\,erg/s/cm$^2$, which is approximately 13 times lower than the integrated H$\alpha$ emission of \texttt{f1}. 
In the magnetospheric accretion model, this flux corresponds to $L_{acc}=1.21\times10^{-6} - 1.30\times10^{-5}\,L_{\odot}$ and $\dot{M}_{acc}=2.79\times10^{-8} - 2.86\times10^{-7}$\,M$_{Jup}$/yr.
The time required to accrete $1\,M_{Jup}$ is longer than 3.7\,Myrs and potentially exceeds the system age.  
Therefore, it is unlikely that the H$\alpha$ emission corresponds to a planetary-mass object given the age of the system. However, the results differ under the shock model of \citet{aoyama_comparison_2021}, which yields an equivalent mass of 0.45-1.81\,M$_{Jup}$.

\subsection{Discussion on accretion}

\paragraph{Accretion rate and mass estimations}
Tab. \ref{tab:fluxhalpha} compiles all flux measurements for the various features. Overall, the integrated H$\alpha$ fluxes for \texttt{f1} and the knot in \texttt{f1} are consistent with accretion rates expected for objects in the planetary mass regime, but can vary significantly (by up to two orders of magnitude) depending on the model used to convert the H$\alpha$ luminosity into mass. 
Tab. \ref{tab:fluxhalpha} also lists the luminosity and mass derived using the boundary layer model \citep{szulagyi_hydrogen_2020}.
We find significantly larger accretion luminosities with the boundary layer model. However, the estimated masses remain in the same range, typically $6-10$\,M$_J$, regardless of the H$\alpha$ flux.

\citet{huelamo_searching_2022} report similar dispersions of $L_{acc}$ for other targets. To illustrate the differences, we compare the accretion luminosity as a function of the H$\alpha$ luminosity for different models in Fig. \ref{fig:comparemodels}. The magnetospheric accretion model exhibits a steep dependence; whereas the boundary layer model represents the opposite extreme, with a weak dependence of  accretion on H$\alpha$ luminosity. 
All models provide consistent results for H$\alpha$ luminosities larger than about $2\times10^{-5}\,L_{\odot}$, but the features of interest are fainter, leading to a strong dispersion in the accretion rates. 
However, the models of \cite{szulagyi_hydrogen_2020} and \cite{aoyama_comparison_2021} yield consistent mass estimates for \texttt{f1}, although they originate from different physical conditions. Therefore, the models are difficult to disentangle using H$\alpha$ luminosity measurements alone. \citet{aoyama_spectral_2021} studied the strengths of  hydrogen lines in the visible and near IR as accretion tracers and highlight additional lines such as Pa$\beta$ (1.282\,$\muup$m), Br$\gamma$ (2.166\,$\muup$m), and Br$\alpha$ (4.052\,$\muup$m), which help break degeneracies and better constrain the accretion mechanism. However, AB Aur b has not been detected in Pa$\beta$ \citep{biddle_deep_2024}, suggesting that extending observations further into the IR could be useful to take advantage of lower opacities in the main disk.

\paragraph{Detection and non-detection of the features} 
Interestingly, the very high angular resolution capacity of ZIMPOL allows us to resolve the extended structure \texttt{f1} as a series of knots, consistently in both filters. Whether this morphology is affected by the data reduction process, which subtracts the starlight and may introduce self-subtraction, or by partial obscuration from the coronagraphic mask spiders, remains to be determined.
These observations require confirmation and/or finer observations. At present, ZIMPOL is the only instrument capable of such observations with this level of contrast and proximity to the star. We note that \texttt{f1} has not been reported in SCExAO observations \citep{currie_images_2022} or with HST/WFPC3 \citep{zhou_hstwfc3_2022}, as it lies beyond the reach of these instruments. Conversely, the non-detection of \texttt{f2}, which is farther from the star than \texttt{f1}, supports the hypothesis of a scattered-light feature, since the scattered light decreases with the square of the physical distance to the star. Nevertheless, this is not the case for  \texttt{f3}, which lies at the same separation as \texttt{f1}, so its non-detection requires another explanation.
High-contrast imaging instruments operating at visible wavelengths on the \anthony{Extremely Large Telescope} 
will be transformative in this regard.

\paragraph{Non-detection of AB Aur b} 
The very low level of detection of AB\,Aur\,b with ZIMPOL is at odds with previous observations obtained in H$\alpha$ filters, even after accounting for the coronagraph transmission \citep{currie_images_2022, zhou_hstwfc3_2022}, which report detections at levels about an order of magnitude stronger than ours. 
However, comparison with other data sets can be complicated by the variable nature of the source, as measured by \cite{bowler_h_2025}, who report a variability amplitude of 330\% using HST/WFPC3. Another effect that can attenuate the observed flux of AB\,Aur\,b is related to the line profile itself. Using VLT/MUSE, \cite{currie_vltmuse_2025} recently identified an inverse P Cygni profile centered on the H$\alpha$ line at 6562\,$\AA$, with a positive flux in the blue wing and a stronger negative flux in the red wing, which could result in a very faint or even negative flux if integrated in a narrow-band filter. However, observations with SCExAO/VAMPIRES, HST/STIS, and HST/WPC3 detect the object.
To reconcile these spectroscopic and photometric observations, \cite{currie_vltmuse_2025} argue that the H$\alpha$ emission may arise from a combination of several sources including scattered light from the disk, and Balmer continuum plus line emission. In this respect, we note that the main difference is that the ZIMPOL filter is narrower (7.5\,$\AA$) than those of other instruments: VAMPIRES (10\,$\AA$),  WFPC3 (17.9\,$\AA$), while STIS is unfiltered, and the MUSE line width is about 5\,$\AA$.
Therefore, ZIMPOL images could  be more affected by the P Cygni profile, yielding a nearly zero flux. It would be relevant to test this hypothesis with other accretion lines in future observations, in particular Br$\alpha$ and Br$\gamma$ tracers in the IR, which may be less affected by disk opacity. 

\subsection{Constraints on binarity}
The first hint of a possible companion star around AB~Aur was reported by \citet{baines_binarity_2006} based on spectroscopic constraints (positional shifts across the \Ha line and variations in the FWHM). However, the authors could only infer a binary separation in the range $0.5-3.0''$, with uncertain constraints on the position angle.
\citet{poblete_binary-induced_2020} produced hydrodynamical simulations of a disk perturbed by a binary star to account for the spiral arms detected  with ALMA inside the dust-continuum cavity. They find that a binary star on an eccentric orbit (e=0.5), perpendicular to the disk midplane, with a semi-major axis of 40\,au (0.25$''$) and a mass ratio of 0.25, can qualitatively reproduce the spiral arms observed in CO. With a mass of 0.6\,M$_\odot$, this hypothetical companion corresponds to a K4-5 star and would have flux ratios relative to the primary of $\Delta H=3.25$ and $\Delta Ks=3.07$. Such a component would be very bright unless it is hidden by some obscuring material due to a combination of disk midplane opacity and orbital configuration, or if it lies behind SPHERE's coronagraphic mask (0.2$''$ diameter, equivalent to 33\,au). 
Given that SPHERE achieves a contrast better than 10\,mag at this angular separation \citep{boccaletti_possible_2020}, the extinction must be larger than 7\,mag for such a binary component to remain undetected. Although no such object has been detected during several years of observations of AB\,Aur, we cannot completely exclude a binary system, even if it appears rather unlikely. 

\begin{figure}
    \centering
    \includegraphics[width=9cm]{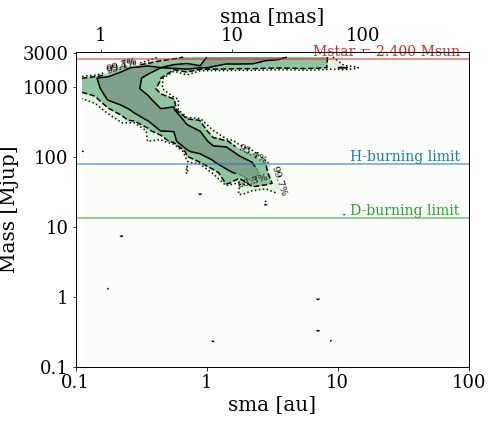}
    \caption{GaiaPMEX analysis for AB\,Aur showing the solutions (green shades) at 1, 2 and 3 $\sigma$ in a mass/semi-major axis (sma) diagram.}
    \label{fig:gaia}
\end{figure}

To investigate closer-in perturbers, we used the GaiaPMEX 
tool \anthony{(standing for Gaia DR3 proper motion anomaly and astrometric noise excess)} developed by \citet{kiefer_searching_2025-1},  which computes the Gaia astrometric excess (based on the renormalized unit weight error and the astrometric excess noise) and accounts for the Gaia-Hipparcos proper motion anomaly. Under the assumption that the excesses are due to a companion, GaiaPMEX provides solutions in a mass--semi-major axis diagram.  \cite{lagrange_searching_2025} and Lagrange et al. (in rev.) present examples of its use for the detection of companions.
By constraining the inclination to $49-51\degb$ (to match the analysis in section \ref{sec:dynamic_f1f2f3}) and the eccentricity to $0.0-0.3$, the astrometric constraints would favor a brown dwarf or stellar component closer than $\sim1-2$\,au and more massive than $\sim50-100$\,M$_{Jupiter}$ (Fig. \ref{fig:gaia}), with a degeneracy toward higher masses at smaller separations.  However, we note that a massive companion on such a close orbit would gravitationally disturb, or even disrupt, any inner CSD, which appears inconsistent with the IR interferometric detections of a warm dust disk or ring in the $0.5-10$\,au inner region of AB~Aur \citep{eisner_resolved_2004, tannirkulam_tale_2008, di_folco_flared_2009,lazareff_structure_2017}. 
This analysis does not exclude the solution proposed by \citet{poblete_binary-induced_2020} because Gaia and Hipparcos do not have the sensitivity to detect an astrometric signature for a companion orbiting at 40\,au. 
\citet[][submitted to AJ]{blakely_dynamical_2026} obtained comparable results in terms of mass and separation solutions.
However, it is important to note that accretion onto the star adds degeneracy in the interpretation of the residuals as discussed in \citet[][submitted to A\&A]{lagrange_searching_2025}.
The amplitude of this effect depends on both the accretion luminosity and the geometry of the accretion region. In such cases, we argue that attributing the astrometric excess solely to the presence of a companion requires caution, until detailed modeling of the accretion feature is available.

\section{Conclusions}
\label{sec:conclusions}
In this paper we present new observations of the AB\,Aur system in near-IR bands with SPHERE/IRDIS at three epochs spanning about 3.85 years, and with SPHERE/ZIMPOL in \Ha filters, with the aim of studying the dynamics of the disk and the accretion regions. In the following, we summarize our findings.
\begin{itemize}
    \item The disk morphology observed over three epochs shows strong consistency, both in large-scale structures and small-scale structures. This high level of stability is an important asset for dual polarimetry observations with SPHERE/IRDIS, enabling detailed and robust dynamical analyses. 
    \item We recovered all features reported in \citet{boccaletti_possible_2020}, and we identify new ones, namely the inner spiral arms S1 and S2, the pinhole in S1,  the bridge, and the localized features \texttt{f1}, \texttt{f2}, and \texttt{f3}. The three localized features are compact structures or point-like sources. The feature \texttt{f1}, initially identified as a near-IR twist, is resolved as an assembly of knots at shorter wavelengths. Therefore, their morphology contrasts with the broad spatial extent of AB\,Aur b, possibly indicating a fundamentally different nature.  
    \item Differential rotation of the disk is an obvious feature when comparing the three epochs. We developed a method to measure the rotation rate as a function of the radius and find that the disk globally follows a Keplerian rotation, except in the inner regions within $\sim$60\,au where the velocity is slower. We observe the same behavior in parts of the brightest spiral arms within the millimeter cavity. 
    \item Motivated by the very small motion of \texttt{f2}, we used \texttt{orbitize!} to constrain the orbital solutions for \texttt{f1}, \texttt{f2}, and \texttt{f3}, despite the limited orbital coverage and the spatial extent of \texttt{f1}. 
    We obtain posterior distributions for the orbital inclinations, which 
     favor orbital motions outside the disk plane. For each feature, we find two possible inclinations from the disk midplane: 75$^{\circ}$ and 32$^{\circ}$ for \texttt{f1} , 38$^{\circ}$ and 77 for $^{\circ}$ \texttt{f2}, and 34$^{\circ}$ and 78$^{\circ}$ for \texttt{f3}. We note that the dispersion on these values can range from 15$\degb$ to 20$\degb$.
    \item Several radial shadows are visible in the images and we can roughly follow their rotation across the epochs. These shadows are very narrow, likely indicating the presence of optically thick regions at stellocentric distances of 30 to 70\,au.
    \item The departure from Keplerian motion within $\sim$60\,au could be related to difficulties in precisely measuring the innermost separations as the disk morphology is intrinsically variable. However, in that case, the departure should be random, whereas it is systematically negative in our data. Out-of-plane structures, and more generally the vertical structure of the disk, could also bias the measurements toward slower velocities. Nevertheless, a reasonable interpretation is that the inner 60\,au of the disk is  gravitationally perturbed by several bodies, consistent with the dynamical behavior of the shadows and with the complexity of the spiral structures.  
    \item While the velocity of a spiral arm has been considered as a way to constrain its origin (self-gravity versus gravitational perturbers), in  real systems the multiplicity, as well as noncircular and noncoplanar orbits of hypothetical perturbers, complicate this simplistic framework. In this context, we note that given their velocities, the spiral S1 may have been triggered by gravitational instability rather than being launched by \texttt{f1}, although this conclusion relies on the simplified case of a single perturber on circular orbit. However, a dynamical link between the spiral S2 and AB\,Aur\,b cannot be ruled out at this stage.
    \item The excess noise in Gaia astrometric measurements, together with the Hipparcos-Gaia proper motion, could suggest a potential binarity of AB\,Aur. However, we cannot draw firm conclusion since stellar accretion is neglected in astrometric analyses and can also affect the astrometric signal.
    \item \Ha imaging with SPHERE/ZIMPOL reveals a strong feature matching \texttt{f1} in the near-IR. This is the first time this structure is observed in the visible, highlighting the performance of SPHERE. For the \Ha line emission, we measure \anthony{$8.22\times10^{-15}$ erg/s/cm$^2$ for the \texttt{f1} }   
    $6.23\times10^{-6}$ erg/s/cm$^2$  for  \texttt{f1}. 

    In contrast, AB\,Aur\,b is likely at the limit of detection, with a flux  of \anthony{$6.46\times10^{-16}$ erg/s/cm$^2$ } 
    \item If \texttt{f1} were a point-source, its flux  would correspond to the \Ha emission expected for planetary-mass objects, although the conversion to accretion rates or masses is strongly model-dependent.
    \item This study argues in favor of obtaining further follow-up observations to (1) analyze the dynamics of the spiral arms and better understand the implication of the presence of  protoplanets, and (2) derive accretion rates from other hydrogen lines at longer wavelengths, in order to break degeneracies in the accretion mechanism and clarify the nature of AB\,Aur\,b. 
\end{itemize}



\begin{acknowledgement}
This work has made use of the High Contrast Data Centre, jointly operated by OSUG/IPAG (Grenoble), PYTHEAS/LAM/CeSAM (Marseille), OCA/Lagrange (Nice), Observatoire de Paris/LIRA (Paris), and Observatoire de Lyon/CRAL, and supported by a grant from Labex OSUG@2020 (Investissements d’avenir – ANR10 LABX56).
This work was supported by the "Programme National de Planétologie" (PNP) and the "Programme National de Physique Stellaire" (PNPS) of CNRS/INSU co-funded by CEA and CNES.
NH is funded by the Spanish grant MCIN/AEI/10.13039/501100011033 PID2023-150468NB-I00 
JV, is funded from the Hungarian NKFIH OTKA project no. K-147380. This work was also supported by the NKFIH NKKP grant ADVANCED 149943. Project no.149943 has been implemented with the support provided by the Ministry of Culture and Innovation of Hungary from the National Research, Development and Innovation Fund, financed under the NKKP ADVANCED funding scheme. 

\end{acknowledgement}

\bibliographystyle{aa}
\bibliography{references}

@article{kozdon_12_2026,
	title = {$^{\textrm{12}}\apj$ {CO} {Rovibrational} {Spectroscopy} of {AB} {Aurigae}—{A} {Potential} {Point} {Source} {Is} {Present}},
	volume = {171},
	issn = {0004-6256, 1538-3881},
	url = {https://iopscience.iop.org/article/10.3847/1538-3881/ae48f6},
	doi = {10.3847/1538-3881/ae48f6},
	number = {4},
	urldate = {2026-04-28},
	journal = {\aj},
	author = {Kozdon, Janus and Fung, Jeffrey and Brittain, Sean D. and Jensen, Stanley and Kern, Josh and Padgett, Cory and Hasegawa, Yasuhiro},
	month = apr,
	year = {2026},
	pages = {250},
}

@article{akansoy_modelling_2025,
	title = {Modelling shadows in scattered light observations as signals from companions in protoplanetary discs},
	volume = {540},
	copyright = {https://creativecommons.org/licenses/by/4.0/},
	issn = {0035-8711, 1365-2966},
	url = {https://academic.oup.com/mnras/article/540/4/3186/8157926},
	doi = {10.1093/mnras/staf925},
	language = {en},
	number = {4},
	urldate = {2025-12-17},
	journal = {\mnras},
	author = {Akansoy, Deniz and Petrou, Helen and Ballabio, Giulia and Penzlin, Anna},
	month = jun,
	year = {2025},
	pages = {3186--3203},
}

@article{dong_how_2016,
	title = {{HOW} {SPIRALS} {AND} {GAPS} {DRIVEN} {BY} {COMPANIONS} {IN} {PROTOPLANETARY} {DISKS} {APPEAR} {IN} {SCATTERED} {LIGHT} {AT} {ARBITRARY} {VIEWING} {ANGLES}},
	volume = {826},
	issn = {0004-637X, 1538-4357},
	url = {https://iopscience.iop.org/article/10.3847/0004-637X/826/1/75},
	doi = {10.3847/0004-637X/826/1/75},
	language = {en},
	number = {1},
	urldate = {2026-04-07},
	journal = {\apj},
	author = {Dong, Ruobing and Fung, Jeffrey and Chiang, Eugene},
	month = jul,
	year = {2016},
	pages = {75},
}

@article{delorme_sphere_2017,
	title = {The {SPHERE} data center: a reference for high contrast imaging processing},
	url = {http://adsabs.harvard.edu/cgi-bin/nph-data_query?bibcode=2017sf2a.conf..347D&link_type=ARTICLE},
	journal = {SF2A-2017: Proceedings of the Annual meeting of the French Society of A\&A},
	author = {Delorme, Ph and Meunier, N and Albert, D and Lagadec, E and Coroller, H Le and Galicher, R and Mouillet, D and Boccaletti, A and Mesa, D and Meunier, J -C and Beuzit, J -L and Lagrange, A -M and Chauvin, G and Sapone, A and Langlois, M and Maire, A -L and Montargès, M and Gratton, R and Vigan, A and Surace, C and Moreau, C and Fenouillet, Th},
	month = dec,
	year = {2017},
	pages = {347 -- 361},
}

@inproceedings{maire_sphere_2016,
	address = {Edinburgh, United Kingdom},
	title = {{SPHERE} {IRDIS} and {IFS} astrometric strategy and calibration},
	url = {http://proceedings.spiedigitallibrary.org/proceeding.aspx?doi=10.1117/12.2233013},
	doi = {10.1117/12.2233013},
	urldate = {2025-12-17},
	booktitle = {Society of {Photo}-{Optical} {Instrumentation} {Engineers} ({SPIE}) {Conference} {Series}},
	publisher = {SPIE},
	author = {Maire, Anne-Lise and Langlois, Maud and Dohlen, Kjetil and Lagrange, Anne-Marie and Gratton, Raffaele and Chauvin, Gaël and Desidera, Silvano and Girard, Julien  H. and Milli, Julien and Vigan, Arthur and Zins, Gerard and Delorme, Philippe and Beuzit, Jean-Luc and Claudi, Riccardo U. and Feldt, Markus and Mouillet, David and Puget, Pascal and Turatto, Massimo and Wildi, François},
	editor = {Evans, Christopher J. and Simard, Luc and Takami, Hideki},
	month = aug,
	year = {2016},
	pages = {990834},
}

@inproceedings{bacon_muse_2010,
	address = {San Diego, California, USA},
	title = {The {MUSE} second-generation {VLT} instrument},
	url = {http://proceedings.spiedigitallibrary.org/proceeding.aspx?doi=10.1117/12.856027},
	doi = {10.1117/12.856027},
	urldate = {2025-10-23},
	booktitle = {Proc. {SPIE} 7735, {Ground}-based and {Airborne} {Instrumentation} for {Astronomy} {III}},
	author = {Bacon, R. and Accardo, M. and Adjali, L. and Anwand, H. and Bauer, S. and Biswas, I. and Blaizot, J. and Boudon, D. and Brau-Nogue, S. and Brinchmann, J. and Caillier, P. and Capoani, L. and Carollo, C. M. and Contini, T. and Couderc, P. and Daguisé, E. and Deiries, S. and Delabre, B. and Dreizler, S. and Dubois, J. and Dupieux, M. and Dupuy, C. and Emsellem, E. and Fechner, T. and Fleischmann, A. and François, M. and Gallou, G. and Gharsa, T. and Glindemann, A. and Gojak, D. and Guiderdoni, B. and Hansali, G. and Hahn, T. and Jarno, A. and Kelz, A. and Koehler, C. and Kosmalski, J. and Laurent, F. and Le Floch, M. and Lilly, S. J. and Lizon, J.-L. and Loupias, M. and Manescau, A. and Monstein, C. and Nicklas, H. and Olaya, J.-C. and Pares, L. and Pasquini, L. and Pécontal-Rousset, A. and Pelló, R. and Petit, C. and Popow, E. and Reiss, R. and Remillieux, A. and Renault, E. and Roth, M. and Rupprecht, G. and Serre, D. and Schaye, J. and Soucail, G. and Steinmetz, M. and Streicher, O. and Stuik, R. and Valentin, , H and Vernet, J. and Weilbacher, P. and Wisotzki, L. and Yerle, N.},
	editor = {McLean, Ian S. and Ramsay, Suzanne K. and Takami, Hideki},
	month = jul,
	year = {2010},
	pages = {773508},
}

@misc{blakely_dynamical_2026,
	title = {Dynamical {Mass} {Constraints} on {Transition} {Disk} {Perturbers} with the {G23H} {Catalog}},
	url = {http://arxiv.org/abs/2602.07731},
	doi = {10.48550/arXiv.2602.07731},
	language = {en},
	urldate = {2026-03-05},
	publisher = {arXiv},
	author = {Blakely, Dori and Thompson, William and Johnstone, Doug and Speedie, Jessica and Xuan, Jerry W. and Blouin, Simon and Zhang, Jingwen and Ruffio, Jean-Baptiste and Nielsen, Eric and Bowler, Brendan P. and Franson, Kyle and Roberson, William and Cloutier, Ryan and Fogal, Andre and Hessel, Kaitlyn and Marois, Christian and Rochon, Alexandra},
	month = feb,
	year = {2026},
	note = {arXiv:2602.07731 [astro-ph]},
}

@article{fuente_probing_2017,
	title = {Probing the {Cold} {Dust} {Emission} in the {AB} {Aur} {Disk}: {A} {Dust} {Trap} in a {Decaying} {Vortex}?*},
	volume = {846},
	issn = {2041-8205, 2041-8213},
	shorttitle = {Probing the {Cold} {Dust} {Emission} in the {AB} {Aur} {Disk}},
	url = {https://iopscience.iop.org/article/10.3847/2041-8213/aa8558},
	doi = {10.3847/2041-8213/aa8558},
	number = {1},
	urldate = {2026-02-12},
	journal = {\apjl},
	author = {Fuente, Asunción and Baruteau, Clément and Neri, Roberto and Carmona, Andrés and Agúndez, Marcelino and Goicoechea, Javier R. and Bachiller, Rafael and Cernicharo, José and Berné, Olivier},
	month = sep,
	year = {2017},
	pages = {L3},
}

@article{dewarf_intrinsic_2003,
	title = {Intrinsic {Properties} of the {Young} {Stellar} {Object} {SU} {Aurigae}},
	volume = {590},
	issn = {0004-637X, 1538-4357},
	url = {https://iopscience.iop.org/article/10.1086/374979},
	doi = {10.1086/374979},
	language = {en},
	number = {1},
	urldate = {2026-01-23},
	journal = {\apj},
	author = {DeWarf, L. E. and Sepinsky, J. F. and Guinan, E. F. and Ribas, I. and Nadalin, I.},
	month = jun,
	year = {2003},
	pages = {357--367},
}

@article{xie_disk_2024,
	address = {circular},
	title = {Disk {Evolution} {Study} {Through} {Imaging} of {Nearby} {Young} {Stars} ({DESTINYS}): {Dynamical} {Evidence} of a {Spiral}-{Arm}-{Driving} and {Gap}-{Opening} {Protoplanet} from {SAO} 206462 {Spiral} {Motion}},
	volume = {10},
	issn = {2218-1997},
	shorttitle = {Disk {Evolution} {Study} {Through} {Imaging} of {Nearby} {Young} {Stars} ({DESTINYS})},
	url = {https://www.mdpi.com/2218-1997/10/12/465},
	doi = {10.3390/universe10120465},
	language = {en},
	number = {12},
	urldate = {2025-12-17},
	journal = {Universe},
	author = {Xie, Chen and Xie, Chengyan and Ren, Bin B. and Benisty, Myriam and Ginski, Christian and Fang, Taotao and Casassus, Simon and Bae, Jaehan and Facchini, Stefano and Ménard, François and Van Holstein, Rob G.},
	month = dec,
	year = {2024},
	pages = {465},
}

@article{jorquera_large_2022,
	title = {Large {Binocular} {Telescope} {Search} for {Companions} and {Substructures} in the ({Pre})transitional {Disk} of {AB} {Aurigae}},
	volume = {926},
	issn = {0004-637X, 1538-4357},
	url = {https://iopscience.iop.org/article/10.3847/1538-4357/ac4be4},
	doi = {10.3847/1538-4357/ac4be4},
	language = {en},
	number = {1},
	urldate = {2025-12-17},
	journal = {\apj},
	author = {Jorquera, Sebastián and Bonnefoy, Mickaël and Betti, Sarah and Chauvin, Gaël and Buenzli, Esther and Pérez, Laura M. and Follette, Katherine B. and Hinz, Philip M. and Boccaletti, Anthony and Bailey, Vanessa and Biller, Beth and Defrère, Denis and Eisner, Josh and Henning, Thomas and Klahr, Hubert and Leisenring, Jarron and Olofsson, Johan and Schlieder, Joshua E. and Skemer, Andrew J. and Skrutskie, Michael F. and Van Boekel, Roy},
	month = feb,
	year = {2022},
	pages = {71},
}

@article{szulagyi_meridional_2022,
	title = {Meridional {Circulation} of {Dust} and {Gas} in the {Circumstellar} {Disk}: {Delivery} of {Solids} onto the {Circumplanetary} {Region}},
	volume = {924},
	issn = {0004-637X, 1538-4357},
	shorttitle = {Meridional {Circulation} of {Dust} and {Gas} in the {Circumstellar} {Disk}},
	url = {https://iopscience.iop.org/article/10.3847/1538-4357/ac32d1},
	doi = {10.3847/1538-4357/ac32d1},
	language = {en},
	number = {1},
	urldate = {2025-12-17},
	journal = {\apj},
	author = {Szulágyi, J. and Binkert, F. and Surville, C.},
	month = jan,
	year = {2022},
	pages = {1},
}

@article{speedie_gravitational_2024,
	title = {Gravitational instability in a planet-forming disk},
	volume = {633},
	issn = {0028-0836, 1476-4687},
	url = {https://www.nature.com/articles/s41586-024-07877-0},
	doi = {10.1038/s41586-024-07877-0},
	language = {en},
	number = {8028},
	urldate = {2025-12-17},
	journal = {Nature},
	author = {Speedie, Jessica and Dong, Ruobing and Hall, Cassandra and Longarini, Cristiano and Veronesi, Benedetta and Paneque-Carreño, Teresa and Lodato, Giuseppe and Tang, Ya-Wen and Teague, Richard and Hashimoto, Jun},
	month = sep,
	year = {2024},
	pages = {58--62},
}

@article{speedie_mapping_2025,
	title = {Mapping the {Merging} {Zone} of {Late} {Infall} in the {AB} {Aur} {Planet}-forming {System}},
	volume = {981},
	issn = {2041-8205, 2041-8213},
	url = {https://iopscience.iop.org/article/10.3847/2041-8213/adb7d5},
	doi = {10.3847/2041-8213/adb7d5},
	language = {en},
	number = {2},
	urldate = {2025-12-17},
	journal = {\apjl},
	author = {Speedie, Jessica and Dong, Ruobing and Teague, Richard and Segura-Cox, Dominique and Pineda, Jaime E. and Calcino, Josh and Longarini, Cristiano and Hall, Cassandra and Tang, Ya-Wen and Hashimoto, Jun and Paneque-Carreño, Teresa and Lodato, Giuseppe and Veronesi, Bennedetta},
	month = mar,
	year = {2025},
	pages = {L30},
}

@article{currie_vltmuse_2025,
	title = {{VLT}/{MUSE} {Detection} of the {AB} {Aurigae} b {Protoplanet} with {H}$_{\textrm{ \textit{α} }}$ {Spectroscopy}},
	volume = {990},
	issn = {2041-8205, 2041-8213},
	url = {https://iopscience.iop.org/article/10.3847/2041-8213/adf7a0},
	doi = {10.3847/2041-8213/adf7a0},
	language = {en},
	number = {2},
	urldate = {2025-12-17},
	journal = {\apjl},
	author = {Currie, Thayne and Hashimoto, Jun and Aoyama, Yuhiko and Dong, Ruobing and Fukagawa, Misato and Muto, Takayuki and Dykes, Erica and El Morsy, Mona and Tamura, Motohide},
	month = sep,
	year = {2025},
	pages = {L42},
}

@article{calcino_anatomy_2025,
	title = {Anatomy of a fall: stationary and super-{Keplerian} spiral arms generated by accretion streamers in protostellar discs},
	volume = {537},
	copyright = {https://creativecommons.org/licenses/by/4.0/},
	issn = {0035-8711, 1365-2966},
	shorttitle = {Anatomy of a fall},
	url = {https://academic.oup.com/mnras/article/537/3/2695/7991275},
	doi = {10.1093/mnras/staf135},
	language = {en},
	number = {3},
	urldate = {2025-12-17},
	journal = {\mnras},
	author = {Calcino, Josh and Price, Daniel J and Hilder, Thomas and Christiaens, Valentin and Speedie, Jessica and Ormel, Chris W},
	month = feb,
	year = {2025},
	pages = {2695--2707},
}

@article{biddle_deep_2024,
	title = {Deep {Paβ} {Imaging} of the {Candidate} {Accreting} {Protoplanet} {AB} {Aur} b},
	volume = {167},
	issn = {0004-6256, 1538-3881},
	url = {https://iopscience.iop.org/article/10.3847/1538-3881/ad2a52},
	doi = {10.3847/1538-3881/ad2a52},
	language = {en},
	number = {4},
	urldate = {2025-12-17},
	journal = {\aj},
	author = {Biddle, Lauren I. and Bowler, Brendan P. and Zhou, Yifan and Franson, Kyle and Zhang, Zhoujian},
	month = apr,
	year = {2024},
	pages = {172},
}

@article{xie_dynamical_2023,
	title = {Dynamical detection of a companion driving a spiral arm in a protoplanetary disk},
	volume = {675},
	copyright = {https://creativecommons.org/licenses/by/4.0},
	issn = {0004-6361, 1432-0746},
	url = {https://www.aanda.org/10.1051/0004-6361/202346305},
	doi = {10.1051/0004-6361/202346305},
	language = {en},
	urldate = {2025-12-15},
	journal = {A\&A},
	author = {Xie, Chen and Ren, Bin B. and Dong, Ruobing and Choquet, Élodie and Vigan, Arthur and Gonzalez, Jean-François and Wagner, Kevin and Fang, Taotao and Ubeira-Gabellini, Maria Giulia},
	month = jul,
	year = {2023},
	pages = {L1},
}

@article{lazareff_structure_2017,
	title = {Structure of {Herbig} {AeBe} disks at the milliarcsecond scale: {A} statistical survey in the \textit{{H}} band using {PIONIER}-{VLTI}},
	volume = {599},
	copyright = {https://www.edpsciences.org/en/authors/copyright-and-licensing},
	issn = {0004-6361, 1432-0746},
	shorttitle = {Structure of {Herbig} {AeBe} disks at the milliarcsecond scale},
	url = {http://www.aanda.org/10.1051/0004-6361/201629305},
	doi = {10.1051/0004-6361/201629305},
	language = {en},
	urldate = {2025-12-12},
	journal = {A\&A},
	author = {Lazareff, B. and Berger, J.-P. and Kluska, J. and Le Bouquin, J.-B. and Benisty, M. and Malbet, F. and Koen, C. and Pinte, C. and Thi, W.-F. and Absil, O. and Baron, F. and Delboulbé, A. and Duvert, G. and Isella, A. and Jocou, L. and Juhasz, A. and Kraus, S. and Lachaume, R. and Ménard, F. and Millan-Gabet, R. and Monnier, J. D. and Moulin, T. and Perraut, K. and Rochat, S. and Soulez, F. and Tallon, M. and Thiébaut, E. and Traub, W. and Zins, G.},
	month = mar,
	year = {2017},
	pages = {A85},
}

@article{tannirkulam_tale_2008,
	title = {A {Tale} of {Two} {Herbig} {Ae} {Stars}, {MWC} 275 and {AB} {Aurigae}: {Comprehensive} {Models} for {Spectral} {Energy} {Distribution} and {Interferometry}},
	volume = {689},
	issn = {0004-637X, 1538-4357},
	shorttitle = {A {Tale} of {Two} {Herbig} {Ae} {Stars}, {MWC} 275 and {AB} {Aurigae}},
	url = {https://iopscience.iop.org/article/10.1086/592346},
	doi = {10.1086/592346},
	language = {en},
	number = {1},
	urldate = {2025-12-12},
	journal = {\apj},
	author = {Tannirkulam, A. and Monnier, J. D. and Harries, T. J. and Millan‐Gabet, R. and Zhu, Z. and Pedretti, E. and Ireland, M. and Tuthill, P. and Ten Brummelaar, T. and McAlister, H. and Farrington, C. and Goldfinger, P. J. and Sturmann, J. and Sturmann, L. and Turner, N.},
	month = dec,
	year = {2008},
	pages = {513--531},
}

@article{baines_binarity_2006,
	title = {On the binarity of {Herbig} {Ae}/{Be} stars},
	volume = {367},
	issn = {0035-8711, 1365-2966},
	url = {https://academic.oup.com/mnras/article-lookup/doi/10.1111/j.1365-2966.2006.10006.x},
	doi = {10.1111/j.1365-2966.2006.10006.x},
	language = {en},
	number = {2},
	urldate = {2025-12-12},
	journal = {\mnras},
	author = {Baines, D. and Oudmaijer, R. D. and Porter, J. M. and Pozzo, M.},
	month = apr,
	year = {2006},
	pages = {737--753},
}

@article{kiefer_searching_2025-1,
	title = {Searching for substellar companion candidates with \textit{{Gaia}}: {I}. {Introducing} the {GaiaPMEX} tool},
	volume = {702},
	copyright = {https://creativecommons.org/licenses/by/4.0},
	issn = {0004-6361, 1432-0746},
	shorttitle = {Searching for substellar companion candidates with \textit{{Gaia}}},
	url = {https://www.aanda.org/10.1051/0004-6361/202449335},
	doi = {10.1051/0004-6361/202449335},
	language = {en},
	urldate = {2025-12-08},
	journal = {A\&A},
	author = {Kiefer, F. and Lagrange, A.-M. and Rubini, P. and Philipot, F.},
	month = oct,
	year = {2025},
	pages = {A76},
}

@misc{lagrange_searching_2025,
	title = {Searching for substellar companion candidates with {Gaia}. {III}. {Search} for companions to members of young associations},
	url = {http://arxiv.org/abs/2501.10488},
	doi = {10.48550/arXiv.2501.10488},
	language = {en},
	urldate = {2025-12-08},
	publisher = {arXiv},
	author = {Lagrange, A.-M. and Kiefer, F. and Rubini, P. and Squicciarini, V. and Chomez, A. and Milli, J. and Zurlo, A. and Bouvier, J. and Delorme, P. and Beust, H. and Mazoyer, J. and Flasseur, O. and Meunier, N. and Mignon, L. and Chauvin, G. and Palma-Bifani, P.},
	month = jan,
	year = {2025},
	note = {arXiv:2501.10488 [astro-ph]},
}

@article{yoshida_winding_2025,
	title = {Winding motion of spirals in a gravitationally unstable protoplanetary disk},
	issn = {2397-3366},
	url = {https://www.nature.com/articles/s41550-025-02639-y},
	doi = {10.1038/s41550-025-02639-y},
	language = {en},
	urldate = {2025-10-31},
	journal = {Nat. Astron.},
	author = {Yoshida, Tomohiro C. and Nomura, Hideko and Doi, Kiyoaki and Barraza-Alfaro, Marcelo and Teague, Richard and Furuya, Kenji and Yamato, Yoshihide and Tsukagoshi, Takashi},
	month = sep,
	year = {2025},
}

@article{cossins_characterizing_2009,
	title = {Characterizing the gravitational instability in cooling accretion discs},
	volume = {393},
	issn = {00358711, 13652966},
	url = {https://academic.oup.com/mnras/article-lookup/doi/10.1111/j.1365-2966.2008.14275.x},
	doi = {10.1111/j.1365-2966.2008.14275.x},
	language = {en},
	number = {4},
	urldate = {2025-10-31},
	journal = {\mnras},
	author = {Cossins, Peter and Lodato, Giuseppe and Clarke, C. J.},
	month = mar,
	year = {2009},
	pages = {1157--1173},
}

@article{bowler_h_2025,
	title = {H \textit{α} {Variability} of {AB} {Aur} b with the {Hubble} {Space} {Telescope}: {Probing} the {Nature} of a {Protoplanet} {Candidate} with {Accretion} {Light} {Echoes}},
	volume = {169},
	issn = {0004-6256, 1538-3881},
	shorttitle = {H \textit{α} {Variability} of {AB} {Aur} b with the {Hubble} {Space} {Telescope}},
	url = {https://iopscience.iop.org/article/10.3847/1538-3881/adb6a1},
	doi = {10.3847/1538-3881/adb6a1},
	language = {en},
	number = {5},
	urldate = {2025-10-23},
	journal = {\aj},
	author = {Bowler, Brendan P. and Zhou, Yifan and Biddle, Lauren I. and Jiang, Lillian Yushu and Bae, Jaehan and Close, Laird M. and Follette, Katherine B. and Franson, Kyle and Kraus, Adam L. and Sanghi, Aniket and Tran, Quang and Ward-Duong, Kimberly and Wu, Ya-Lin and Zhu, Zhaohuan},
	month = may,
	year = {2025},
	pages = {258},
}

@inproceedings{groff_first_2017,
	address = {San Diego, United States},
	title = {First light of the {CHARIS} high-contrast integral-field spectrograph},
	isbn = {978-1-5106-1257-0 978-1-5106-1258-7},
	url = {https://www.spiedigitallibrary.org/conference-proceedings-of-spie/10400/2273525/First-light-of-the-CHARIS-high-contrast-integral-field-spectrograph/10.1117/12.2273525.full},
	doi = {10.1117/12.2273525},
	urldate = {2025-10-23},
	booktitle = {Techniques and {Instrumentation} for {Detection} of {Exoplanets} {VIII}},
	publisher = {SPIE},
	author = {Groff, Tyler D. and Rizzo, Maxime and Currie, Thayne and Chilcote, Jeffrey K. and Brandt, Timothy and Kasdin, N. Jeremy and Galvin, Michael B. and Loomis, Craig and Knapp, Gillian and Guyon, Olivier and Jovanovic, Nemanja and Lozi, Julien and Takato, Naruhisa and Hayashi, Masahiko},
	editor = {Shaklan, Stuart},
	month = sep,
	year = {2017},
	pages = {39},
}

@article{zurlo_widest_2020,
	title = {The widest {H} \textit{α} survey of accreting protoplanets around nearby transition disks},
	volume = {633},
	copyright = {http://creativecommons.org/licenses/by/4.0},
	issn = {0004-6361, 1432-0746},
	url = {https://www.aanda.org/10.1051/0004-6361/201936891},
	doi = {10.1051/0004-6361/201936891},
	urldate = {2025-10-23},
	journal = {A\&A},
	author = {Zurlo, A. and Cugno, G. and Montesinos, M. and Perez, S. and Canovas, H. and Casassus, S. and Christiaens, V. and Cieza, L. and Huelamo, N.},
	month = jan,
	year = {2020},
	pages = {A119},
}

@inproceedings{males_magao-x_2024,
	address = {Yokohama, Japan},
	title = {{MagAO}-{X}: commissioning results and status of ongoing upgrades},
	isbn = {978-1-5106-7517-9 978-1-5106-7518-6},
	shorttitle = {{MagAO}-{X}},
	url = {https://www.spiedigitallibrary.org/conference-proceedings-of-spie/13097/3019464/MagAO-X-commissioning-results-and-status-of-ongoing-upgrades/10.1117/12.3019464.full},
	doi = {10.1117/12.3019464},
	urldate = {2025-10-23},
	booktitle = {Adaptive {Optics} {Systems} {IX}},
	publisher = {SPIE},
	author = {Males, Jared R. and Close, Laird M. and Haffert, Sebastiaan Y. and Kautz, Maggie Y. and Kueny, Jay and Long, Joseph D. and McEwen, Eden A. and Swimmer, Noah and Bailey, John I. and Foster, Warren B. and Mazin, Ben and Pearce, Logan A. and Liberman, Joshua and Twitchell, Katie and Weinberger, Alycia J. and Guyon, Olivier and Hedglen, Alexander D. and McLeod, Avalon L. and Roberts, Roswell and Van Gorkom, Kyle and Li, Jialin and Doty, Isabella and Gasho, Victor},
	editor = {Schmidt, Dirk and Vernet, Elise and Jackson, Kathryn J.},
	month = aug,
	year = {2024},
	pages = {8},
}

@article{norris_vampires_2015,
	title = {The {VAMPIRES} instrument: imaging the innermost regions of protoplanetary discs with polarimetric interferometry},
	volume = {447},
	issn = {1365-2966, 0035-8711},
	shorttitle = {The {VAMPIRES} instrument},
	url = {http://academic.oup.com/mnras/article/447/3/2894/986000/The-VAMPIRES-instrument-imaging-the-innermost},
	doi = {10.1093/mnras/stu2529},
	language = {en},
	number = {3},
	urldate = {2025-10-23},
	journal = {\mnras},
	author = {Norris, Barnaby and Schworer, Guillaume and Tuthill, Peter and Jovanovic, Nemanja and Guyon, Olivier and Stewart, Paul and Martinache, Frantz},
	month = mar,
	year = {2015},
	pages = {2894--2906},
}

@article{currie_images_2022,
	title = {Images of embedded {Jovian} planet formation at a wide separation around {AB} {Aurigae}},
	volume = {6},
	doi = {10.1038/s41550-022-01634-x},
	journal = {Nat. Astron.},
	author = {Currie, Thayne and Lawson, Kellen and Schneider, Glenn and Lyra, Wladimir and Wisniewski, John and Grady, Carol and Guyon, Olivier and Tamura, Motohide and Kotani, Takayuki and Kawahara, Hajime and Brandt, Timothy and Uyama, Taichi and Muto, Takayuki and Dong, Ruobing and Kudo, Tomoyuki and Hashimoto, Jun and Fukagawa, Misato and Wagner, Kevin and Lozi, Julien and Chilcote, Jeffrey and Tobin, Taylor and Groff, Tyler and Ward-Duong, Kimberly and Januszewski, William and Norris, Barnaby and Tuthill, Peter and Marel, Nienke van der and Sitko, Michael and Deo, Vincent and Vievard, Sebastien and Jovanovic, Nemanja and Martinache, Frantz and Skaf, Nour},
	month = apr,
	year = {2022},
	pages = {751--759},
}

@article{alcala_x-shooter_2017,
	title = {X-shooter spectroscopy of young stellar objects in {Lupus}: {Accretion} properties of class {II} and transitional objects⋆},
	volume = {600},
	issn = {0004-6361, 1432-0746},
	shorttitle = {X-shooter spectroscopy of young stellar objects in {Lupus}},
	url = {http://www.aanda.org/10.1051/0004-6361/201629929},
	doi = {10.1051/0004-6361/201629929},
	language = {en},
	urldate = {2025-08-29},
	journal = {A\&A},
	author = {Alcalá, J. M. and Manara, C. F. and Natta, A. and Frasca, A. and Testi, L. and Nisini, B. and Stelzer, B. and Williams, J. P. and Antoniucci, S. and Biazzo, K. and Covino, E. and Esposito, M. and Getman, F. and Rigliaco, E.},
	month = apr,
	year = {2017},
	pages = {A20},
}

@article{close_wide_2025,
	title = {Wide {Separation} {Planets} in {Time} ({WISPIT}): {Discovery} of a {Gap} {H} \textit{α} {Protoplanet} {WISPIT} 2b with {MagAO}-{X}},
	volume = {990},
	issn = {2041-8205, 2041-8213},
	shorttitle = {Wide {Separation} {Planets} in {Time} ({WISPIT})},
	url = {https://iopscience.iop.org/article/10.3847/2041-8213/adf7a5},
	doi = {10.3847/2041-8213/adf7a5},
	language = {en},
	number = {1},
	urldate = {2025-08-29},
	journal = {\apjl},
	author = {Close, Laird M. and Van Capelleveen, Richelle F. and Weible, Gabriel and Wagner, Kevin and Haffert, Sebastiaan Y. and Males, Jared R. and Ilyin, Ilya and Kenworthy, Matthew A. and Li, Jialin and Long, Joseph D. and Ertel, Steve and Ginski, Christian and Weinberger, Alycia J. and Follette, Kate and Liberman, Joshua and Twitchell, Katie and Johnson, Parker and Kueny, Jay and Apai, Daniel and Doyon, Rene and Foster, Warren and Gasho, Victor and Van Gorkom, Kyle and Guyon, Olivier and Kautz, Maggie Y. and McLeod, Avalon and McEwen, Eden and Pearce, Logan and Schatz, Lauren and Hedglen, Alexander D. and Wu, Ya-Lin and Isbell, Jacob and Power, Jenny and Carlson, Jared and Close, Emmeline and Tonucci, Elena and Mars, Matthijs},
	month = sep,
	year = {2025},
	pages = {L9},
}

@article{tang_circumstellar_2012,
	title = {The circumstellar disk of {AB} {Aurigae}: evidence for envelope accretion at late stages of star formation?},
	volume = {547},
	issn = {0004-6361, 1432-0746},
	shorttitle = {The circumstellar disk of {AB} {Aurigae}},
	url = {http://www.aanda.org/10.1051/0004-6361/201219414},
	doi = {10.1051/0004-6361/201219414},
	language = {en},
	urldate = {2025-08-28},
	journal = {A\&A},
	author = {Tang, Y.-Wen and Guilloteau, S. and Piétu, V. and Dutrey, A. and Ohashi, N. and Ho, P. T. P.},
	month = nov,
	year = {2012},
	pages = {A84},
}

@article{van_capelleveen_wide_2025,
	title = {{WIde} {Separation} {Planets} {In} {Time} ({WISPIT}): {A} {Gap}-clearing {Planet} in a {Multi}-ringed {Disk} around the {Young} {Solar}-type {Star} {WISPIT} 2},
	volume = {990},
	issn = {2041-8205, 2041-8213},
	shorttitle = {{WIde} {Separation} {Planets} {In} {Time} ({WISPIT})},
	url = {https://iopscience.iop.org/article/10.3847/2041-8213/adf721},
	doi = {10.3847/2041-8213/adf721},
	language = {en},
	number = {1},
	urldate = {2025-08-28},
	journal = {\apjl},
	author = {Van Capelleveen, Richelle F. and Ginski, Christian and Kenworthy, Matthew A. and Byrne, Jake and Lawlor, Chloe and McLachlan, Dan and Mamajek, Eric E. and Stolker, Tomas and Benisty, Myriam and Bohn, Alexander J. and Close, Laird M. and Dominik, Carsten and Haffert, Sebastiaan and Landman, Rico and Ma, Jie and Snellen, Ignas and Tazaki, Ryo and Van Der Marel, Nienke and Welzel, Lukas and Zhang, Yapeng},
	month = sep,
	year = {2025},
	pages = {L8},
}

@article{blunt_orbitize_2020,
	title = {orbitize!: {A} {Comprehensive} {Orbit}-fitting {Software} {Package} for the {High}-contrast {Imaging} {Community}},
	volume = {159},
	issn = {0004-6256, 1538-3881},
	shorttitle = {orbitize!},
	url = {https://iopscience.iop.org/article/10.3847/1538-3881/ab6663},
	doi = {10.3847/1538-3881/ab6663},
	language = {en},
	number = {3},
	urldate = {2025-07-07},
	journal = {\aj},
	author = {Blunt, Sarah and Wang, Jason J. and Angelo, Isabel and Ngo, Henry and Cody, Devin and De Rosa, Robert J. and Graham, James R. and Hirsch, Lea and Nagpal, Vighnesh and Nielsen, Eric L. and Pearce, Logan and Rice, Malena and Tejada, Roberto},
	month = mar,
	year = {2020},
	pages = {89},
}

@article{blunt_orbits_2017,
	title = {Orbits for the {Impatient}: {A} {Bayesian} {Rejection}-sampling {Method} for {Quickly} {Fitting} the {Orbits} of {Long}-period {Exoplanets}},
	volume = {153},
	issn = {0004-6256, 1538-3881},
	shorttitle = {Orbits for the {Impatient}},
	url = {https://iopscience.iop.org/article/10.3847/1538-3881/aa6930},
	doi = {10.3847/1538-3881/aa6930},
	language = {en},
	number = {5},
	urldate = {2025-07-07},
	journal = {\aj},
	author = {Blunt, Sarah and Nielsen, Eric L. and De Rosa, Robert J. and Konopacky, Quinn M. and Ryan, Dominic and Wang, Jason J. and Pueyo, Laurent and Rameau, Julien and Marois, Christian and Marchis, Franck and Macintosh, Bruce and Graham, James R. and Duchêne, Gaspard and Schneider, Adam C.},
	month = may,
	year = {2017},
	pages = {229},
}

@article{dykes_scexaocharis_2024,
	title = {{SCExAO}/{CHARIS} {Near}-infrared {Scattered}-light {Imaging} and {Integral} {Field} {Spectropolarimetry} of the {AB} {Aurigae} {Protoplanetary} {System}},
	volume = {977},
	issn = {0004-637X, 1538-4357},
	url = {https://iopscience.iop.org/article/10.3847/1538-4357/ad8ba0},
	doi = {10.3847/1538-4357/ad8ba0},
	language = {en},
	number = {2},
	urldate = {2025-03-20},
	journal = {\apj},
	author = {Dykes, Erica and Currie, Thayne and Lawson, Kellen and Lucas, Miles and Kudo, Tomoyuki and Chen, Minghan and Guyon, Olivier and Groff, Tyler D. and Lozi, Julien and Chilcote, Jeffrey and Brandt, Timothy D. and Vievard, Sebastien and Skaf, Nour and Deo, Vincent and Morsy, Mona El and Bovie, Danielle and Uyama, Taichi and Grady, Carol and Sitko, Michael and Hashimoto, Jun and Martinache, Frantz and Jovanovic, Nemanja and Tamura, Motohide and Kasdin, N. Jeremy},
	month = dec,
	year = {2024},
	pages = {172},
}

@article{dutrey_sulfur_2024,
	title = {Sulfur monoxide ({SO}) as a shock tracer in protoplanetary disks: {Case} of {AB} {Aurigae}},
	volume = {689},
	copyright = {https://creativecommons.org/licenses/by/4.0},
	issn = {0004-6361, 1432-0746},
	shorttitle = {Sulfur monoxide ({SO}) as a shock tracer in protoplanetary disks},
	url = {https://www.aanda.org/10.1051/0004-6361/202451299},
	doi = {10.1051/0004-6361/202451299},
	language = {en},
	urldate = {2025-03-20},
	journal = {A\&A},
	author = {Dutrey, A. and Chapillon, E. and Guilloteau, S. and Tang, Y. W. and Boccaletti, A. and Bouscasse, L. and Collin-Dufresne, T. and Di Folco, E. and Fuente, A. and Piétu, V. and Rivière-Marichalar, P. and Semenov, D.},
	month = sep,
	year = {2024},
	pages = {L7},
}

@article{aoyama_comparison_2021,
	title = {Comparison of {Planetary} {Hα}-emission {Models}: {A} {New} {Correlation} with {Accretion} {Luminosity}},
	volume = {917},
	issn = {2041-8205, 2041-8213},
	shorttitle = {Comparison of {Planetary} {Hα}-emission {Models}},
	url = {https://iopscience.iop.org/article/10.3847/2041-8213/ac19bd},
	doi = {10.3847/2041-8213/ac19bd},
	language = {en},
	number = {2},
	urldate = {2025-01-29},
	journal = {\apjl},
	author = {Aoyama, Yuhiko and Marleau, Gabriel-Dominique and Ikoma, Masahiro and Mordasini, Christoph},
	month = aug,
	year = {2021},
	pages = {L30},
}

@article{garcia_lopez_accretion_2006,
	title = {Accretion rates in {Herbig} {Ae} stars},
	volume = {459},
	issn = {0004-6361, 1432-0746},
	url = {http://www.aanda.org/10.1051/0004-6361:20065575},
	doi = {10.1051/0004-6361:20065575},
	number = {3},
	urldate = {2025-01-23},
	journal = {A\&A},
	author = {Garcia Lopez, R. and Natta, A. and Testi, L. and Habart, E.},
	month = dec,
	year = {2006},
	pages = {837--842},
}

@article{mendigutia_accretion_2011,
	title = {Accretion rates and accretion tracers of {Herbig} {Ae}/{Be} stars},
	volume = {535},
	issn = {0004-6361, 1432-0746},
	url = {http://www.aanda.org/10.1051/0004-6361/201117444},
	doi = {10.1051/0004-6361/201117444},
	language = {en},
	urldate = {2025-01-23},
	journal = {A\&A},
	author = {Mendigutía, I. and Calvet, N. and Montesinos, B. and Mora, A. and Muzerolle, J. and Eiroa, C. and Oudmaijer, R. D. and Merín, B.},
	month = nov,
	year = {2011},
	pages = {A99},
}

@article{szulagyi_hydrogen_2020,
	title = {Hydrogen {Recombination} {Line} {Luminosities} and {Variability} from {Forming} {Planets}},
	volume = {902},
	issn = {0004-637X, 1538-4357},
	url = {https://iopscience.iop.org/article/10.3847/1538-4357/abb5a2},
	doi = {10.3847/1538-4357/abb5a2},
	language = {en},
	number = {2},
	urldate = {2024-12-03},
	journal = {\apj},
	author = {Szulágyi, Judit and Ercolano, Barbara},
	month = oct,
	year = {2020},
	pages = {126},
}

@article{zhou_hstwfc3_2022,
	title = {{HST}/{WFC3} {Hα} {Direct}-imaging {Detection} of a {Pointlike} {Source} in the {Disk} {Cavity} of {AB} {Aur}},
	volume = {934},
	issn = {2041-8205, 2041-8213},
	url = {https://iopscience.iop.org/article/10.3847/2041-8213/ac7fef},
	doi = {10.3847/2041-8213/ac7fef},
	language = {en},
	number = {1},
	urldate = {2024-11-25},
	journal = {\apjl},
	author = {Zhou, Yifan and Sanghi, Aniket and Bowler, Brendan P. and Wu, Ya-Lin and Close, Laird M. and Long, Feng and Ward-Duong, Kimberly and Zhu, Zhaohuan and Kraus, Adam L. and Follette, Katherine B. and Bae, Jaehan},
	month = jul,
	year = {2022},
	pages = {L13},
}

@article{salyk_measuring_2013,
	title = {{MEASURING} {PROTOPLANETARY} {DISK} {ACCRETION} {WITH} {H} {I} {PFUND} β},
	volume = {769},
	copyright = {http://iopscience.iop.org/info/page/text-and-data-mining},
	issn = {0004-637X, 1538-4357},
	url = {https://iopscience.iop.org/article/10.1088/0004-637X/769/1/21},
	doi = {10.1088/0004-637X/769/1/21},
	language = {en},
	number = {1},
	urldate = {2024-11-25},
	journal = {\apj},
	author = {Salyk, Colette and Herczeg, Gregory J. and Brown, Joanna M. and Blake, Geoffrey A. and Pontoppidan, Klaus M. and Van Dishoeck, Ewine F.},
	month = may,
	year = {2013},
	pages = {21},
}

@article{rigliaco_x-shooter_2012,
	title = {X-shooter spectroscopy of young stellar objects: {I}. {Mass} accretion rates of low-mass {T} {Tauri} stars in \textit{σ} {Orionis}⋆},
	volume = {548},
	issn = {0004-6361, 1432-0746},
	shorttitle = {X-shooter spectroscopy of young stellar objects},
	url = {http://www.aanda.org/10.1051/0004-6361/201219832},
	doi = {10.1051/0004-6361/201219832},
	language = {en},
	urldate = {2024-10-29},
	journal = {A\&A},
	author = {Rigliaco, E. and Natta, A. and Testi, L. and Randich, S. and Alcalà, J. M. and Covino, E. and Stelzer, B.},
	month = dec,
	year = {2012},
	pages = {A56},
}

@article{zhou_uv-optical_2023,
	title = {{UV}-optical {Emission} of {AB} {Aur} b {Is} {Consistent} with {Scattered} {Stellar} {Light}},
	volume = {166},
	issn = {0004-6256, 1538-3881},
	url = {https://iopscience.iop.org/article/10.3847/1538-3881/acf9ec},
	doi = {10.3847/1538-3881/acf9ec},
	language = {en},
	number = {6},
	urldate = {2024-03-15},
	journal = {\aj},
	author = {Zhou, Yifan and Bowler, Brendan P. and Yang, Haifeng and Sanghi, Aniket and Herczeg, Gregory J. and Kraus, Adam L. and Bae, Jaehan and Long, Feng and Follette, Katherine B. and Ward-Duong, Kimberly and Zhu, Zhaohuan and Biddle, Lauren and Close, Laird M. and Jiang, Lillian Yushu and Wu, Ya-Lin},
	month = dec,
	year = {2023},
	pages = {220},
}

@article{zhu_simple_2022,
	title = {A simple time-dependent method for calculating spirals: applications to eccentric planets in protoplanetary discs},
	volume = {510},
	issn = {0035-8711, 1365-2966},
	shorttitle = {A simple time-dependent method for calculating spirals},
	url = {https://academic.oup.com/mnras/article/510/3/3986/6462926},
	doi = {10.1093/mnras/stab3641},
	language = {en},
	number = {3},
	urldate = {2023-11-13},
	journal = {\mnras},
	author = {Zhu, Zhaohuan and Zhang, Raymond M},
	month = jan,
	year = {2022},
	pages = {3986--3999},
}

@article{di_folco_flared_2009,
	title = {The flared inner disk of the {Herbig} {Ae} star {AB} {Aurigae} revealed by {VLTI}/{MIDI} in the \textit{{N}} -band},
	volume = {500},
	issn = {0004-6361, 1432-0746},
	url = {http://www.aanda.org/10.1051/0004-6361/200809902},
	doi = {10.1051/0004-6361/200809902},
	language = {en},
	number = {3},
	urldate = {2023-10-23},
	journal = {A\&A},
	author = {Di Folco, E. and Dutrey, A. and Chesneau, O. and Wolf, S. and Schegerer, A. and Leinert, Ch. and Lopez, B.},
	month = jun,
	year = {2009},
	pages = {1065--1076},
}

@article{eriksson_strong_2020,
	title = {Strong {H} \textit{α} emission and signs of accretion in a circumbinary planetary mass companion from {MUSE}},
	volume = {638},
	issn = {0004-6361, 1432-0746},
	url = {https://www.aanda.org/10.1051/0004-6361/202038131},
	doi = {10.1051/0004-6361/202038131},
	language = {en},
	urldate = {2023-02-28},
	journal = {A\&A},
	author = {Eriksson, Simon C. and Asensio Torres, Rubén and Janson, Markus and Aoyama, Yuhiko and Marleau, Gabriel-Dominique and Bonnefoy, Mickael and Petrus, Simon},
	month = jun,
	year = {2020},
	pages = {L6},
}

@article{xie_searching_2020,
	title = {Searching for proto-planets with {MUSE}},
	volume = {644},
	issn = {0004-6361, 1432-0746},
	url = {https://www.aanda.org/10.1051/0004-6361/202038242},
	doi = {10.1051/0004-6361/202038242},
	language = {en},
	urldate = {2023-02-26},
	journal = {A\&A},
	author = {Xie, C. and Haffert, S. Y. and de Boer, J. and Kenworthy, M. A. and Brinchmann, J. and Girard, J. and Snellen, I. A. G. and Keller, C. U.},
	month = dec,
	year = {2020},
	pages = {A149},
}

@article{huelamo_searching_2022,
	title = {Searching for {H} $_{\textrm{ \textit{α} }}$ -emitting sources in the gaps of five transitional disks: {SPHERE}/{ZIMPOL} high-contrast imaging},
	volume = {668},
	issn = {0004-6361, 1432-0746},
	shorttitle = {Searching for {H} $_{\textrm{ \textit{α} }}$ -emitting sources in the gaps of five transitional disks},
	url = {https://www.aanda.org/10.1051/0004-6361/202243918},
	doi = {10.1051/0004-6361/202243918},
	language = {en},
	urldate = {2023-02-23},
	journal = {A\&A},
	author = {Huélamo, N. and Chauvin, G. and Mendigutía, I. and Whelan, E. and Alcalá, J. M. and Cugno, G. and Schmid, H. M. and de Gregorio-Monsalvo, I. and Zurlo, A. and Barrado, D. and Benisty, M. and Quanz, S. P. and Bouy, H. and Montesinos, B. and Beletsky, Y. and Szulagyi, J.},
	month = dec,
	year = {2022},
	pages = {A138},
}

@article{eisner_resolved_2004,
	title = {Resolved {Inner} {Disks} around {Herbig} {Ae}/{Be} {Stars}},
	volume = {613},
	url = {http://adsabs.harvard.edu/cgi-bin/nph-data_query?bibcode=2004ApJ...613.1049E&link_type=ABSTRACT},
	doi = {10.1086/423314},
	number = {2},
	journal = {\apj},
	author = {Eisner, J. A. and Lane, B. F. and Hillenbrand, L. A. and Akeson, R. L. and Sargent, A. I.},
	month = oct,
	year = {2004},
	pages = {1049 -- 1071},
}

@article{fukagawa_spiral_2004,
	title = {Spiral {Structure} in the {Circumstellar} {Disk} around {AB} {Aurigae}},
	volume = {605},
	url = {http://adsabs.harvard.edu/cgi-bin/nph-data_query?bibcode=2004ApJ...605L..53F&link_type=ABSTRACT},
	doi = {10.1086/420699},
	language = {English},
	number = {1},
	journal = {\apj},
	author = {Fukagawa, Misato and Hayashi, Masahiko and Tamura, Motohide and Itoh, Yoichi and Hayashi, Saeko S. and Oasa, Yumiko and Takeuchi, Taku and Morino, Jun-ichi and Murakawa, Koji and Oya, Shin and Yamashita, Takuya and Suto, Hiroshi and Mayama, Satoshi and Naoi, Takahiro and Ishii, Miki and Pyo, Tae-Soo and Nishikawa, Takayuki and Takato, Naruhisa and Usuda, Tomonori and Ando, Hiroyasu and Iye, Masanori and Miyama, Shoken M. and Kaifu, Norio},
	month = apr,
	year = {2004},
	pages = {L53 -- L56},
}

@article{soummer_apodized_2005,
	title = {Apodized {Pupil} {Lyot} {Coronagraphs} for {Arbitrary} {Telescope} {Apertures}},
	volume = {618},
	url = {http://adsabs.harvard.edu/cgi-bin/nph-data_query?bibcode=2005ApJ...618L.161S&link_type=ABSTRACT},
	doi = {10.1086/427923},
	number = {2},
	journal = {\apj},
	author = {Soummer, Rémi},
	month = jan,
	year = {2005},
	pages = {L161 -- L164},
}

@article{sallum_accreting_2015,
	title = {Accreting protoplanets in the {LkCa} 15 transition disk},
	volume = {527},
	url = {http://dx.doi.org/10.1038/nature15761},
	doi = {10.1038/nature15761},
	number = {7578},
	journal = {Nature},
	author = {Sallum, S and Follette, K B and Eisner, J. A. and Close, L M and Hinz, P. and Kratter, K and Males, J and Skemer, A and Macintosh, B and Tuthill, P. and Bailey, V and Defrère, D and Morzinski, K and Rodigas, T and Spalding, E and Vaz, A and Weinberger, A. J.},
	month = nov,
	year = {2015},
	pages = {342 -- 344},
}

@article{schmid_spherezimpol_2017,
	title = {{SPHERE}/{ZIMPOL} observations of the symbiotic system {R} {Aquarii}⋆},
	volume = {602},
	issn = {0004-6361},
	url = {http://adsabs.harvard.edu/cgi-bin/nph-data_query?bibcode=2017A%26A...602A..53S&link_type=EJOURNAL},
	doi = {10.1051/0004-6361/201629416},
	language = {English},
	journal = {A\&A},
	author = {Schmid, H. M. and Bazzon, A. and Milli, J. and Roelfsema, R. and Engler, N. and Mouillet, D. and Lagadec, E. and Sissa, E. and Sauvage, J.-F. and Ginski, C. and Baruffolo, A. and Beuzit, J. L. and Boccaletti, A. and Bohn, A. J. and Claudi, R. and Costille, A. and Desidera, S. and Dohlen, K. and Dominik, C. and Feldt, M. and Fusco, T. and Gisler, D. and Girard, J. H. and Gratton, R. and Henning, T. and Hubin, N. and Joos, F. and Kasper, M. and Langlois, M. and Pavlov, A. and Pragt, J. and Puget, P. and Quanz, S. P. and Salasnich, B. and Siebenmorgen, R. and Stute, M. and Suarez, M. and Szulágyi, J. and Thalmann, C. and Turatto, M. and Udry, S. and Vigan, A. and Wildi, F.},
	month = mar,
	year = {2017},
	pages = {A53},
}

@article{tang_planet_2017,
	title = {Planet {Formation} in {AB} {Aurigae}: {Imaging} of the {Inner} {Gaseous} {Spirals} {Observed} inside the {Dust} {Cavity}},
	volume = {840},
	url = {http://dx.doi.org/10.3847/1538-4357/aa6af7},
	doi = {10.3847/1538-4357/aa6af7},
	number = {1},
	journal = {\apj},
	author = {Tang, Ya-Wen and Guilloteau, Stephane and Dutrey, Anne and Muto, Takayuki and Shen, Bo-Ting and Gu, Pin-Gao and Inutsuka, Shu-ichiro and Momose, Munetake and Pietu, Vincent and Fukagawa, Misato and Chapillon, Edwige and Ho, Paul T P and Folco, Emmanuel Di and Corder, Stuartt and Ohashi, Nagayoshi and Hashimoto, Jun},
	month = apr,
	year = {2017},
	pages = {0 -- 0},
}

@article{muller_orbital_2018,
	title = {Orbital and atmospheric characterization of the planet within the gap of the {PDS} 70 transition disk},
	volume = {617},
	issn = {0004-6361},
	url = {https://www.aanda.org/10.1051/0004-6361/201833584},
	doi = {10.1051/0004-6361/201833584},
	language = {English},
	journal = {A\&A},
	author = {Müller, A and Keppler, M and Henning, Th and Samland, M and Chauvin, G and Beust, H and Maire, A.-L. and Molaverdikhani, K and Boekel, R van and benisty, M and Boccaletti, A and Bonnefoy, M and Cantalloube, F and Charnay, B and Baudino, J L and Gennaro, M and Long, Z C and Cheetham, A and Desidera, S and Feldt, M. and Fusco, T and Girard, J. and Gratton, R and Hagelberg, J and Janson, M and Lagrange, A M and Langlois, M. and Lazzoni, C and Ligi, R and Menard, F and Mesa, D. and Meyer, M and Mollière, P and Mordasini, C and Moulin, T and Pavlov, A. and Pawellek, N and Quanz, S. P. and Ramos, J and Rouan, D and Sissa, E and Stadler, E and Vigan, A and Wahhaj, Z. and Weber, L and Zurlo, A},
	month = sep,
	year = {2018},
	pages = {L2 -- 11},
}

@article{betti_detection_2022,
	title = {Detection of {Near}-infrared {Water} {Ice} at the {Surface} of the ({Pre}){Transitional} {Disk} of {AB} {Aur}: {Informing} {Icy} {Grain} {Abundance}, {Composition}, and {Size}},
	volume = {163},
	issn = {0004-6256, 1538-3881},
	shorttitle = {Detection of {Near}-infrared {Water} {Ice} at the {Surface} of the ({Pre}){Transitional} {Disk} of {AB} {Aur}},
	url = {https://iopscience.iop.org/article/10.3847/1538-3881/ac4d9b},
	doi = {10.3847/1538-3881/ac4d9b},
	language = {en},
	number = {4},
	urldate = {2023-01-27},
	journal = {\aj},
	author = {Betti, S. K. and Follette, K. and Jorquera, S. and Duchêne, G. and Mazoyer, J. and Bonnefoy, M. and Chauvin, G. and Pérez, L. M. and Boccaletti, A. and Pinte, C. and Weinberger, A. J. and Grady, C. and Close, L. M. and Defrère, D. and Downey, E. C. and Hinz, P. M. and Ménard, F. and Schneider, G. and Skemer, A. J. and Vaz, A.},
	month = apr,
	year = {2022},
	pages = {145},
}

@article{schmid_spherezimpol_2018,
	title = {{SPHERE}/{ZIMPOL} high resolution polarimetric imager},
	volume = {619},
	issn = {0004-6361},
	url = {https://www.aanda.org/10.1051/0004-6361/201833620},
	doi = {10.1051/0004-6361/201833620},
	language = {English},
	journal = {A\&A},
	author = {Schmid, H. M. and Bazzon, A and Roelfsema, R. and Mouillet, D and Milli, J and Menard, F and Gisler, D and Hunziker, S and Pragt, J. and Dominik, C and Boccaletti, A and Ginski, C and Abe, L and Antoniucci, S and Avenhaus, H and Baruffolo, A. and Baudoz, P and Beuzit, J L and Carbillet, M and Chauvin, G and Claudi, R. and Costille, A and Daban, J B and Haan, M de and Desidera, S and Dohlen, K and Downing, M. and Elswijk, E and Engler, N and Feldt, M. and Fusco, T and Girard, J H and Gratton, R and Hanenburg, H and Henning, Th and Hubin, N and Joos, F. and Kasper, M and Keller, C.-U. and Langlois, M. and Lagadec, E and Martinez, P and Mulder, E and Pavlov, A. and Podio, L and Puget, P. and Quanz, S. P. and Rigal, F. and Salasnich, B and Sauvage, J.-F. and Schuil, M and Siebenmorgen, R and Sissa, E and Snik, F and Suarez, M and Thalmann, Ch and Turatto, M. and Udry, S and Duin, A van and Holstein, R G van and Vigan, A and Wildi, F.},
	month = nov,
	year = {2018},
	pages = {A9 -- 37},
}

@article{holstein_polarimetric_2020,
	title = {Polarimetric imaging mode of {VLT}/{SPHERE}/{IRDIS}},
	volume = {633},
	issn = {0004-6361},
	doi = {10.1051/0004-6361/201834996},
	journal = {A\&A},
	author = {Holstein, R. G. van and Girard, J. H. and Boer, J. de and Snik, F. and Milli, J. and Stam, D. M. and Ginski, C. and Mouillet, D. and Wahhaj, Z. and Schmid, H. M. and Keller, C. U. and Langlois, M. and Dohlen, K. and Vigan, A. and Pohl, A. and Carbillet, M. and Fantinel, D. and Maurel, D. and Origné, A. and Petit, C. and Ramos, J. and Rigal, F. and Sevin, A. and Boccaletti, A. and Coroller, H. Le and Dominik, C. and Henning, T. and Lagadec, E. and Ménard, F. and Turatto, M. and Udry, S. and Chauvin, G. and Feldt, M. and Beuzit, J.-L.},
	year = {2020},
	pages = {A64},
}

@article{schmid_limb_2006,
	title = {Limb polarization of {Uranus} and {Neptune}. {I}. {Imaging} polarimetry and comparison with analytic models},
	volume = {452},
	url = {http://adsabs.harvard.edu/cgi-bin/nph-data_query?bibcode=2006A%26A...452..657S&link_type=EJOURNAL},
	doi = {10.1051/0004-6361:20053273},
	number = {2},
	journal = {A\&A},
	author = {Schmid, H. M. and Joos, F. and Tschan, D},
	month = jun,
	year = {2006},
	pages = {657 -- 668},
}

@article{riviere-marichalar_gas_2019,
	title = {Gas {Accretion} within the {Dust} {Cavity} in {AB} {Aur}},
	volume = {879},
	issn = {2041-8205},
	url = {https://iopscience.iop.org/article/10.3847/2041-8213/ab289d},
	doi = {10.3847/2041-8213/ab289d},
	number = {1},
	journal = {\apj},
	author = {Rivière-Marichalar, Pablo and Fuente, Asunción and Baruteau, Clément and Neri, Roberto and Treviño-Morales, Sandra P and Carmona, Andrés and Agúndez, Marcelino and Bachiller, Rafael},
	month = jun,
	year = {2019},
	pages = {L14},
}

@article{wagner_thermal_2019,
	title = {Thermal {Infrared} {Imaging} of {MWC} 758 with the {Large} {Binocular} {Telescope}: {Planetary}- driven {Spiral} {Arms}?},
	volume = {882},
	url = {http://dx.doi.org/10.3847/1538-4357/ab32ea},
	doi = {10.3847/1538-4357/ab32ea},
	number = {1},
	journal = {\apj},
	author = {Wagner, Kevin and Stone, Jordan M and Spalding, Eckhart and Apai, Daniel and Dong, Ruobing and Ertel, Steve and Leisenring, Jarron and Webster, Ryan},
	month = aug,
	year = {2019},
	pages = {0 -- 0},
}

@article{calcino_are_2020,
	title = {Are the spiral arms in the {MWC} 758 protoplanetary disc driven by a companion inside the cavity?},
	volume = {498},
	issn = {0035-8711},
	url = {http://arxiv.org/abs/2007.06155v1},
	doi = {10.1093/mnras/staa2468},
	number = {1},
	journal = {\mnras},
	author = {Calcino, Josh and Christiaens, Valentin and Price, Daniel J and Pinte, Christophe and Davis, Tamara M and Marel, Nienke van der and Cuello, Nicolás},
	month = jul,
	year = {2020},
	pages = {639--650},
}

@article{poblete_binary-induced_2020,
	title = {Binary-induced spiral arms inside the disc cavity of {AB} {Aurigae}},
	volume = {496},
	issn = {0035-8711},
	url = {http://arxiv.org/abs/2005.10722v1},
	doi = {10.1093/mnras/staa1655},
	number = {2},
	journal = {\mnras},
	author = {Poblete, Pedro P and Calcino, Josh and Cuello, Nicolás and Macías, Enrique and Ribas, Álvaro and Price, Daniel J and Cuadra, Jorge and Pinte, Christophe},
	month = may,
	year = {2020},
	pages = {2362--2371},
}

@article{marleau_accreting_2022,
	title = {Accreting protoplanets: {Spectral} signatures and magnitude of gas and dust extinction at {H} \textit{α}},
	volume = {657},
	issn = {0004-6361, 1432-0746},
	shorttitle = {Accreting protoplanets},
	url = {https://www.aanda.org/10.1051/0004-6361/202037494},
	doi = {10.1051/0004-6361/202037494},
	language = {en},
	urldate = {2022-10-07},
	journal = {A\&A},
	author = {Marleau, G.-D. and Aoyama, Y. and Kuiper, R. and Follette, K. and Turner, N. J. and Cugno, G. and Manara, C. F. and Haffert, S. Y. and Kitzmann, D. and Ringqvist, S. C. and Wagner, K. R. and van Boekel, R. and Sallum, S. and Janson, M. and Schmidt, T. O. B. and Venuti, L. and Lovis, Ch. and Mordasini, C.},
	month = jan,
	year = {2022},
	pages = {A38},
}

@misc{aoyama_spectral_2021,
	title = {Spectral appearance of the planetary-surface accretion shock: {Global} spectra and hydrogen-line profiles and fluxes},
	shorttitle = {Spectral appearance of the planetary-surface accretion shock},
	url = {http://arxiv.org/abs/2011.06608},
	language = {en},
	urldate = {2022-10-06},
	publisher = {arXiv},
	author = {Aoyama, Yuhiko and Marleau, Gabriel-Dominique and Mordasini, Christoph and Ikoma, Masahiro},
	month = apr,
	year = {2021},
	note = {arXiv:2011.06608 [astro-ph]},
}

@article{guerri_apodized_2011,
	title = {Apodized {Lyot} coronagraph for {SPHERE}/{VLT}: {II}. {Laboratory} tests and performance},
	volume = {30},
	url = {http://adsabs.harvard.edu/cgi-bin/nph-data_query?bibcode=2011ExA....30...59G&link_type=ABSTRACT},
	doi = {10.1007/s10686-011-9220-y},
	number = {1},
	journal = {Experimental Astronomy},
	author = {Guerri, Geraldine and Daban, Jean-Baptiste and Robbe-Dubois, Sylvie and Douet, Richard and Abe, Lyu and Baudrand, Jacques and Carbillet, Marcel and Boccaletti, Anthony and Bendjoya, Philippe and Gouvret, Carole and Vakili, Farrokh},
	month = may,
	year = {2011},
	pages = {59 -- 81},
}

@article{marino_shadows_2015,
	title = {Shadows {Cast} by a {Warp} in the {HD} 142527 {Protoplanetary} {Disk}},
	volume = {798},
	url = {http://adsabs.harvard.edu/cgi-bin/nph-data_query?bibcode=2015ApJ...798L..44M&link_type=EJOURNAL},
	doi = {10.1088/2041-8205/798/2/l44},
	number = {2},
	journal = {\apjl},
	author = {Marino, S and Perez, S and Casassus, S},
	month = jan,
	year = {2015},
	pages = {L44},
}

@article{hashimoto_direct_2011,
	title = {Direct {Imaging} of {Fine} {Structures} in {Giant} {Planet}-forming {Regions} of the {Protoplanetary} {Disk} {Around} {AB} {Aurigae}},
	volume = {729},
	url = {http://adsabs.harvard.edu/cgi-bin/nph-data_query?bibcode=2011ApJ...729L..17H&link_type=EJOURNAL},
	doi = {10.1088/2041-8205/729/2/l17},
	number = {2},
	journal = {\apjl},
	author = {Hashimoto, J. and Tamura, M and Muto, T and Kudo, T. and Fukagawa, M. and Fukue, T. and Goto, M. and Grady, C A and Henning, T and Hodapp, K and Honda, M. and Inutsuka, S and Kokubo, E and Knapp, G and McElwain, M. W. and Momose, M and Ohashi, N and Okamoto, Y K and Takami, M. and Turner, E. L. and Wisniewski, J and Janson, M and Abe, L and Brandner, W and Carson, J. and Egner, S. and Feldt, M. and Golota, T. and Guyon, O and Hayano, Y. and Hayashi, M. and Hayashi, S. and Ishii, M. and Kandori, R. and Kusakabe, N. and Matsuo, T. and Mayama, S and Miyama, S. and Morino, J.-I. and Moro-Martín, A. and Nishimura, T. and Pyo, T S and Suto, H. and Suzuki, R. and Takato, N. and Terada, H. and Thalmann, C. and Tomono, D. and Watanabe, M. and Yamada, T. and Takami, H. and Usuda, T.},
	month = mar,
	year = {2011},
	pages = {L17},
}

@article{ren_dynamical_2020,
	title = {Dynamical {Evidence} of a {Spiral} {Arm}–driving {Planet} in the {MWC} 758 {Protoplanetary} {Disk}},
	volume = {898},
	url = {http://dx.doi.org/10.3847/2041-8213/aba43e},
	doi = {10.3847/2041-8213/aba43e},
	number = {2},
	journal = {\apj},
	author = {Ren, Bin and Dong, Ruobing and Holstein, Rob G van and Ruffio, Jean-Baptiste and Calvin, Benjamin A and Girard, Julien H and Benisty, Myriam and Boccaletti, Anthony and Esposito, Thomas M and Choquet, Élodie and Mawet, Dimitri and Pueyo, Laurent and Stolker, Tomas and Chiang, Eugene and Boer, Jozua de and Debes, John H and Garufi, Antonio and Grady, Carol A and Hines, Dean C and Maire, Anne-Lise and Ménard, François and Millar-Blanchaer, Maxwell A and Perrin, Marshall D and Poteet, Charles A and Schneider, Glenn},
	month = jul,
	year = {2020},
	pages = {L38},
}

@article{haffert_two_2019,
	title = {Two accreting protoplanets around the young star {PDS} 70},
	volume = {3},
	url = {http://adsabs.harvard.edu/cgi-bin/nph-data_query?bibcode=2019NatAs...3..749H&link_type=EJOURNAL},
	doi = {10.1038/s41550-019-0780-5},
	journal = {Nat. Astron.},
	author = {Haffert, S Y and Bohn, A J and Boer, J De and Snellen, I A G and Brinchmann, J and Girard, J H and Keller, C.-U. and Bacon, R},
	month = jun,
	year = {2019},
	pages = {749 -- 754},
}

@article{soummer_detection_2012,
	title = {Detection and {Characterization} of {Exoplanets} and {Disks} {Using} {Projections} on {Karhunen}-{Loève} {Eigenimages}},
	volume = {755},
	url = {http://adsabs.harvard.edu/cgi-bin/nph-data_query?bibcode=2012ApJ...755L..28S&link_type=ABSTRACT},
	doi = {10.1088/2041-8205/755/2/l28},
	number = {2},
	journal = {\apjl},
	author = {Soummer, Rémi and Pueyo, Laurent and Larkin, James},
	month = aug,
	year = {2012},
	pages = {L28},
}

@article{cugno_search_2019,
	title = {A search for accreting young companions embedded in circumstellar disks. {High}-contrast {Hα} imaging with {VLT}/{SPHERE}},
	volume = {622},
	url = {http://adsabs.harvard.edu/cgi-bin/nph-data_query?bibcode=2019A%26A...622A.156C&link_type=EJOURNAL},
	doi = {10.1051/0004-6361/201834170},
	journal = {A\&A},
	author = {Cugno, G and Quanz, S. P. and Hunziker, S and Stolker, T and Schmid, H. M. and Avenhaus, H and Baudoz, P and Bohn, A J and Bonnefoy, M and Buenzli, E. and Chauvin, G and Cheetham, A and Desidera, S and Dominik, C and Feautrier, P. and Feldt, M. and Ginski, C and Girard, J H and Gratton, R and Hagelberg, J and Hugot, E and Janson, M and Lagrange, A M and Langlois, M. and Magnard, Y and Maire, A.-L. and Menard, F and Meyer, M and Milli, J and Mordasini, C and Pinte, C and Pragt, J. and Roelfsema, R. and Rigal, F. and Szulágyi, J and Boekel, R van and Plas, G van der and Vigan, A and Wahhaj, Z. and Zurlo, A},
	month = feb,
	year = {2019},
	pages = {A156},
}

@article{gratton_blobs_2019,
	title = {Blobs, spiral arms, and a possible planet around {HD} 169142★★★},
	volume = {623},
	issn = {0004-6361},
	url = {http://adsabs.harvard.edu/cgi-bin/nph-data_query?bibcode=2019A%26A...623A.140G&link_type=EJOURNAL},
	doi = {10.1051/0004-6361/201834760},
	language = {English},
	journal = {A\&A},
	author = {Gratton, R. and Ligi, R. and Sissa, E. and Desidera, S. and Mesa, D. and Bonnefoy, M. and Chauvin, G. and Cheetham, A. and Feldt, M. and Lagrange, A. M. and Langlois, M. and Meyer, M. and Vigan, A. and Boccaletti, A. and Janson, M. and Lazzoni, C. and Zurlo, A. and Boer, J. De and Henning, T. and D’Orazi, V. and Gluck, L. and Madec, F. and Jaquet, M. and Baudoz, P. and Fantinel, D. and Pavlov, A. and Wildi, F.},
	month = jan,
	year = {2019},
	pages = {A140},
}

@article{boccaletti_possible_2020,
	title = {Possible evidence of ongoing planet formation in {AB} {Aurigae}},
	volume = {637},
	issn = {0004-6361},
	url = {https://www.aanda.org/10.1051/0004-6361/202038008},
	doi = {10.1051/0004-6361/202038008},
	journal = {A\&A},
	author = {Boccaletti, A. and Folco, E. Di and Pantin, E. and Dutrey, A. and Guilloteau, S. and Tang, Y. W. and Piétu, V. and Habart, E. and Milli, J. and Beck, T. L. and Maire, A.-L.},
	month = may,
	year = {2020},
	pages = {L5},
}

@article{boer_polarimetric_2020,
	title = {Polarimetric imaging mode of {VLT}/{SPHERE}/{IRDIS}},
	volume = {633},
	issn = {0004-6361},
	doi = {10.1051/0004-6361/201834989},
	journal = {A\&A},
	author = {Boer, J. de and Langlois, M. and Holstein, R. G. van and Girard, J. H. and Mouillet, D. and Vigan, A. and Dohlen, K. and Snik, F. and Keller, C. U. and Ginski, C. and Stam, D. M. and Milli, J. and Wahhaj, Z. and Kasper, M. and Schmid, H. M. and Rabou, P. and Gluck, L. and Hugot, E. and Perret, D. and Martinez, P. and Weber, L. and Pragt, J. and Sauvage, J.-F. and Boccaletti, A. and Coroller, H. Le and Dominik, C. and Henning, T. and Lagadec, E. and Ménard, F. and Turatto, M. and Udry, S. and Chauvin, G. and Feldt, M. and Beuzit, J.-L.},
	year = {2020},
	pages = {A63},
}

\begin{appendix}

\section{IRDIS DPI images}
\anthony{
Figure \ref{fig:irdap_1epoch_JHKband} presents the images of AB\,Aur in three near-IR broad band filters, and Fig. \ref{fig:dpi_uphi} the U$_{\phi}$ signal in the H band for the three epochs. }

\begin{figure*}
    \centering
    \includegraphics[width=18cm]{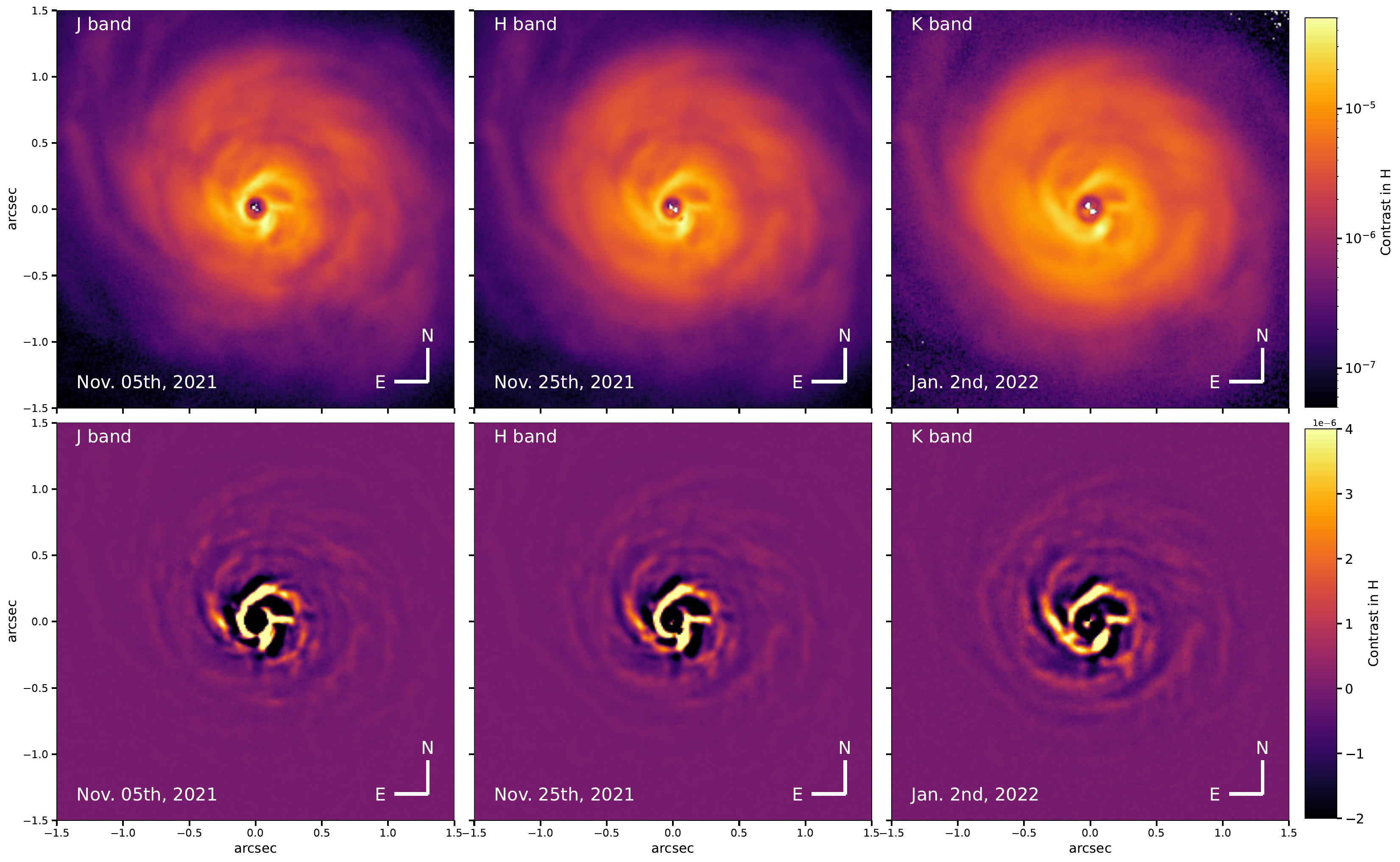}
    \caption{Same as Fig. \ref{fig:irdap_3epochs_Hband} but at one period of observations (Nov. 2021 through Jan. 2022) and for the J, H and Ks bands. For visualisation purpose the contrasts of the J and K bands were renormalized to that of the H band. 
    }
    \label{fig:irdap_1epoch_JHKband}
\end{figure*}

\begin{figure*}
    \centering
    \includegraphics[width=18cm, clip]{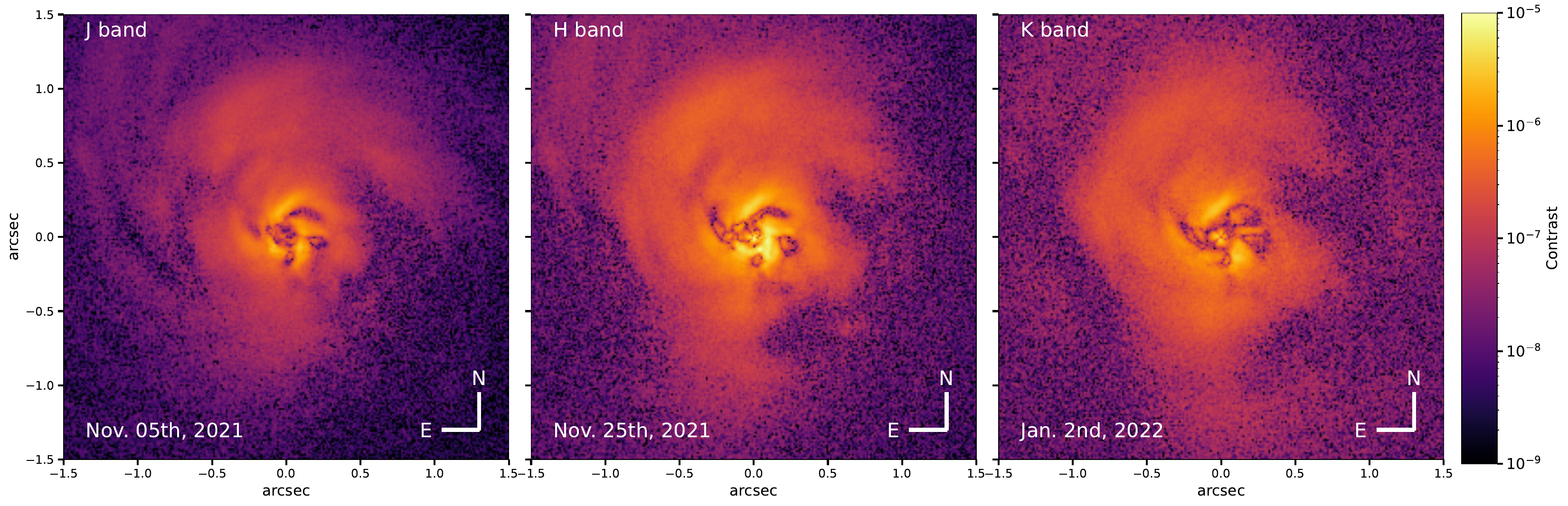}
    \caption{AB Aur U$_{\phi}$ images in J, H, and Ks bands in a $3''\times3''$ field of view as obtained in P108. The color bar reports the contrast obtained by normalizing with the out of mask PSF image and is displayed in logarithmic scale. North is up and east is left.}
    \label{fig:dpi_uphi}
\end{figure*}


\section{Dynamical analysis}

\subsection{Dynamics of the spiral arms in the disk}
\label{appendix:dynam_spirals}
We repeated the analysis of section \ref{sec:dynamical} to estimate the rotation of the disk between the three epochs, but focusing on the main spiral arms and the bridge. In Fig. \ref{fig:rotation_withmasks}, we present the results for five different masks. As for the spiral S1, the mask S1a includes the feature \texttt{f1} and the brightest part of S1, while the mask S1b excludes \texttt{f1} but extend to the northern part of the spiral. For what concern S2, similarly, we considered two masks, S2a located at the brightest northern tip, and S2b which follows S2a all the way to the southern part. The last mask is encompassing the bridge only. 

\subsection{Dynamics of the features}
\label{appendix:dynam_features}
Further analysis of the dynamics of the features requires to explore more orbital solutions, in particular those that are out of the disk plane. To account for such solutions, we use the package \texttt{orbitize!} \citep{blunt_orbitize_2020}. Measuring precisely the position of the features is a complicated task, the Gaussian fitting being systematically biased by the other surrounding spiral features as well as the evolution of the feature themselves (see Fig. \ref{fig:rotation_features}). Therefore, we proceed with a coarse visual centroiding, assigning a conservative error bar of 10\,mas per axis (about 1 pixel).
These astrometric measurements for \texttt{f1}, \texttt{f2}, and \texttt{f3} are provided in Tab. \ref{tab:astromf2}. 
For \texttt{f2}, this is in good agreement with Gaussian fitting within error bars, while because of the above mentioned biases, for \texttt{f3}, and particularly \texttt{f1} (since the feature is extended) this is more discrepant.
Because of the small orbital coverage and the few data points, we opt for the OFTI method instead of MCMC as it is more relevant in that particular case \citep{blunt_orbits_2017}. In addition, we impose prior constraints, restricting the eccentricity to a range of $0.0-0.3$ with a uniform distribution (to minimize biases), leaving the other parameters free. 
The results are shown in Fig. \ref{fig:astromf1f2f3} and Fig. \ref{fig:cornerf1f2f3}. 
Interestingly, even if the orbit sampling is very poor, we note that the orbits of the three features have their inclination from the sky plane peaking at about $50\degb-52\degb$ for \texttt{f1} and \texttt{f2}, and $57\degb$ for \texttt{f3}, but with large dispersion of typically $\pm20\degb$. Calculating the mutual inclinations, $\phi$, with the disk plane requires to take into account the position angle of the disk, $\Omega_d$, and the position angle of the nodes of the features' orbits, $\Omega_f$, as well as the inclination of the planes, $i_d$ and $i_f$, as follows:
\begin{equation}
    \mathrm{cos}\,\phi = \mathrm{cos}\,i_d.\mathrm{cos}\,i_f +\mathrm{sin}\,i_d.\mathrm{sin}\,i_f.\mathrm{cos}\,(\Omega_d-\Omega_f)
    \label{eq:incmut}
\end{equation}
As seen in the corner plots (Fig. \ref{fig:cornerf1f2f3}) the orbital fits provide systematically two solutions for $\Omega_f$ which differ by $\sim180\degb$. The corresponding mutual inclinations for \texttt{f1}, \texttt{f2}, and \texttt{f3} are : $75\degb$, $38\degb$, $34\degb$, or $32\degb$, $77\degb$, $78\degb$.
Although, the posterior distributions of the inclinations are rather broad, the mutual inclinations would always be larger than $\gtrsim15\degb$, so not compatible with perfectly coplanar orbits.
Furthermore, this experiment shows that the apparent stability of \texttt{f2} can be interpreted as an object orbiting out of the disk plane. 

\begin{table}   
\caption{Astrometry of \texttt{f1}, \texttt{f2} and \texttt{f3}.} 
\begin{tabular}{l l l l}     
\hline\hline                             
Date UT & MJD (day) & RA (mas) & DEC (mas)   \\ 
\hline                    
   \texttt{f1} \\ \hline
   2019-12-17 & 58835 & $-64.9 \pm 10$  & $-160.6 \pm 10$     \\  
   2021-11-04 & 59543 & $-80.4 \pm 10$  & $-151.5 \pm 10$    \\
   2023-10-23 & 60240 & $-91.5 \pm 10$ & $-151.7 \pm 10$  \\ \hline
   \texttt{f2} \\ \hline
   2019-12-17 & 58835 & $87.7 \pm 10$  & $667.1 \pm 10$     \\  
   2021-11-04 & 59543 & $94.9 \pm 10$  & $662.7 \pm 10$    \\
   2023-10-23 & 60240 & $106.7 \pm 10$ & $663.2 \pm 10$  \\ \hline
   \texttt{f3} \\ \hline
   2019-12-17 & 58835 & $154.9 \pm 10$  & $125.3 \pm 10$     \\  
   2021-11-04 & 59543 & $156.7 \pm 10$  & $115.4 \pm 10$    \\
   2023-10-23 & 60240 & $157.8 \pm 10$ & $96.0 \pm 10$  \\ \hline

\hline   

\end{tabular}
\tablefoot{Date of the observations in UT and MJD, with the corresponding astrometric position in RA, DEC relative to the star.}
\label{tab:astromf2}
\end{table}

\begin{figure*}[ht]
    \centering
    \includegraphics[height=3.5cm, clip]{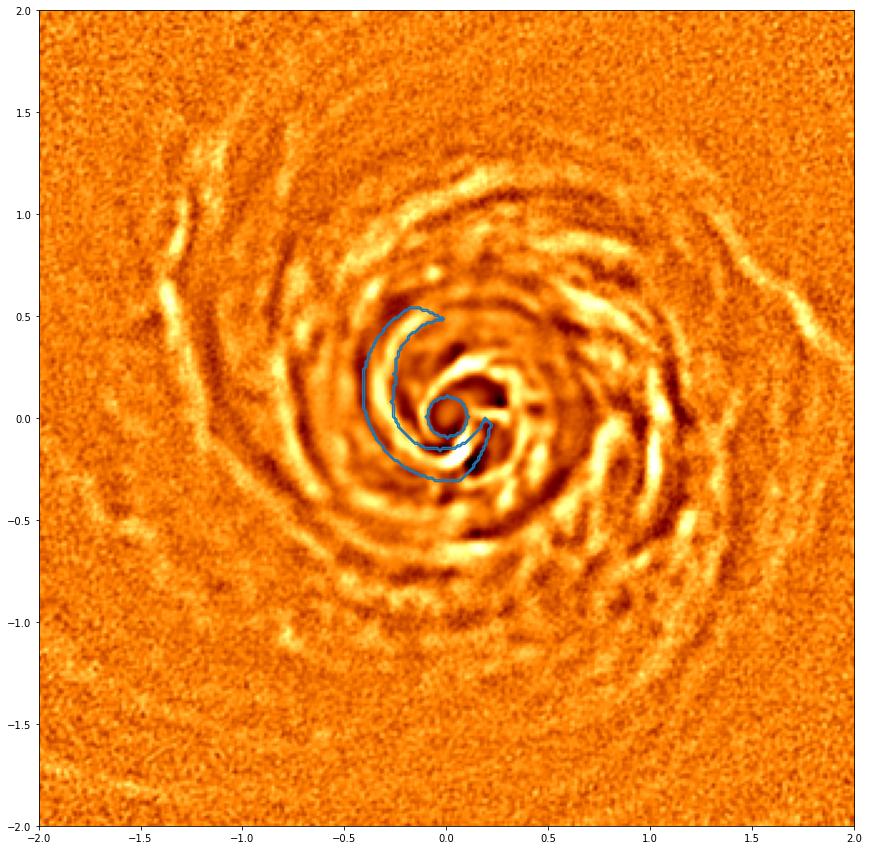}
    \includegraphics[height=3.5cm, clip]{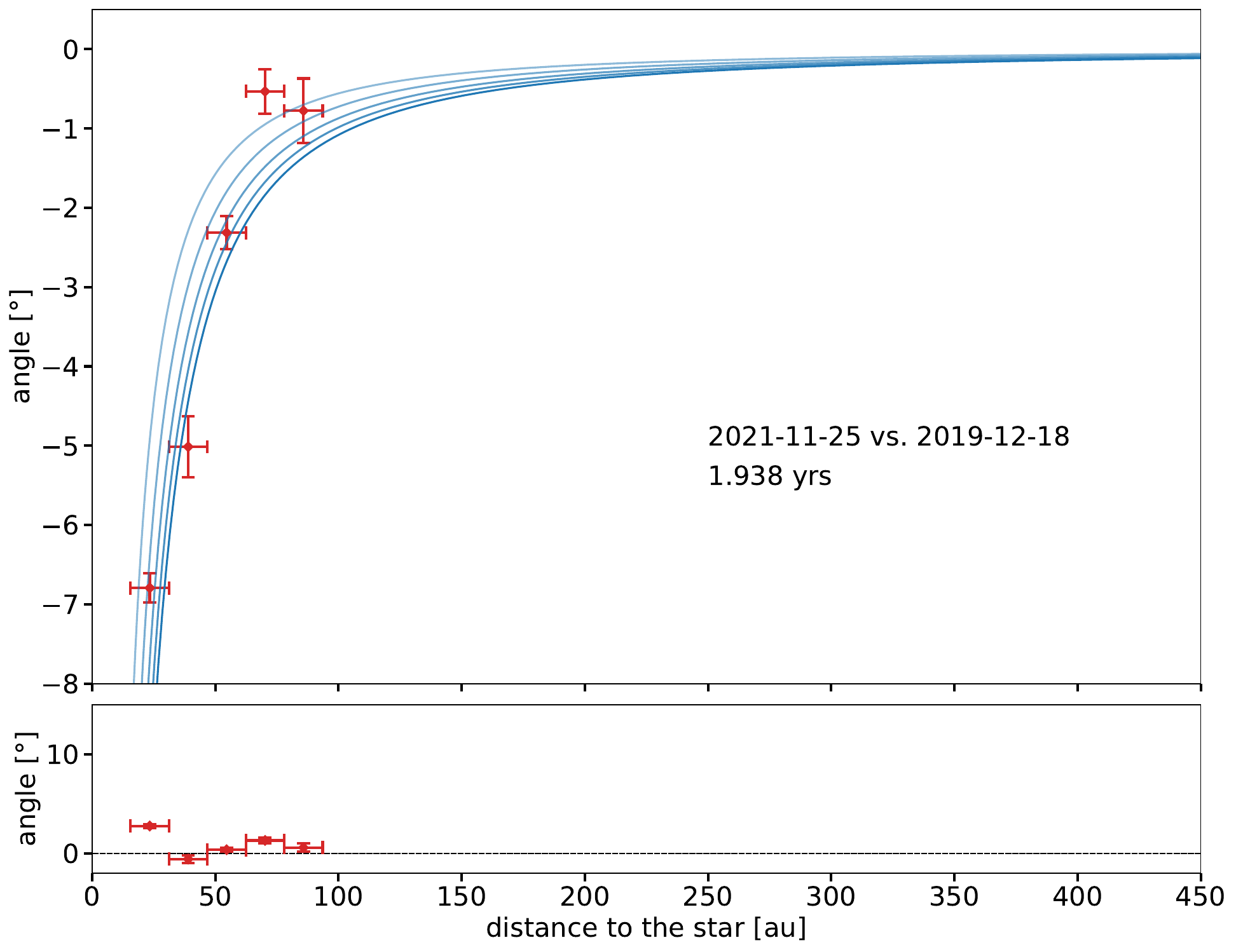}
    \includegraphics[height=3.5cm, clip]{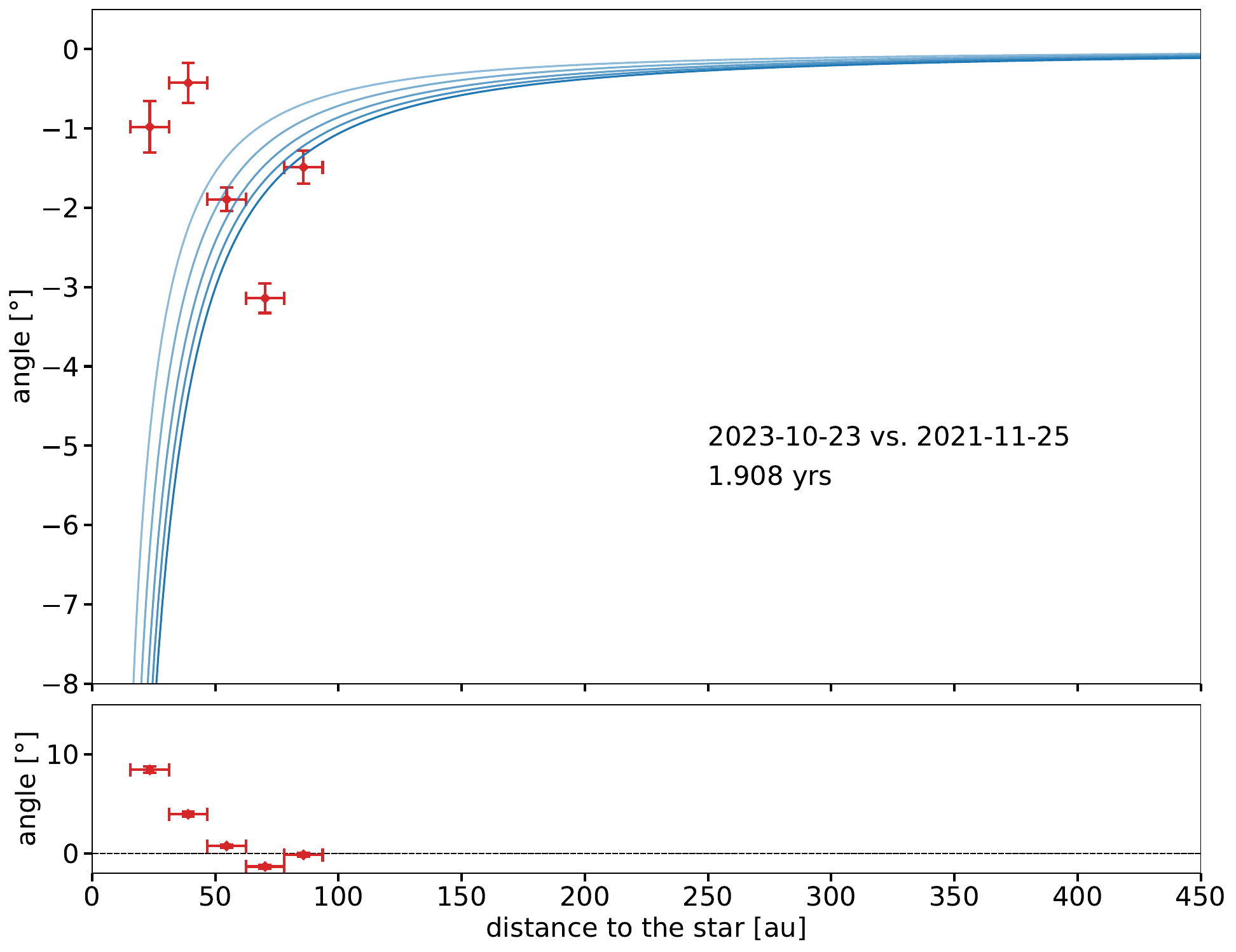}
    \includegraphics[height=3.5cm, clip]{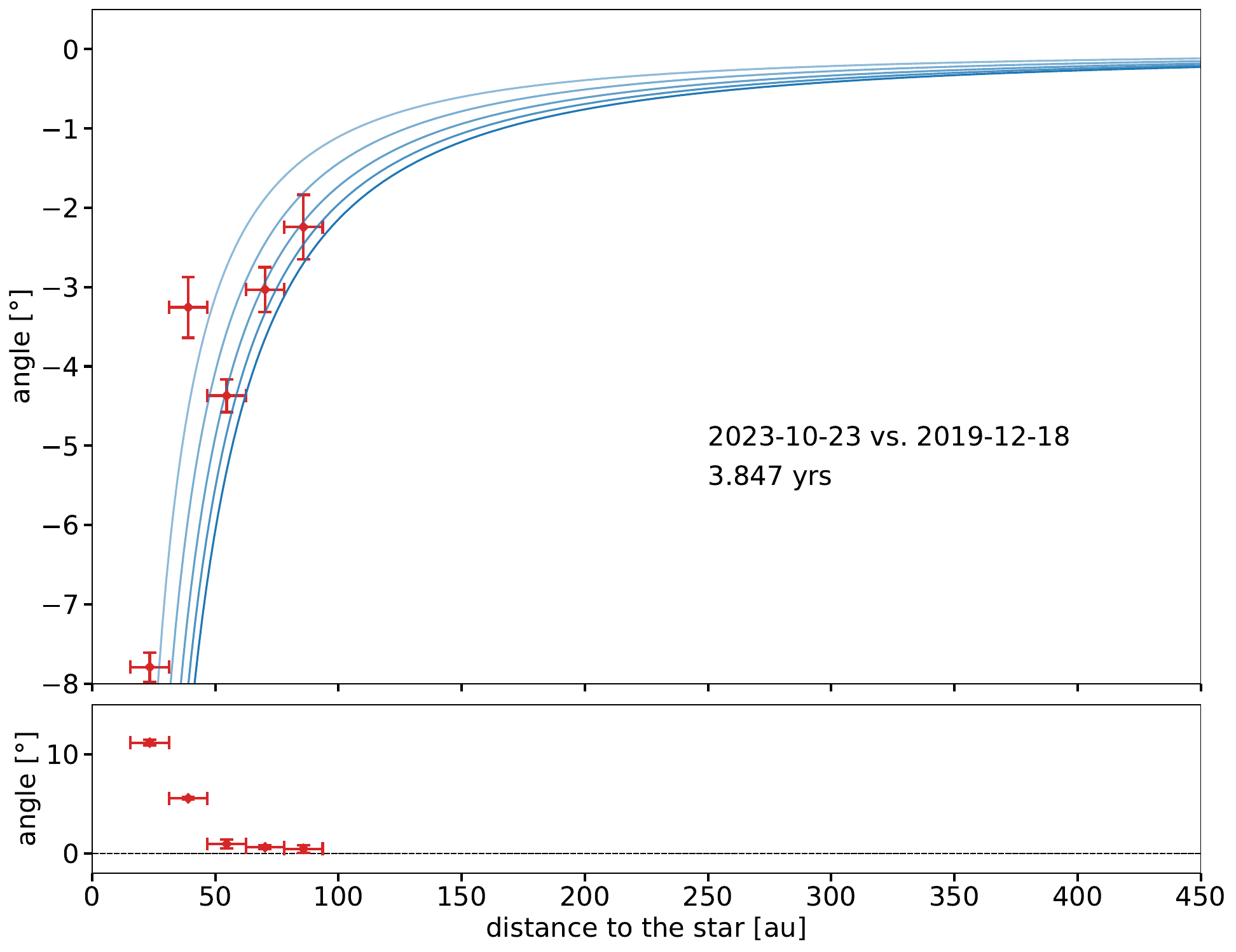}

    \includegraphics[height=3.5cm, clip]{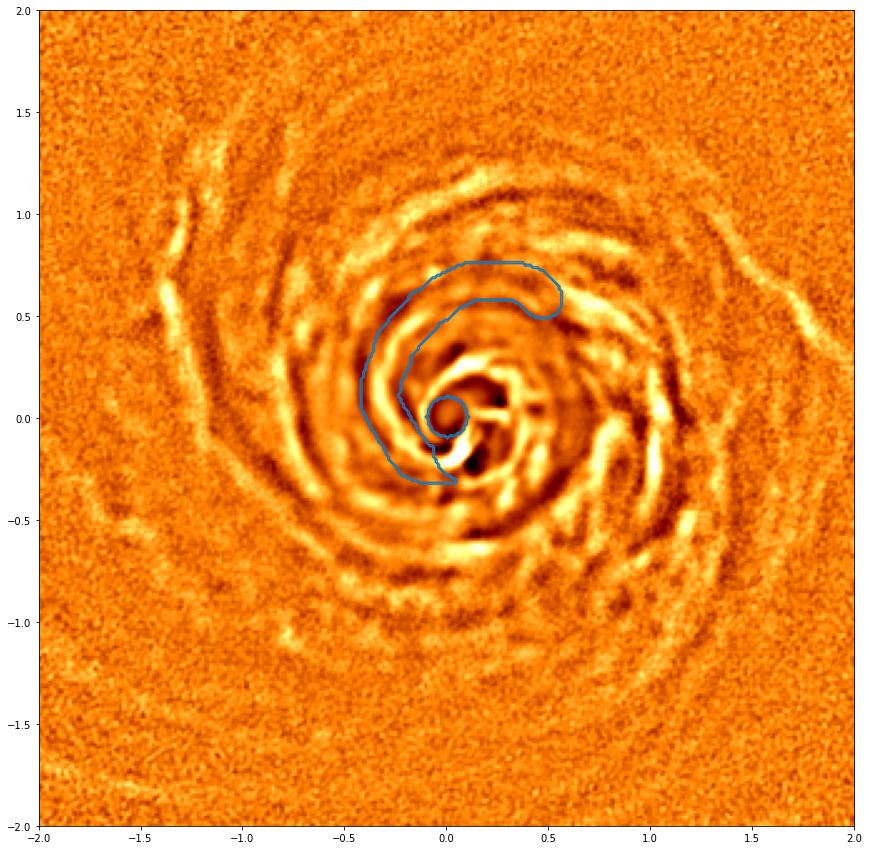}
    \includegraphics[height=3.5cm, clip]{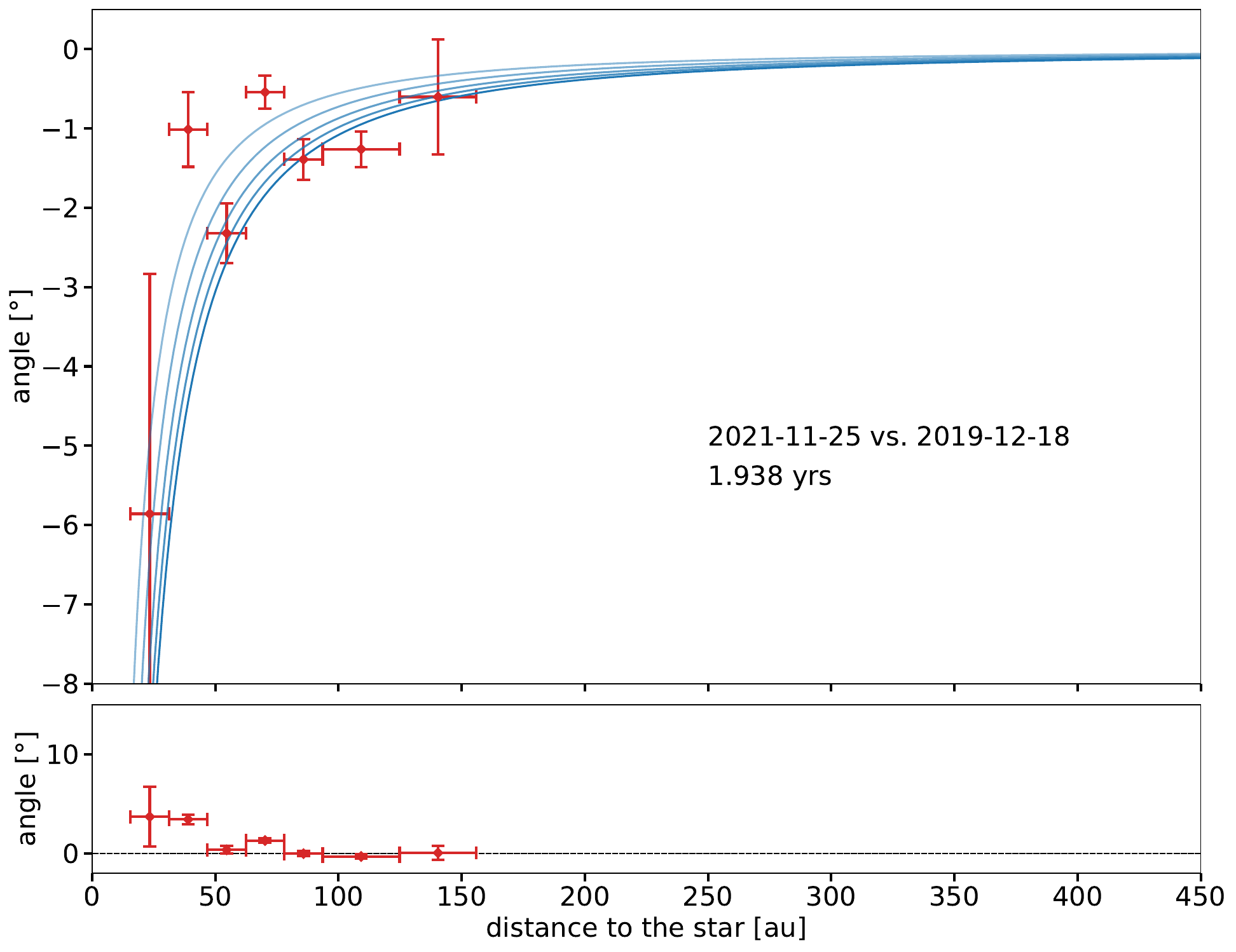}
    \includegraphics[height=3.5cm, clip]{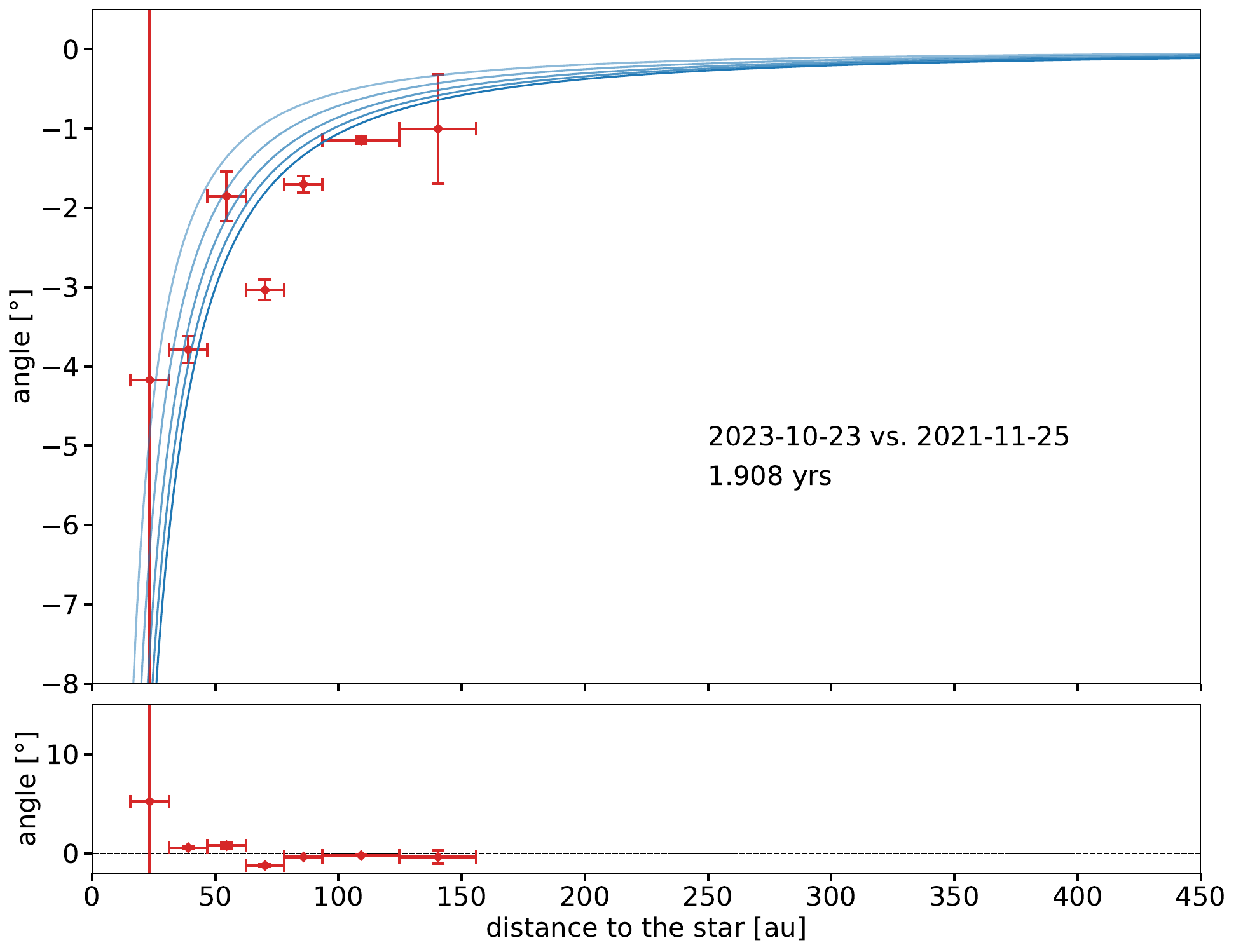}
    \includegraphics[height=3.5cm, clip]{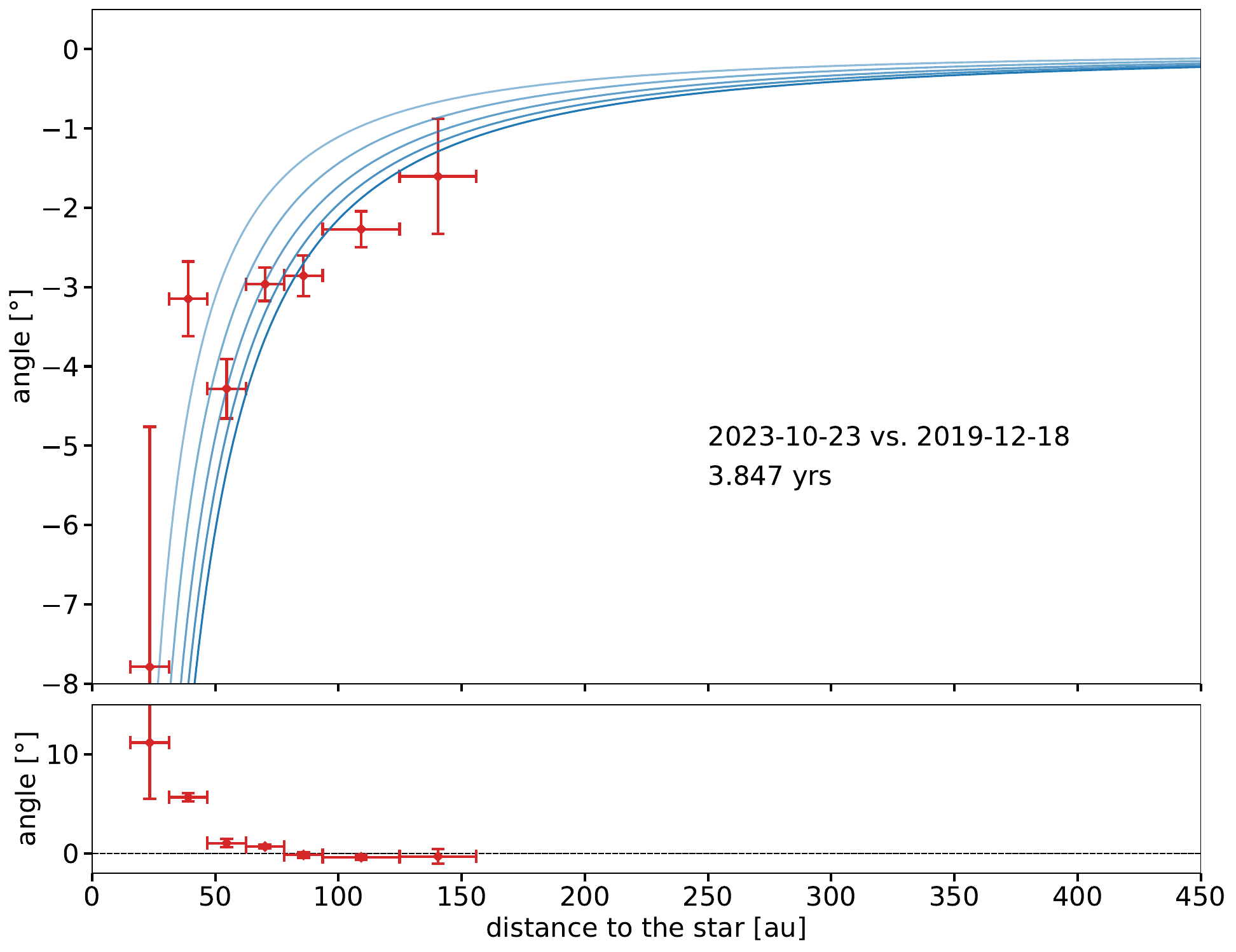}
   
    \includegraphics[height=3.5cm, clip]{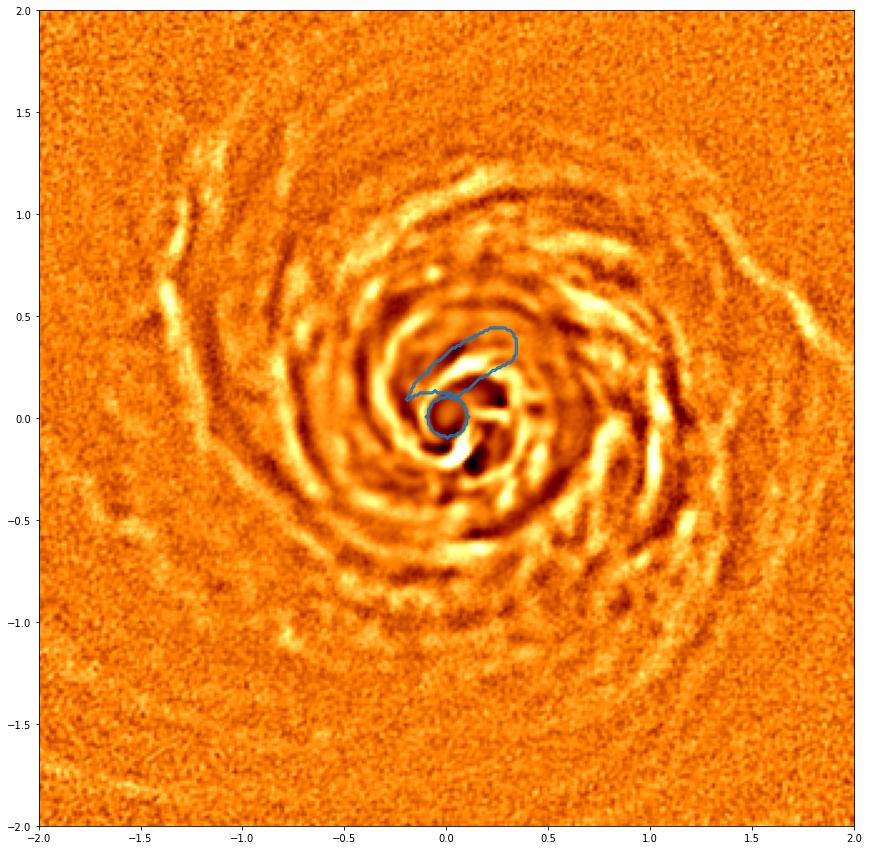}
    \includegraphics[height=3.5cm, clip]{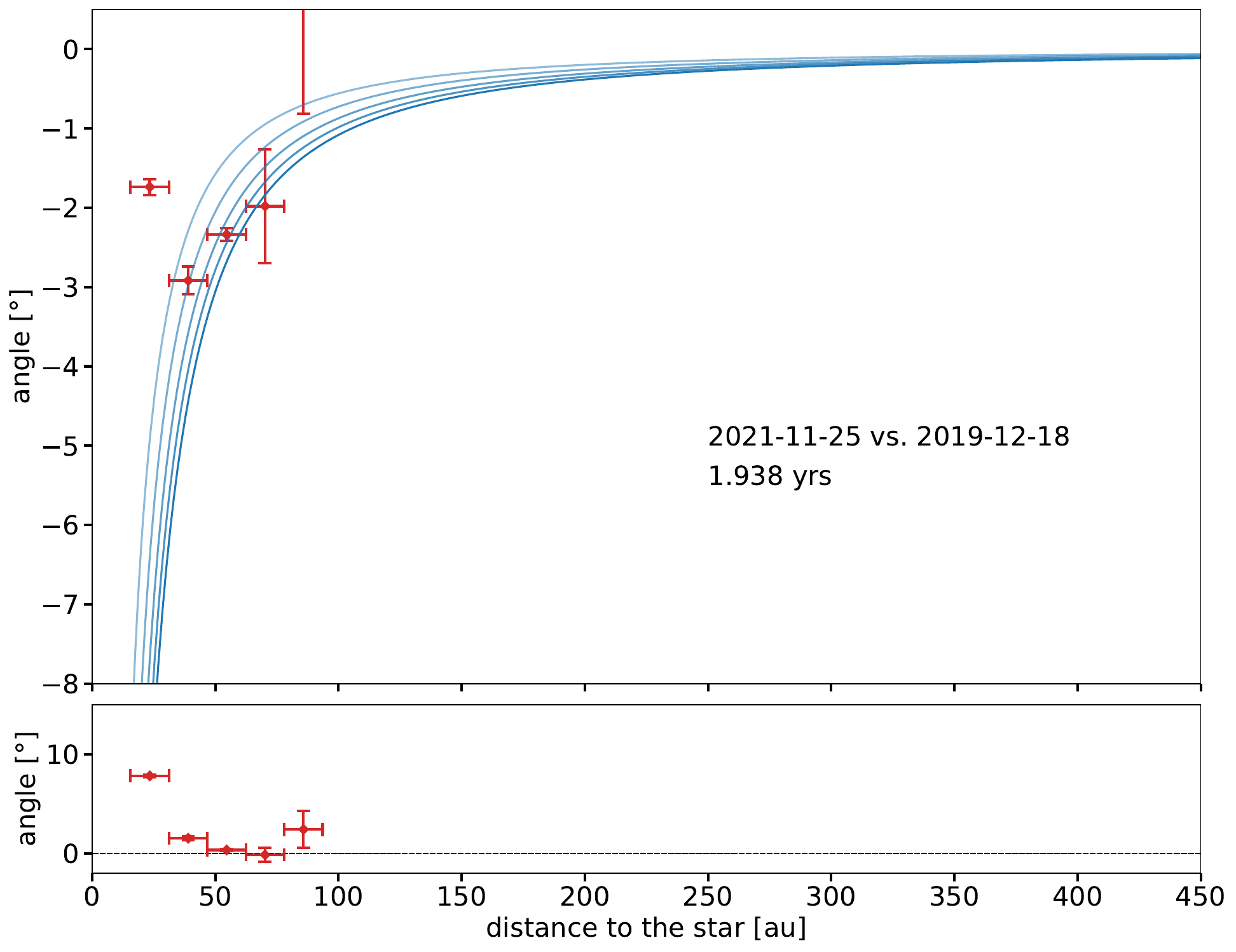}
    \includegraphics[height=3.5cm, clip]{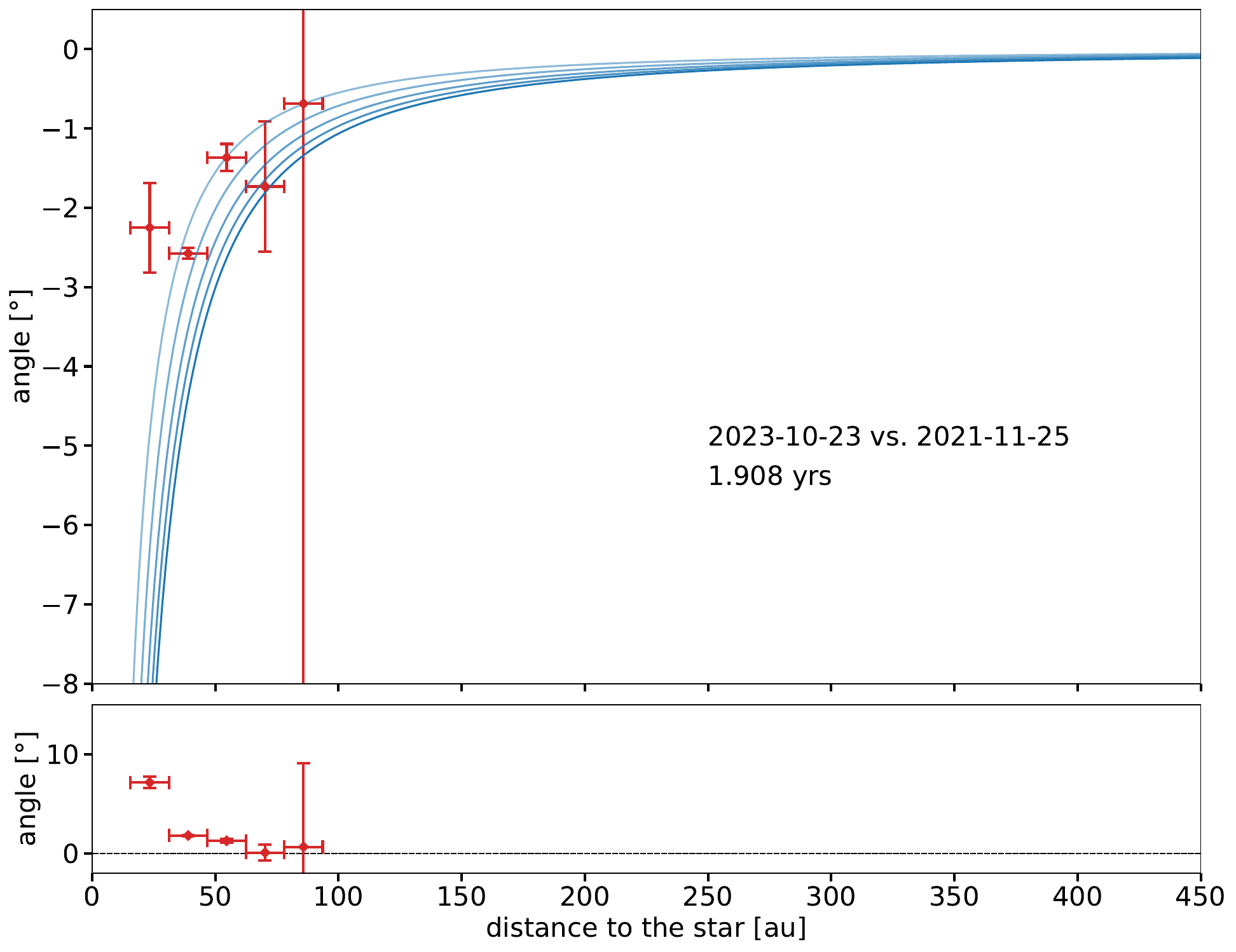}
    \includegraphics[height=3.5cm, clip]{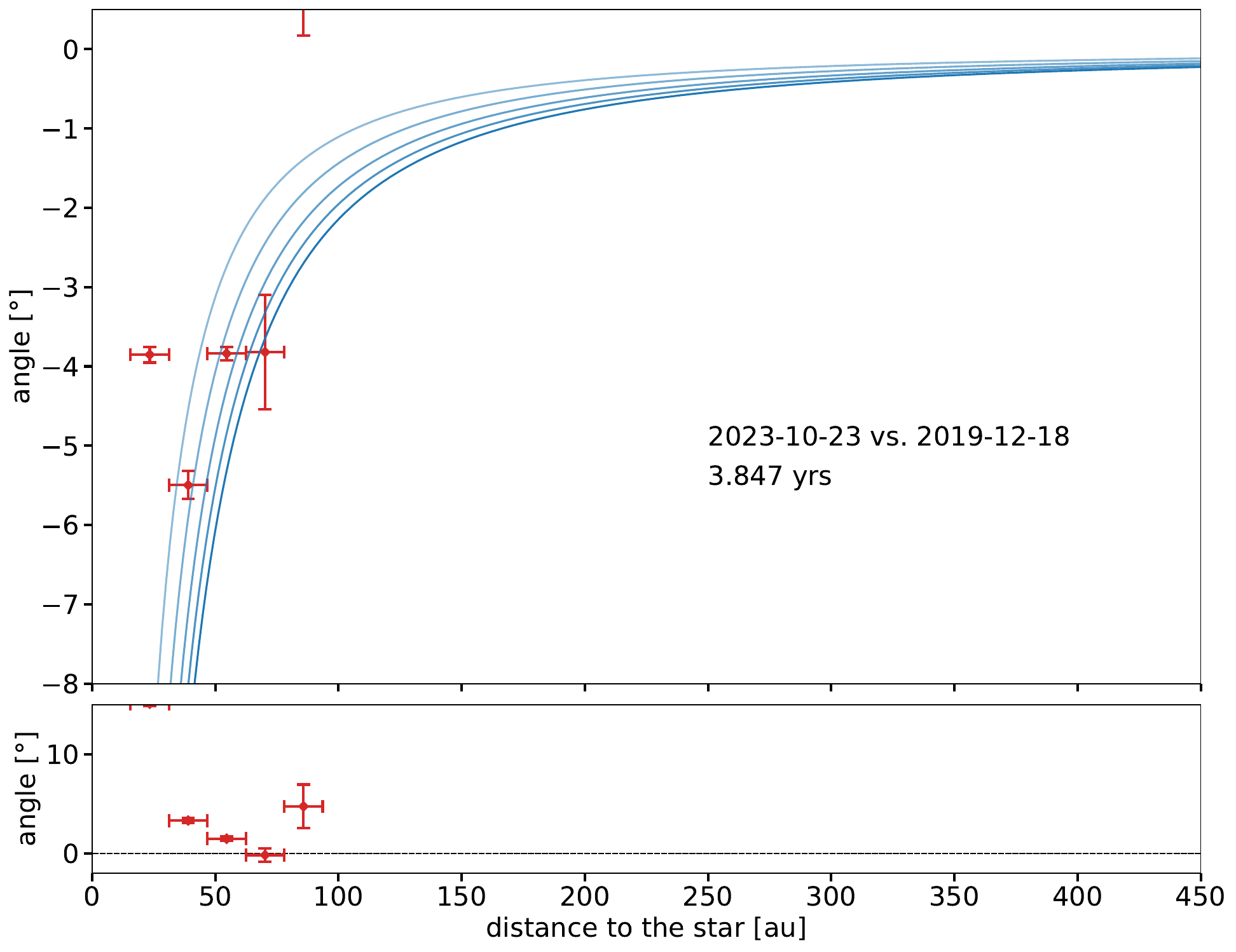}

    \includegraphics[height=3.5cm, clip]{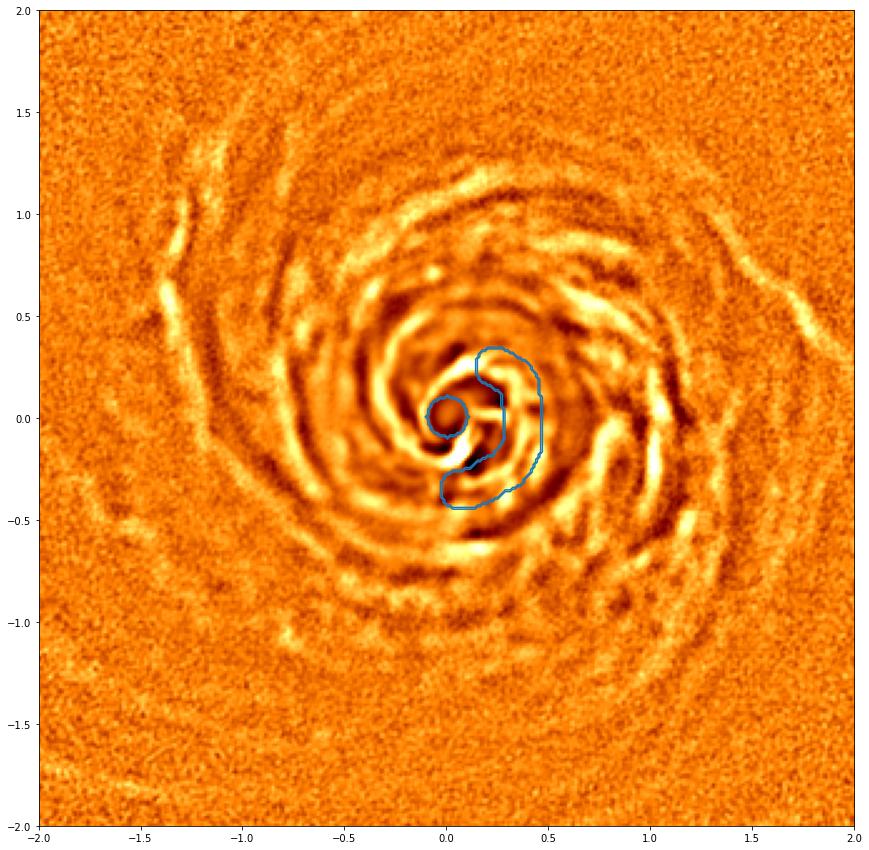}
    \includegraphics[height=3.5cm, clip]{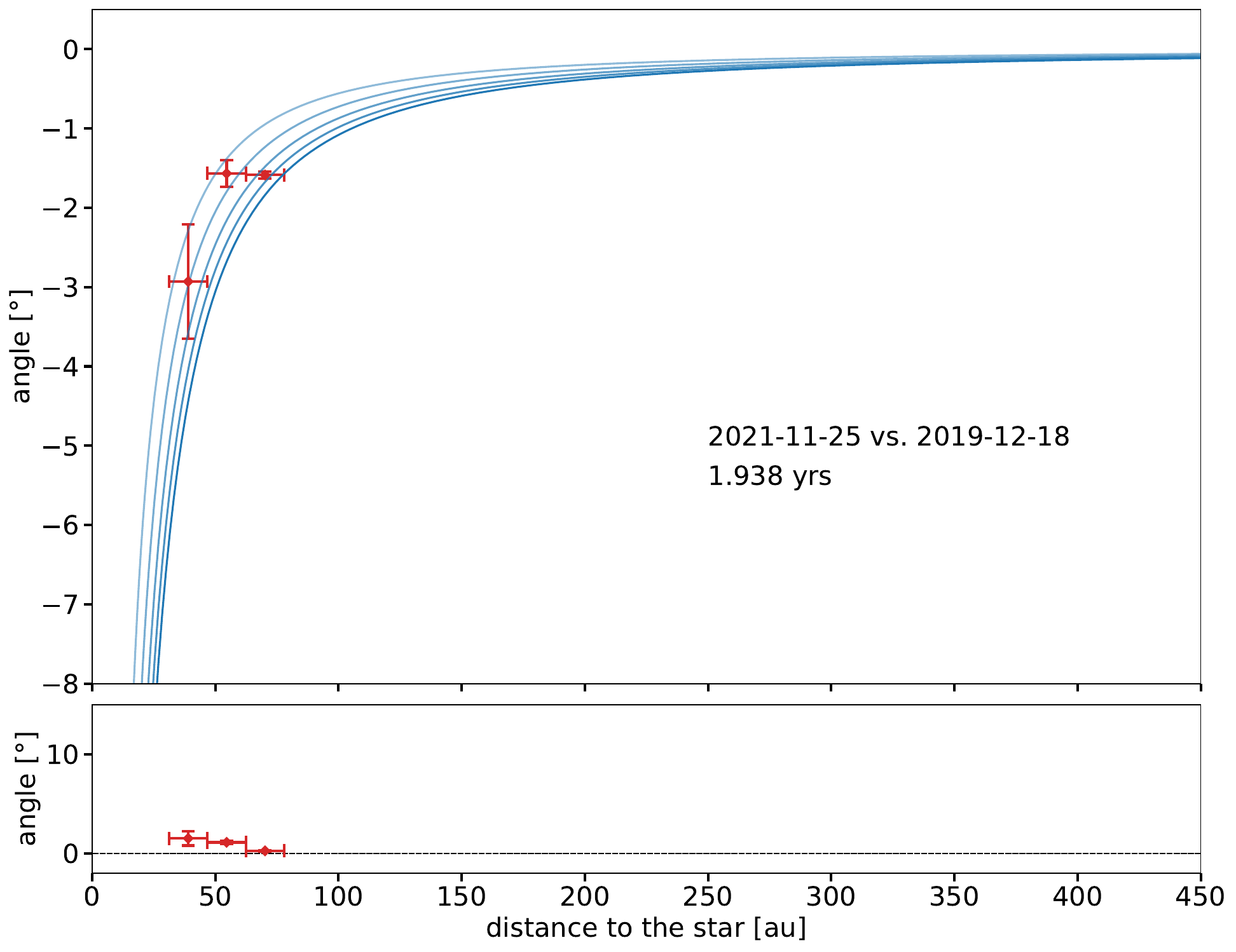}
    \includegraphics[height=3.5cm, clip]{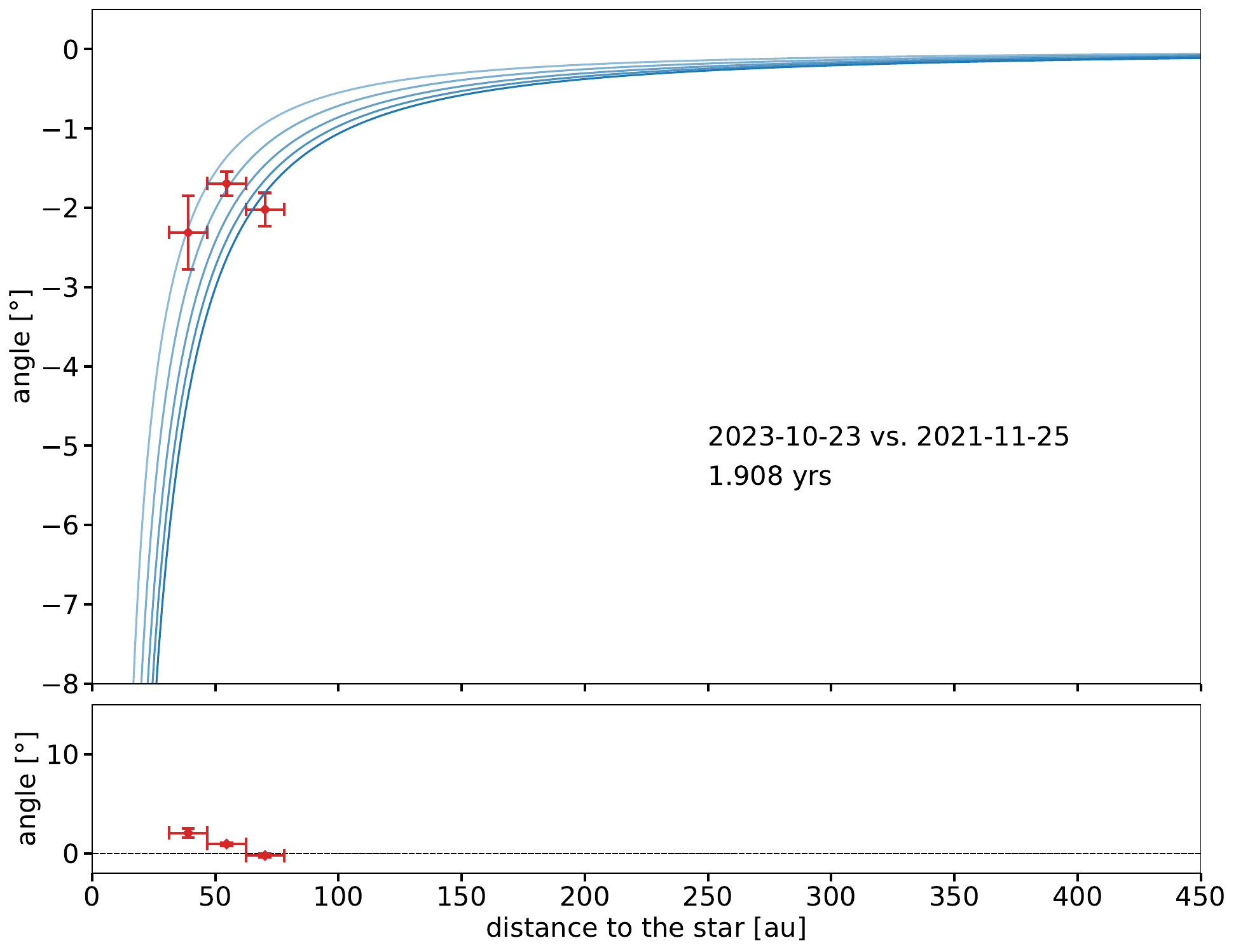}
    \includegraphics[height=3.5cm, clip]{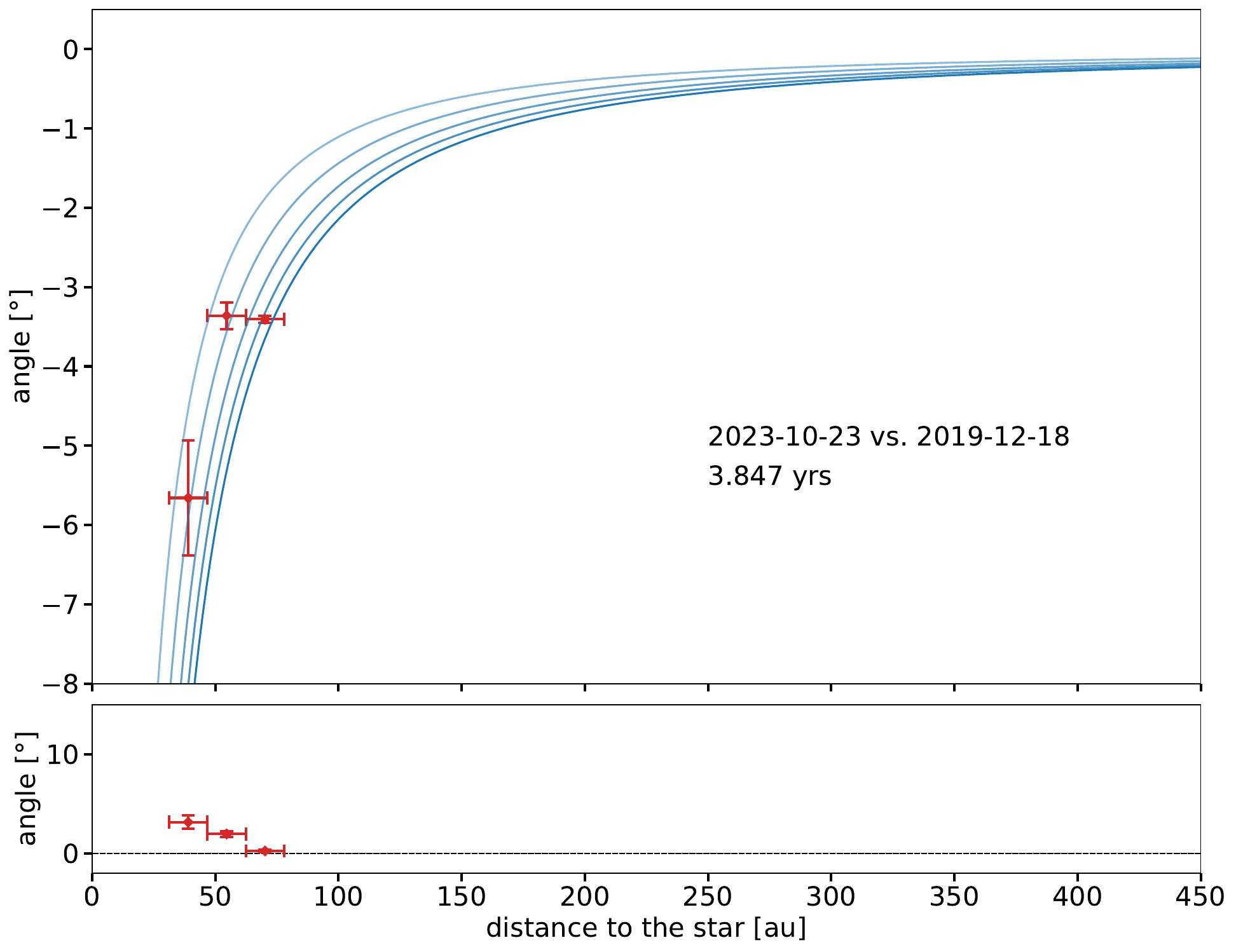}

    \includegraphics[height=3.5cm, clip]{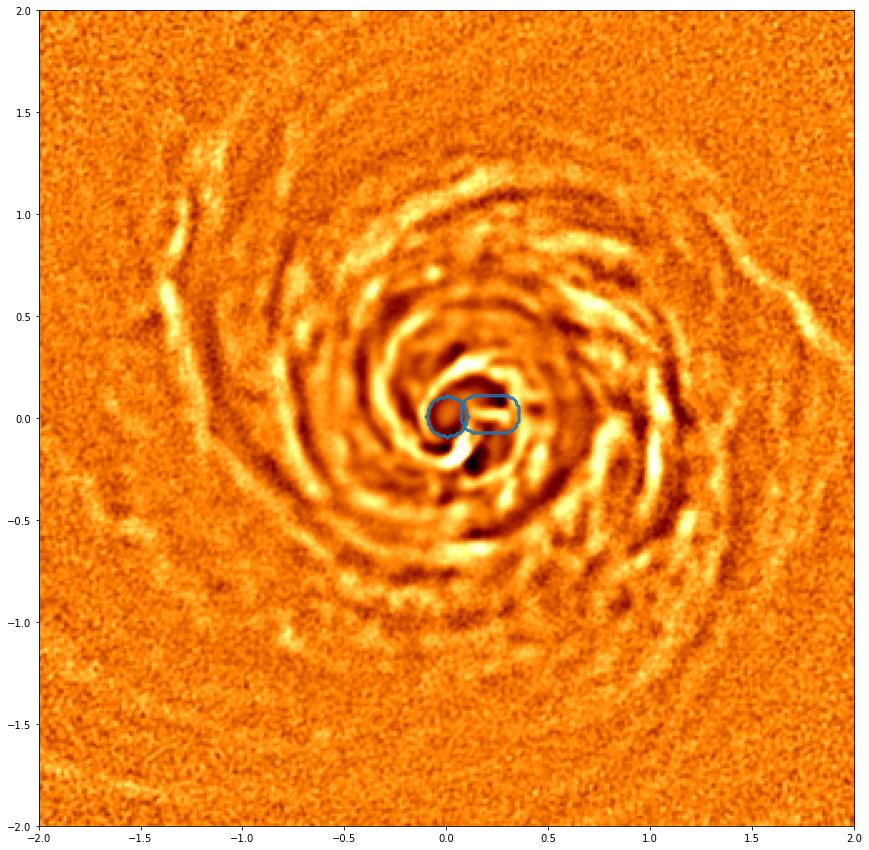}
    \includegraphics[height=3.5cm, clip]{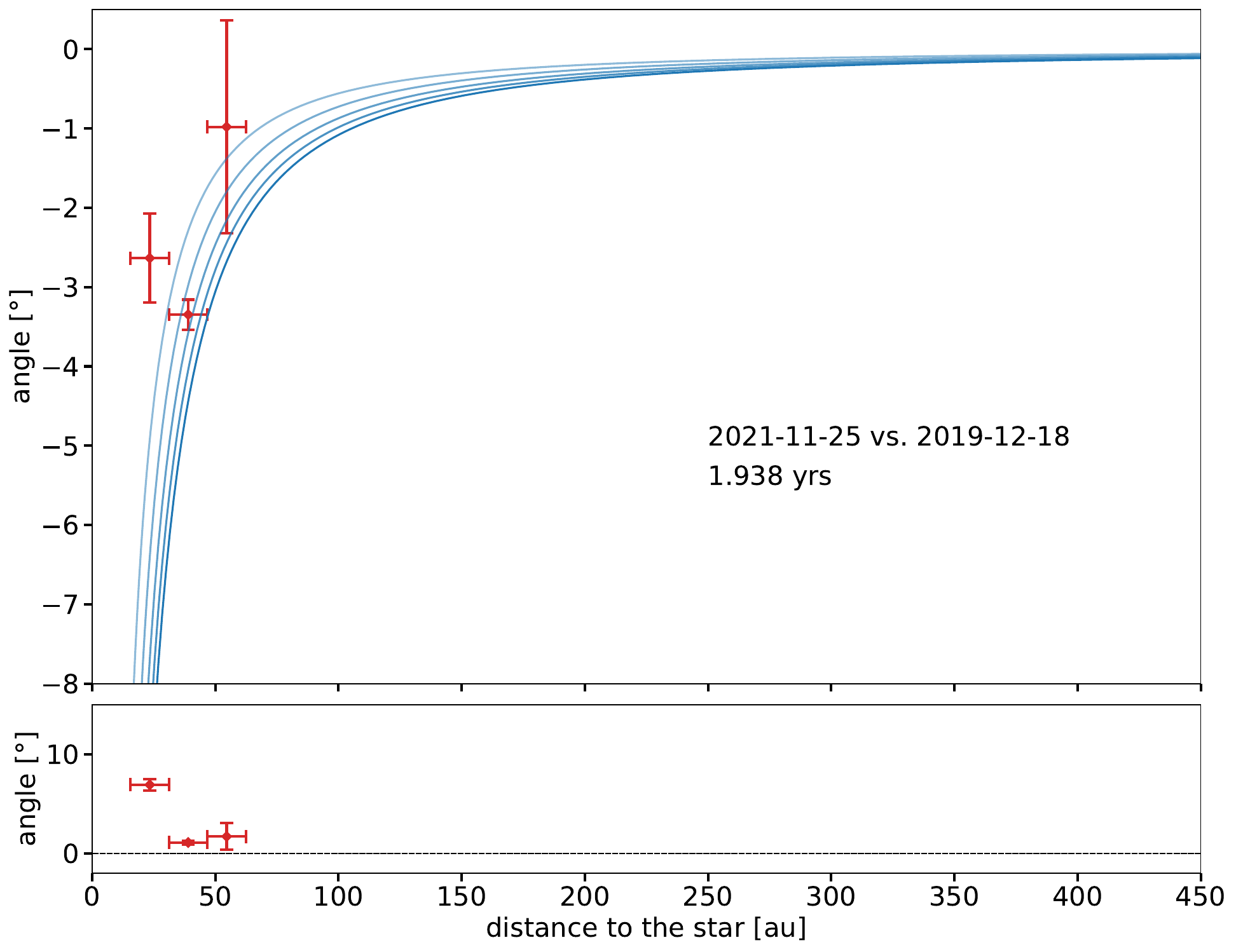}
    \includegraphics[height=3.5cm, clip]{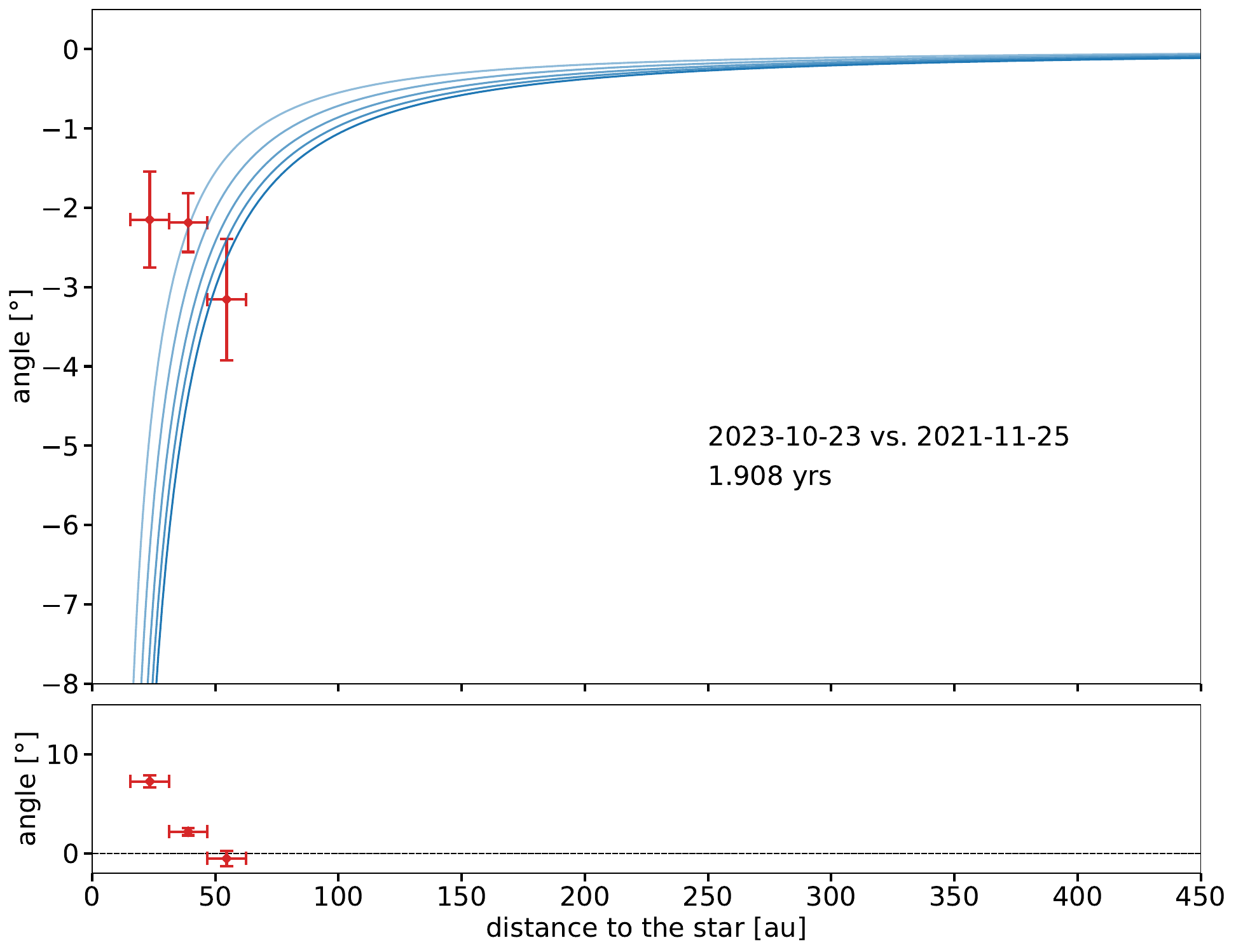}
    \includegraphics[height=3.5cm, clip]{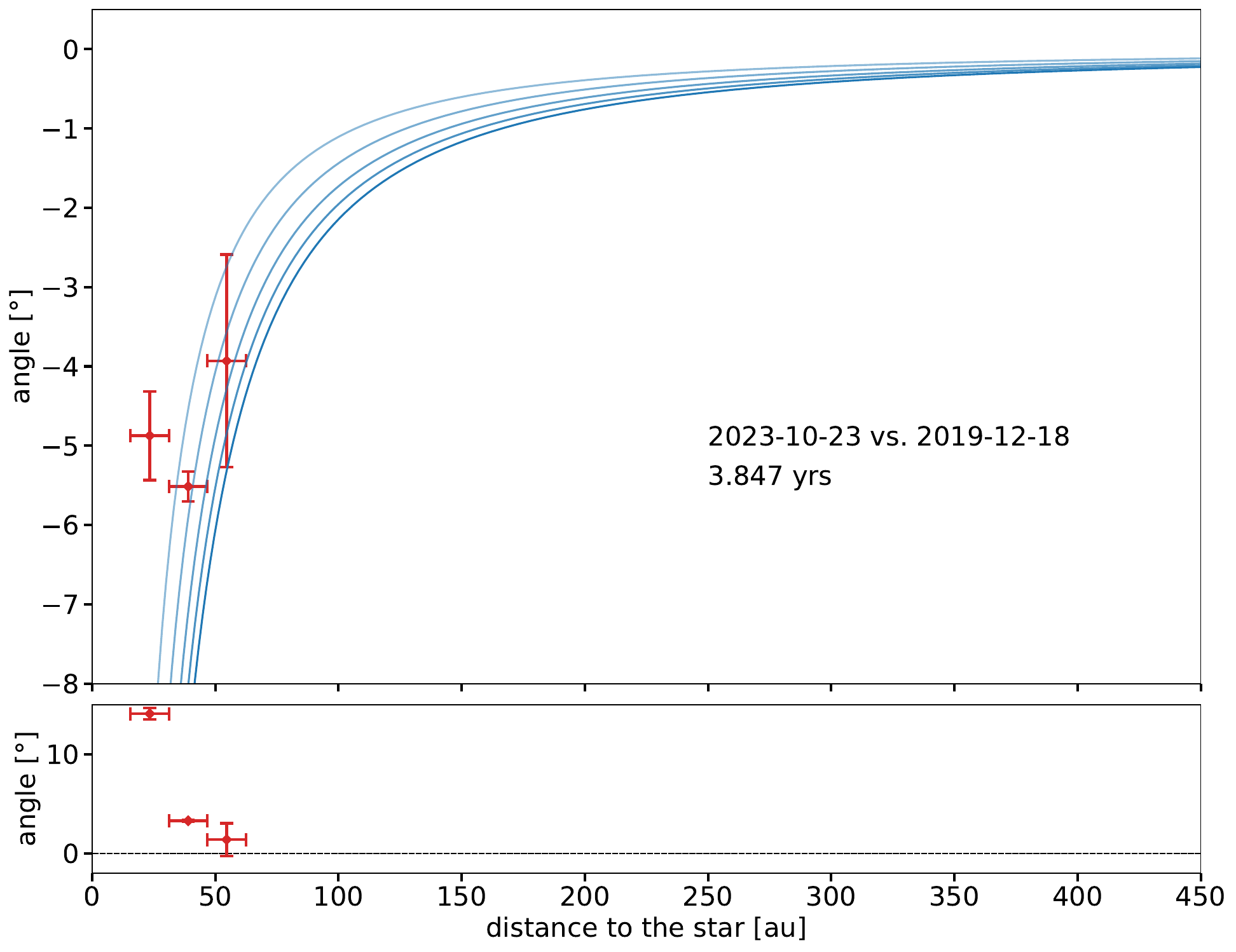}
    \caption{Same as Fig. \ref{fig:rotation_nomask} for the various masks for the spirals and the bridge shown in the left column: from top to bottom: S1a, S1b, S2a, S2b, and the bridge.}
    \label{fig:rotation_withmasks}
\end{figure*}

\begin{figure*}[ht]
    \centering
    \includegraphics[width=12.cm, trim={1cm 0cm 2cm 2cm}, clip]{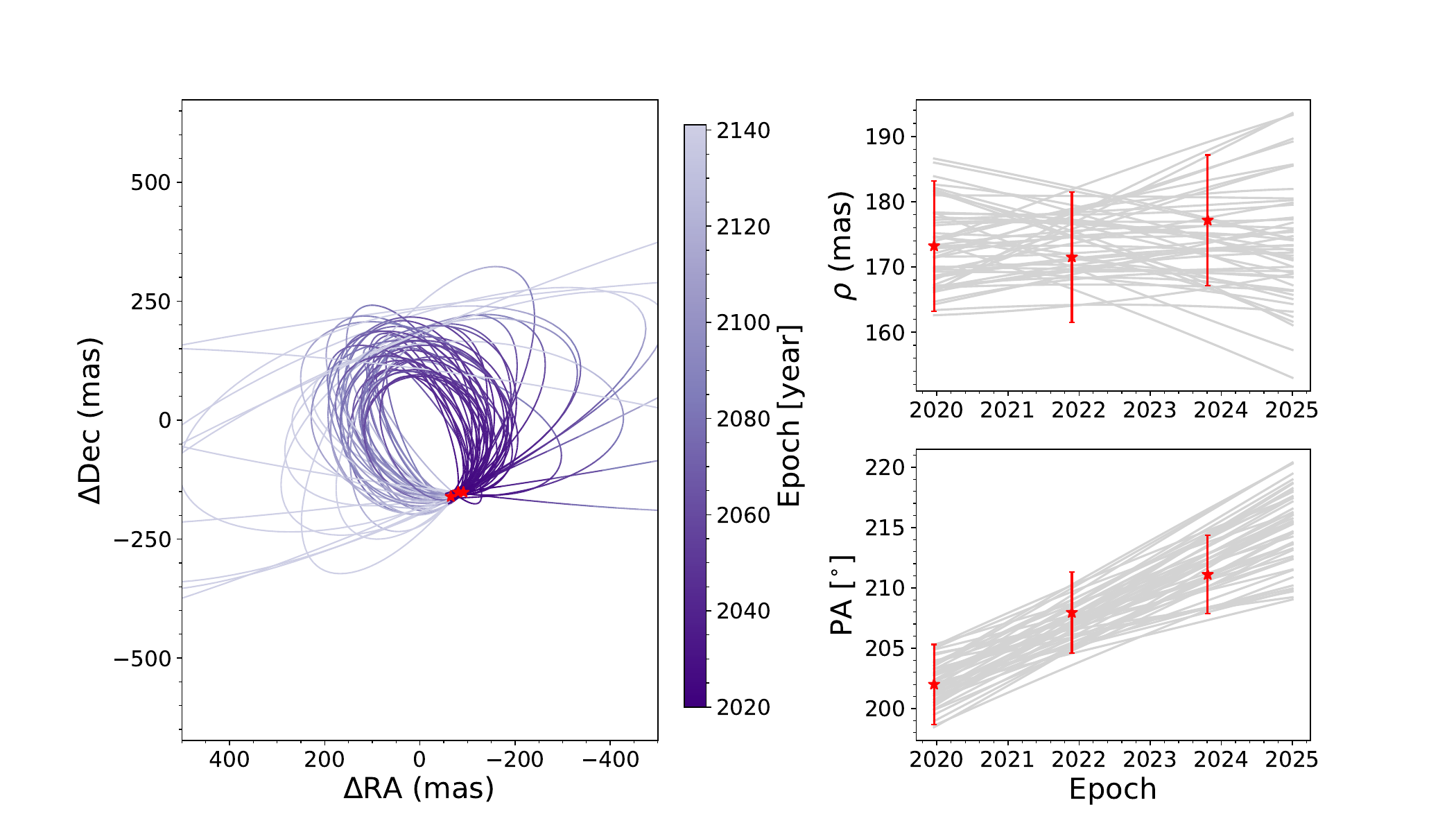}
    \includegraphics[width=12.cm, trim={1cm 0cm 2cm 2cm}, clip]{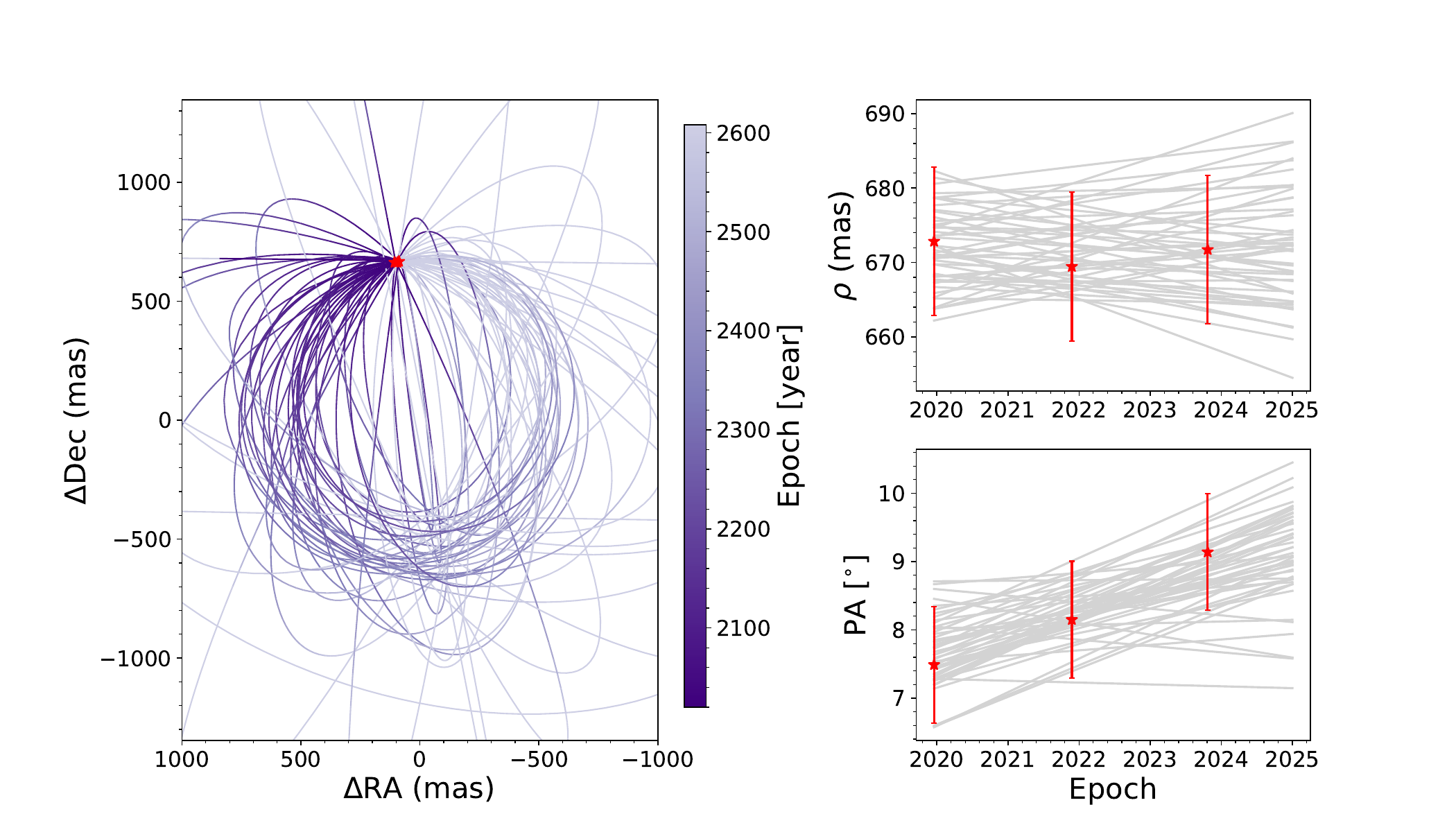}
    \includegraphics[width=12.cm, trim={1cm 0cm 2cm 2cm}, clip]{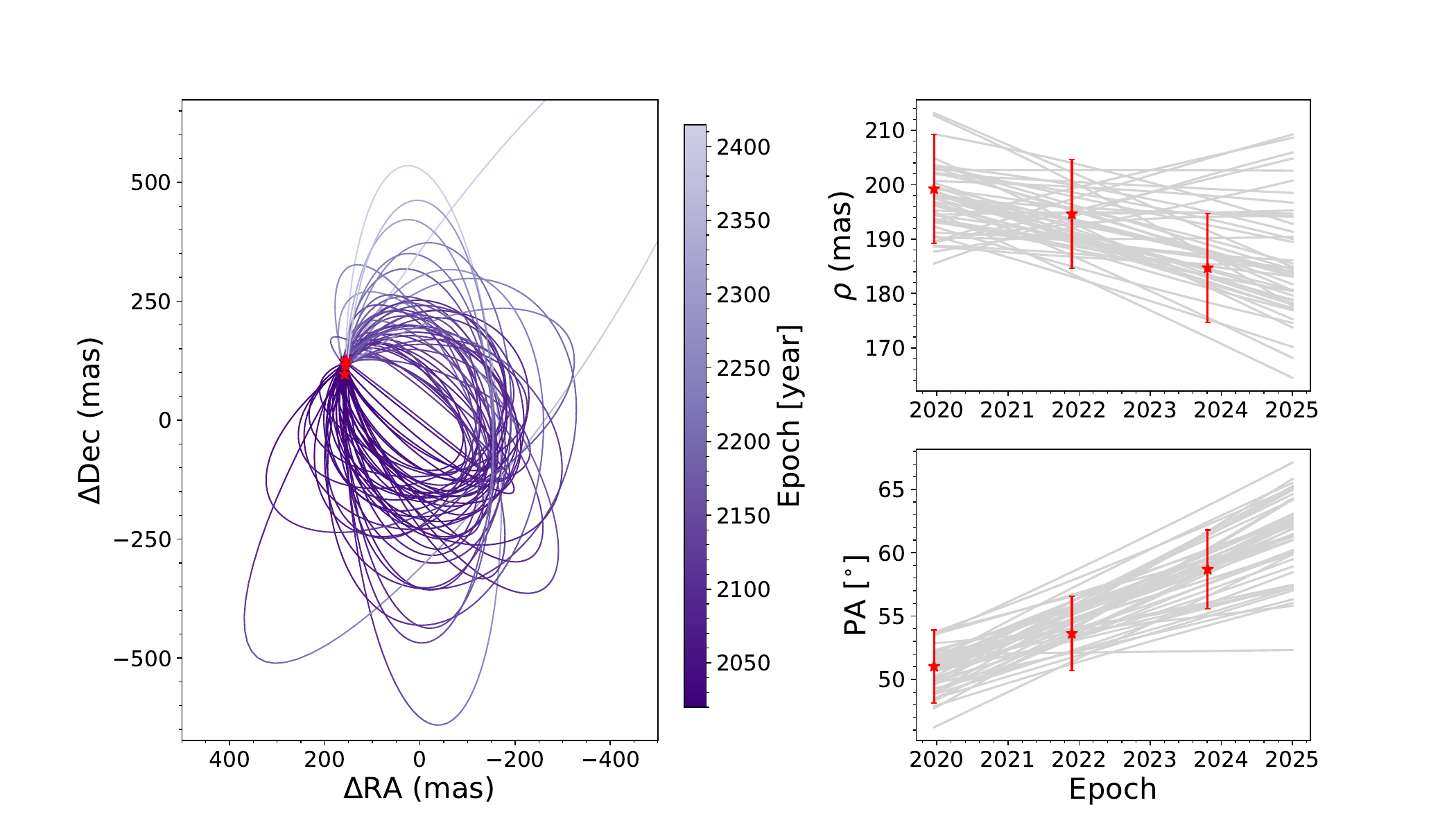}
    \caption{Orbital solutions displayed in the sky plane, calculated with \texttt{orbitize!}, for \texttt{f1},  \texttt{f2} and \texttt{f3} (top to bottom) .}
    \label{fig:astromf1f2f3}
\end{figure*}

\begin{figure*}[ht]
    \includegraphics[width=9cm, trim={0.5cm 0cm 1cm 1cm}, clip]{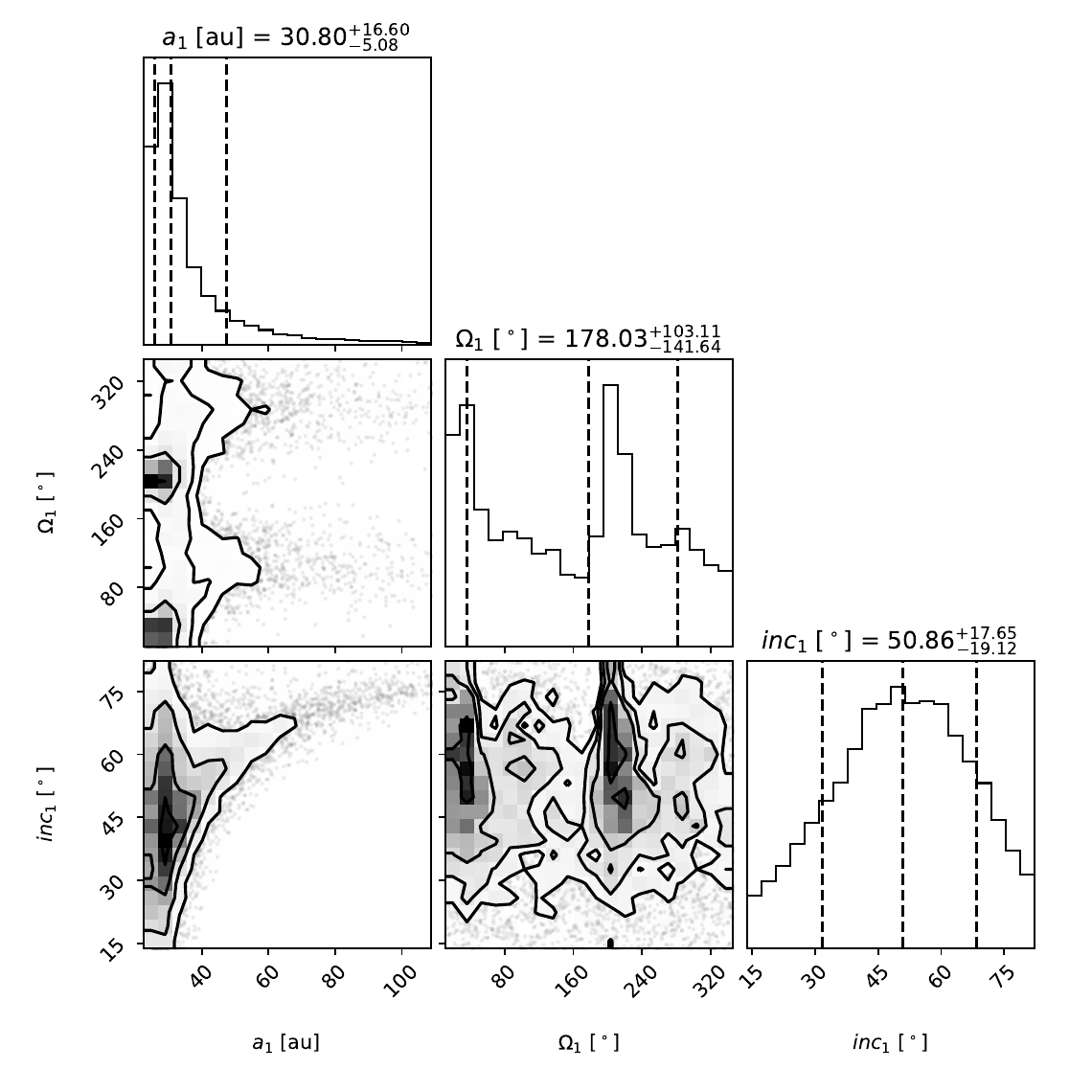}
    \includegraphics[width=9cm, trim={0.5cm 0cm 1cm 1cm}, clip]{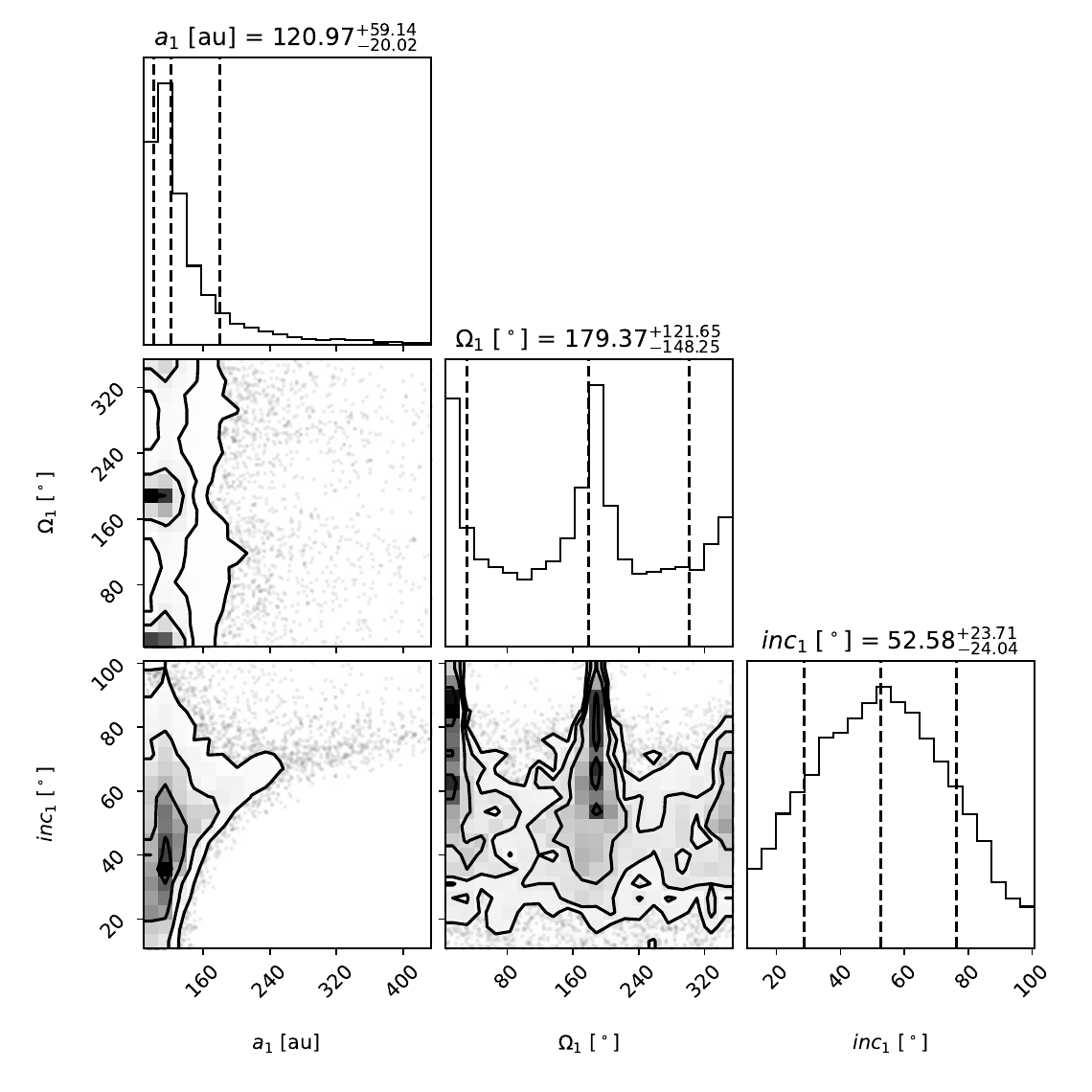}
     \includegraphics[width=9cm, trim={0.5cm 0cm 1cm 1cm}, clip]{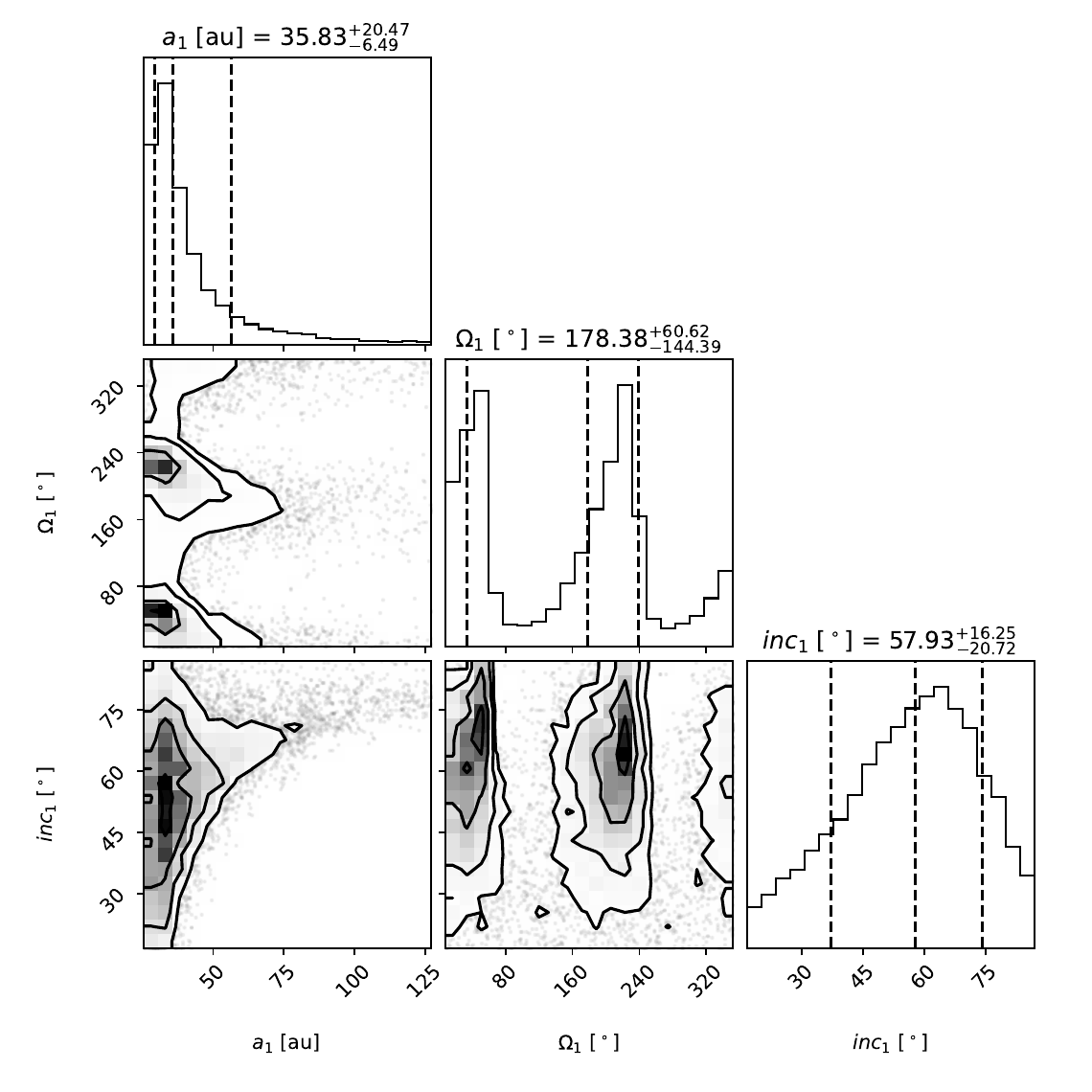}
    \caption{Posterior distributions for the semi-major axis ($a_1$), the position angle of the ascending node ($\Omega_1$), and the inclination ($inc_1$) of the orbital solutions, calculated with \texttt{orbitize!}, for \texttt{f1} (top left),  \texttt{f2} (top right) and \texttt{f3} (bottom left).}
    \label{fig:cornerf1f2f3}
\end{figure*}


\section{Width of radial shadows}
\label{appendix:shadows}

With the objective of understanding whether the shadows could have a planetary origin, we compare the width of the shadow with the predictions of the model by \citet{akansoy_modelling_2025}. A shadow cast by a planet is expected to be thinner than shadows due to an inclined inner disk. Similarly to section \ref{sec:dynamical} we perform the analysis in the disk plane, using both the images of Fig. \ref{fig:irdap_shadows} (Nov. 25th, 2021), and the PCA (PC\#2, Fig. \ref{fig:irdap_shadows_pca}). We focus on the two main shadows \texttt{sh1} and \texttt{sh2} (respectively oriented at $PA\approx275\degb$ and $PA\approx348\degb$ in the sky plane) since they are sufficiently well visible.
We use a rectangular aperture to isolate the shadows from 33 to 74 pixels in the deprojected image coordinates (corresponding to physical distances of $\sim$64 to 144\,au), with a width of 16 pixels ($\sim$31\,au), and fit a Gaussian profile perpendicularly to the shadow. Then, to reproduce the findings of \citet{akansoy_modelling_2025} the width is converted to an angle and scaled with the ratio of the Hill's radius of a planet to the distance of the planet, which corresponds to a factor $({M_p}/{3M_*})^{-1/3}$. We assume an object of 30\,$M_{jupiter}$ (but the dependence with the mass is small). The results are presented in Fig. \ref{fig:shadowwidths}. Although it is difficult to match our width measurements with the model, we end up with values that are qualitatively consistent with those presented in \citet{akansoy_modelling_2025} (their Fig. 7). Therefore, this exercise demonstrates the difficulty to measure the width of a shadow and to unambiguously connect this value to the mass of the source casting the shadow.

\begin{figure}
    \centering
    \includegraphics[width=9cm, trim={2cm 1cm 2cm 2cm}, clip]{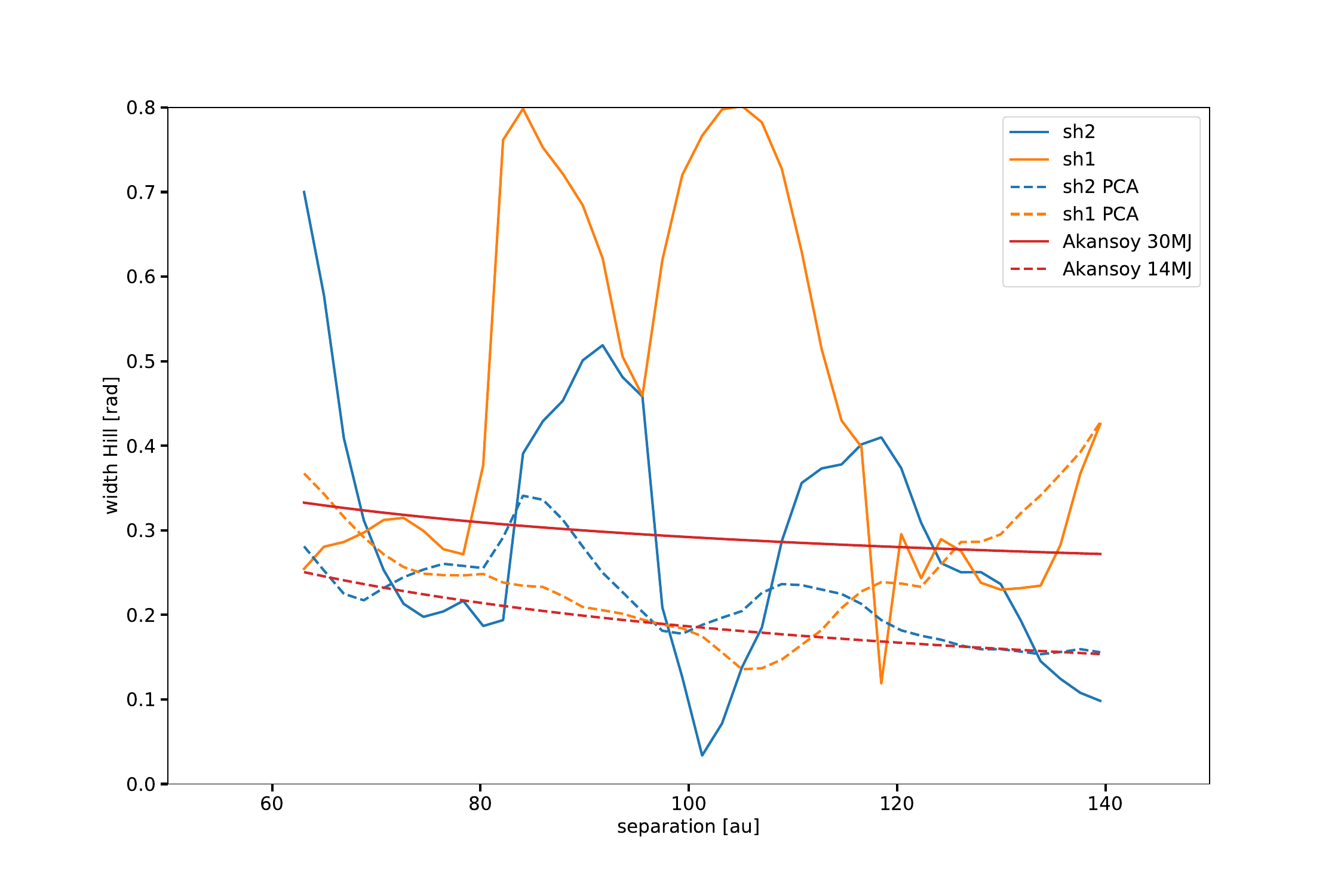}
    \caption{Angular width of the shadows \texttt{sh1} (orange line) and \texttt{sh2} (blue line) versus the separation to star, measured in the disk plane for the second epoch in H band (Nov. 25th, 2021, solid lines) and the second principal component (dashed lines), compared to \citet{akansoy_modelling_2025}. }
    \label{fig:shadowwidths}
\end{figure}


\section{ZIMPOL data reduction}
\anthony{In this section, we present the raw PSFs and coronagraphic images (Fig. \ref{fig:zim_raw}), the measured contrasts (Fig. \ref{fig:zim_contrast}), and the on-sky transmission of the coronagraphic mask  (Fig. \ref{fig:zim_corospider}). }

\begin{figure}[ht]
    \centering
    \includegraphics[width=9cm]{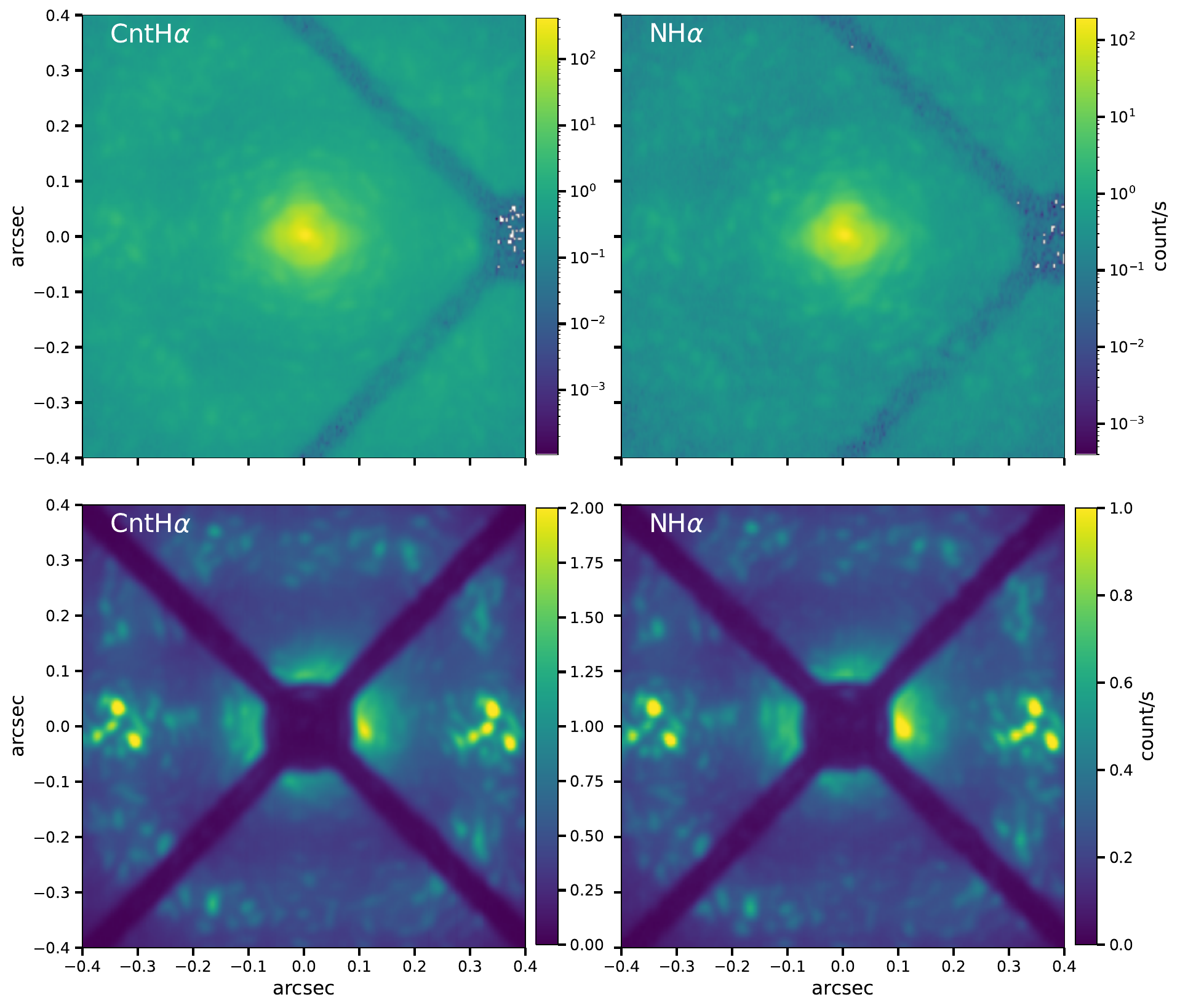}
    \caption{PSFs (log scale) and raw coronagraphic (linear scale) images with ZIMPOL in both filters CntH${\alpha}$ and N\_H$\alpha$. The field of view is $0.8''\times0.8''$. Intensities are in counts per second. Corresponding contrasts are plotted in Fig. \ref{fig:zim_contrast}.}
    \label{fig:zim_raw}
\end{figure}

\begin{figure}[ht]
    \centering
    \includegraphics[width=9cm, trim={2cm 1cm 2cm 2cm}, clip]{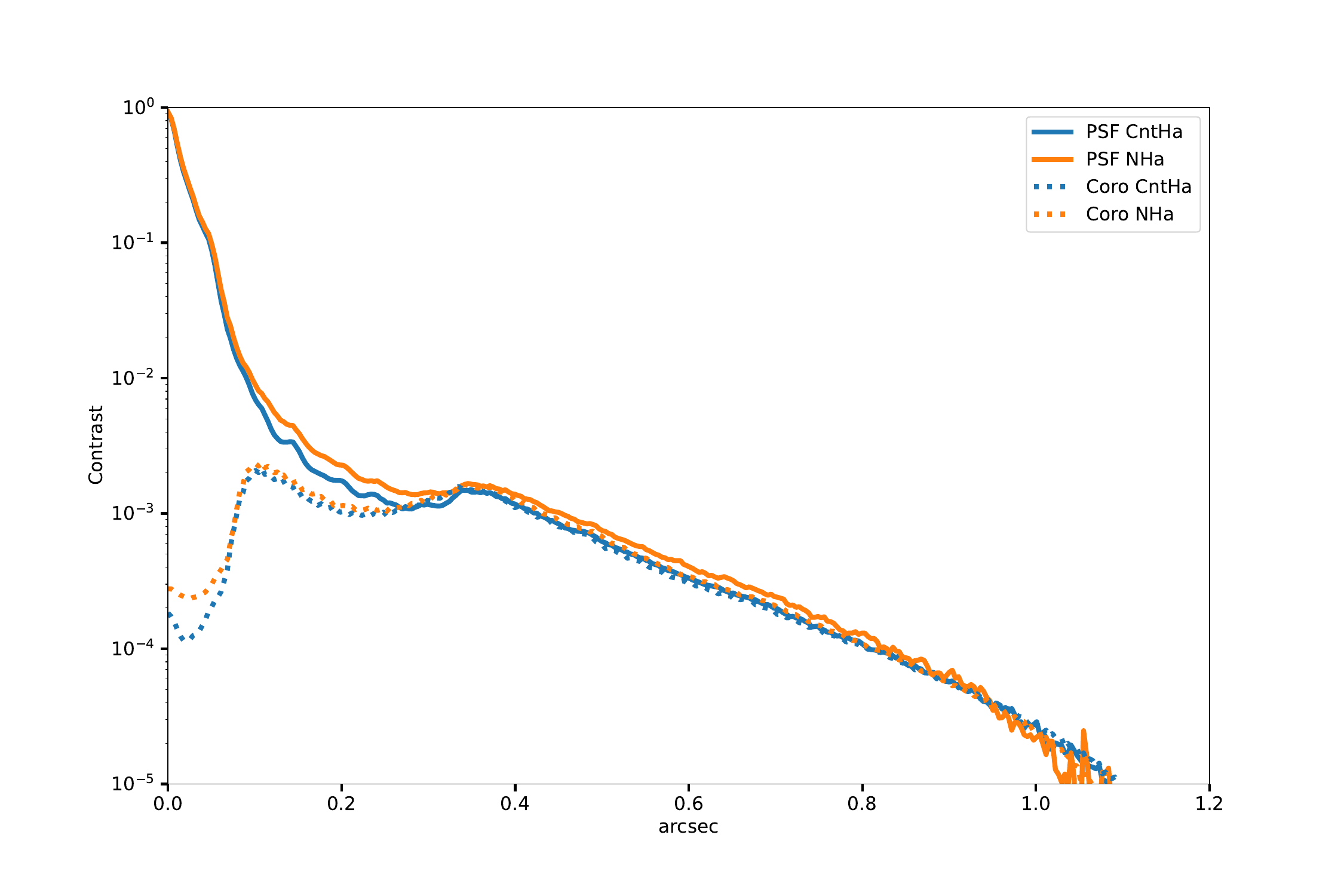}
    \caption{Contrasts measured for the PSF and coronagraphic images displayed in Fig. \ref{fig:zim_raw}.}
    \label{fig:zim_contrast}
\end{figure}

\begin{figure*}[ht]
    \centering
    \includegraphics[width=18cm, clip]{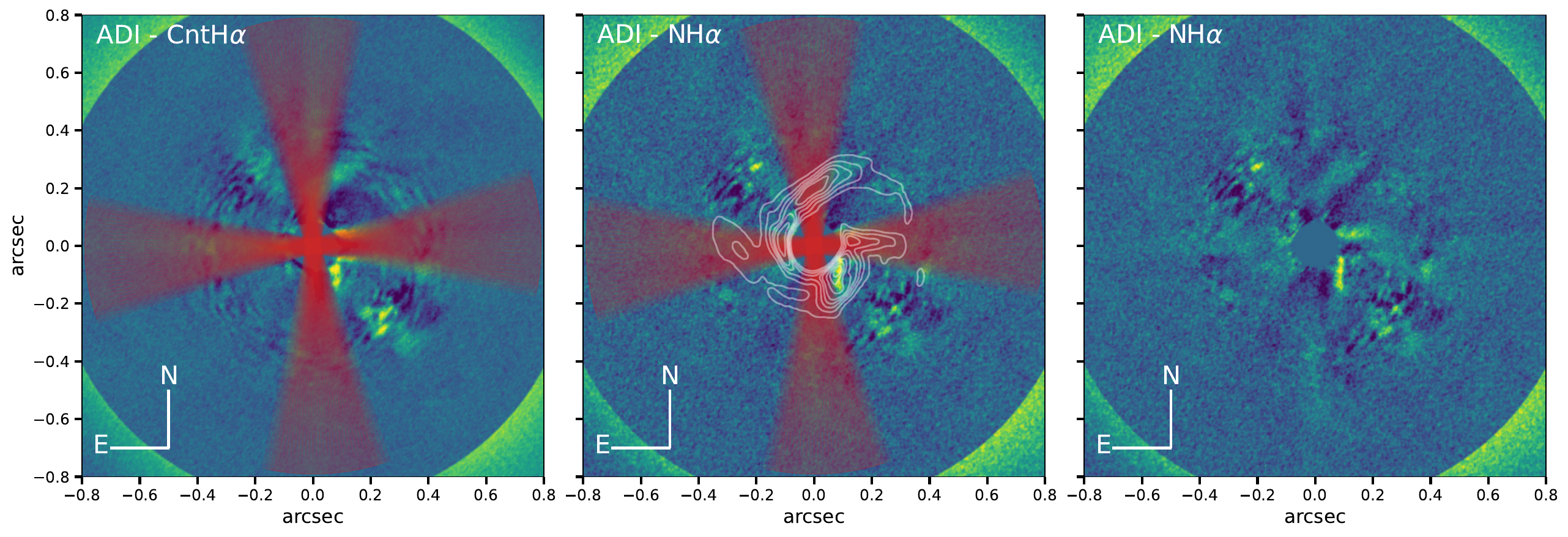}
    \caption{Traces of the spiders coronagraphic mask during the observation (red shaded area) superimposed with the image in filters CntH${\alpha}$ (left) and N\_H$\alpha$ (middle). For comparison we also display the N\_H$\alpha$ alone (right), and the contours of the J band image (middle).}
    \label{fig:zim_corospider}
\end{figure*}


\section{ZIMPOL \Ha photometry}
\label{appendix:zimphotom}

\subsection{Self-subtraction in ZIMPOL ADI images}

The features we are interested in to derive the photometry, in particular \texttt{f1}, the bridge and AB\,Aur\,b, undergo self-subtraction due to the ADI processing. In contrary to a point source, the self-subtraction of an extended source cannot be fully predicted because it depends on the morphology of the source which itself is modified by the self-subtraction. Here, we proceed with a forward modeling approach which applies the very same principal components calculated for the data onto the model image. We assume each feature to be a Gaussian 2D function and we derive the flux ratio before and after the ADI processing in the same apertures used for the photometry. The size of the Gaussian and the amplitude of the self-subtraction ($\tau$) are provided below: \\
- \texttt{f1}: $35\times123$\,mas, $\tau=3.75$\\
- bridge: $123\times53$\,mas, $\tau=4.3$\\
- \texttt{f1} knot: $7\times7$\,mas, $\tau=2.37$\\
- AB\,Aur\,b: $90\times60$\,mas, $\tau=1.44$\\
An example is shown in Fig. \ref{fig:zim_apertures}

\subsection{Flux  in \Ha}
\label{appendix:fluxdensityHa}

To derive thoroughly the H$\alpha$ emission we follow the method exposed in \citet{cugno_search_2019}, instead of using the ASDI image from Fig. \ref{fig:zim_pca}.  It also relies on the calibration of ZIMPOL zero points carried out by \citet{schmid_spherezimpol_2017}. 
The principle is to measure the count rates ($cts$) of a particular feature in the image, then to convert it to flux (erg/s/cm$^2$) for both filters, and to estimate the contribution of the continuum emission in the H$\alpha$ filter to  derive only the flux in the emission line. The latter is the relevant for estimating the accretion luminosity and the mass accretion rate. We describe below the generic approach, in which we first estimate the flux in the two filters according to Eq. 4 of \citet{schmid_spherezimpol_2017} : 

\begin{subequations}
\begin{align}
& F_{\mathrm{CntH{\alpha}}} =  cts_{\mathrm{CntH{\alpha}}}.10^{0.4(am.k_{\mathrm{CntH{\alpha}}}+m_{mode})}.Czp^{cnt}_{\mathrm{CntH{\alpha}}}\\
& F_{\mathrm{NH{\alpha}}} =  cts_{\mathrm{NH{\alpha}}}.10^{0.4(am.k_{\mathrm{NH{\alpha}}}+m_{mode})}.Czp^{line}_{\mathrm{NH{\alpha}}}
\label{eq:fluxNHa}
\end{align}
\end{subequations}
where $Czp^{cnt}$ and $Czp^{line}$ are the zero points in the continuum and in the line, tabulated in \citet{schmid_spherezimpol_2017}. The extinction parameter $k$ varies with wavelength and was measured for Paranal to be 0.085\,mag/airmass, respectively 0.082\,mag/airmass, for filters CntH$\mathrm{\alpha}$, respectively NH$\mathrm{\alpha}$. The airmass is on average 1.85 for PSF observations, and 1.84 for coronagraphic observations. The parameter $m_{mode}$ is a transmission correction factor which depends on the dichroic between the science path and the AO path (-0.23\,mag for DIC\_HA).
To estimate the count rate due to the continuum inside the \filterHa filter, we use Eq. 2 from \citet{cugno_search_2019} which renormalizes the count rate in \filterCnt to the zero point of the continuum in \filterHa:

\begin{equation}
cts_{\mathrm{cnt\_in\_NH{\alpha}}} = cts_{\mathrm{CntH{\alpha}}}\frac{Czp^{cnt}_{\mathrm{CntH{\alpha}}}}{Czp^{cnt}_{\mathrm{NH{\alpha}}}}
\end{equation}
and then subtracting the count rate of the continuum to the total count rate in the \filterHa filter yields:

\begin{equation}
cts_{\mathrm{H{\alpha}}}=    cts_{\mathrm{NH{\alpha}}} - cts_{\mathrm{cnt\_in\_NH{\alpha}}} 
\end{equation}

We can estimate the line to continuum ratio as $cts_{\mathrm{H{\alpha}}}/cts_{\mathrm{NH{\alpha}}}$.
Then, by analogy with Eq. \ref{eq:fluxNHa} the flux  of the H$\alpha$ line reads: 
\setlength{\abovedisplayskip}{12pt}
\begin{equation}
F_{\mathrm{H{\alpha}}} = cts_{\mathrm{H{\alpha}}}.10^{0.4(am.k_{\mathrm{NH{\alpha}}}+m_{mode})}.Czp^{line}_{\mathrm{NH{\alpha}}}
\end{equation}

\subsection{\Ha accretion models}
\label{appendix:accretionmodels}

For each accretion model the accretion luminosity $L_{acc}$ relates to the \Ha luminosity in a log-log relationship (Fig. \ref{fig:comparemodels}), that is either an experimental result or a theoretical result, as follows:

\begin{subequations}
\begin{align}
& \mathrm{log}(L_{acc}^{Al2017}) = (1.13\pm0.05)\times\mathrm{log}(L_{\mathrm{H{\alpha}}})+(1.74\pm0.16) \label{eq:lacc_r2012}
 \\
& \mathrm{log}(L_{acc}^{Ao2021}) = 0.95\times\mathrm{log}(L_{\mathrm{H{\alpha}}})+(1.61\pm0.3) \label{eq:lacc_a2021}
 \\
& \mathrm{log}(L_{acc}^{Sz2020}) = \frac{\mathrm{log}(L_{\mathrm{H{\alpha}}})-38.9}{17.06\pm5.9} \label{eq:lacc_se2020}
\end{align}
\end{subequations}

We note that the dispersions of the accretion luminosity calculated from the $\mathrm{H{\alpha}}$ luminosity is not estimated the same way for the three cases. 
Equation \ref{eq:lacc_se2020} stands for a specific composition which has a gas-to-dust ratio of 100, and a dust mixture of 40\% silicates + 40\% water-ice + 20\% carbon, and a grain size of 1\,$\muup$m, but it has been also derived for 3 other compositions with different coefficients 
\citep[see Tables 3 and 4 in][]{szulagyi_hydrogen_2020}. For each mixture the coefficients can also vary depending on the planet position on its orbit.

\begin{figure}
    \centering
    \includegraphics[width=9cm]{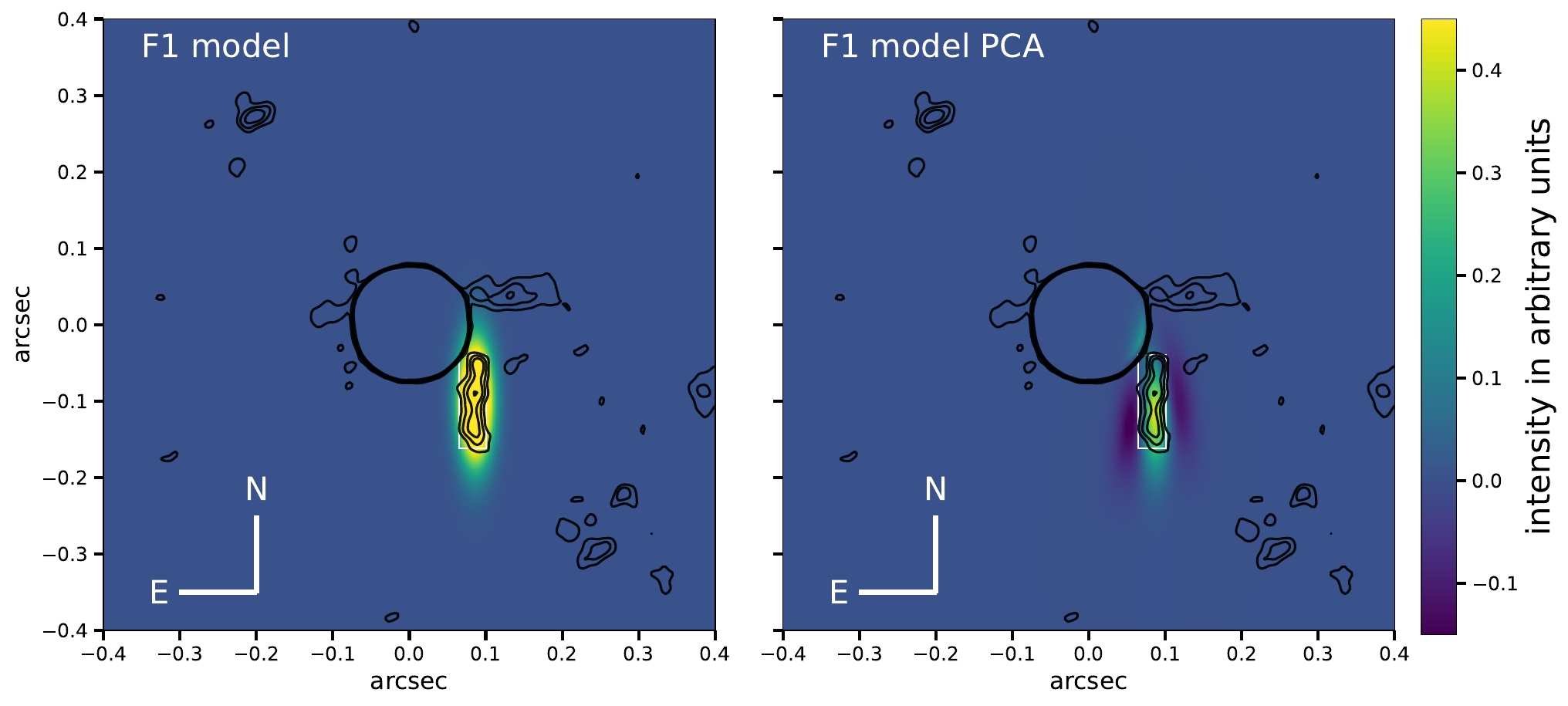}
    \caption{Forward modeling of \texttt{f1}: a Gaussian features matching \texttt{f1} size (left) and the corresponding image with PCA (right). Black contours depict the \filterHa PCA image, and the photometric aperture is also indicated in white.}
    \label{fig:zim_apertures}
\end{figure}

\begin{figure}
    \centering
    \includegraphics[width=9cm]{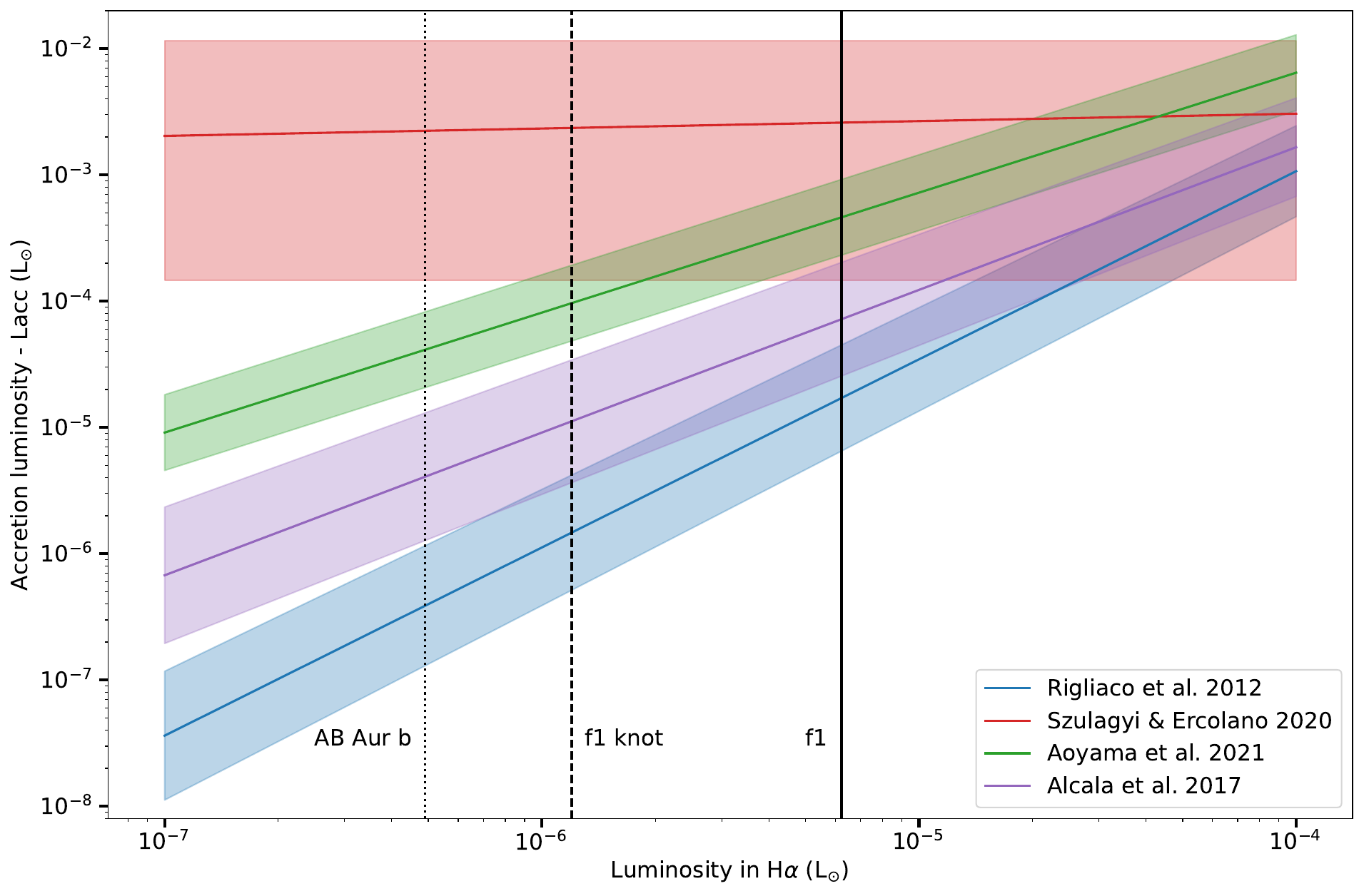}
    \caption{Comparison of models \citep{szulagyi_hydrogen_2020, aoyama_comparison_2021, rigliaco_x-shooter_2012, alcala_x-shooter_2017} predicting the accretion luminosity  as a function of the H$\alpha$ luminosity. Vertical lines stand for the luminosity measured for AB\,Aur\,b, \texttt{f1}, and the knot.}
    \label{fig:comparemodels}
\end{figure}

\begin{table*}
\caption{Photometry and accretion rate. Fluxes are in erg/s/cm$^2$, luminosities in $L_{\odot}$, accretion masses in $M_{Jup}/yr$ and masses in $M_{Jup}$.}
\label{tab:fluxhalpha}
\begin{tabular}{lcccc}     
\hline\hline                      
                            & AB Aur                & \texttt{f1}           & \texttt{f1} knot      &  AB Aur b   \\ 
\hline       \\
$F_{\mathrm{CntH{\alpha}}}$ & $7.10\times10^{-11}$  & $2.00\times10^{-14}$  & $4.38\times10^{-15}$  &  $8.82\times10^{-16}$ \\
$F_{\mathrm{NH{\alpha}}}$   & $5.78\times10^{-11}$  & $1.30\times10^{-14}$  & $2.63\times10^{-15}$  &  $8.57\times10^{-16}$  \\
$F_{\mathrm{H{\alpha}}}$    & $4.08\times10^{-11}$  & $8.22\times10^{-15}$  &  $1.58\times10^{-15}$ &  $6.46\times10^{-16}$ \\
$L_{\mathrm{H{\alpha}}}$    & 0.03                  & $6.23\times10^{-6}$   &  $1.20\times10^{-6}$  &  $4.90\times10^{-7}$ \\
$L_{acc}^*$                 & 4.32                  & -                     & -                     &  - \\
$\dot{M}_{acc}^*$           & $1.80\times10^{-7}$   & -                     & -                     &  - \\
\hline       
$L_{acc}^{Al2017}$          & -                     & $(2.55 -  20.33)\times10^{-5}$    & $(3.65 -  34.25)\times10^{-6}$ & $(1.27 - 13.05)\times10^{-6}$\\
$\dot{M}_{acc}^{Al2017}$    & -                     & $(5.61 -  44.63)\times10^{-7}$     & $(0.80 -  7.53)\times10^{-7}$ & $(2.79 - 28.64)\times10^{-8}$\\
$M_{1Myr}^{Al2017}$         & -                     & $0.56 - 4.46$                     & $0.08 - 0.75$                 & $0.03 - 0.28$ \\
\hline       
$L_{acc}^{Ao2021}$          & -                     & $(2.31 - 9.22)\times10^{-4}$      & $(0.48- 1.93)\times10^{-4}$  & $(2.07 - 8.24)\times10^{-5}$ \\
$\dot{M}_{acc}^{Ao2021}$    & -                     & $(5.08 - 20.25)\times10^{-6}$     & ($1.06 - 4.23)\times10^{-6}$ & $(4.54 - 18.08)\times10^{-7}$\\
$M_{1Myr}^{Ao2021}$         & -                     & $5.09 - 20.25$                    & $1.06 - 4.23$                & $0.45 - 1.81$\\
\hline       
$L_{acc}^{Sz2020}$          & -                     & $(1.11 - 119.95)\times10^{-4}$    & $(0.96 -   111.64)\times10^{-4}$  & $(8.89 - 1073.82)\times10^{-5}$\\
$M^{Sz2020}$                & -                     & $6.36 - 10.33$                    & $6.10 - 9.90$                       & $5.96 - 9.67$\\
\hline       
\end{tabular}
\end{table*}

\end{appendix}

\end{document}